\documentclass[english,a4paper,twoside,openright,12pt, preprint]{aastex}
\usepackage[english]{babel}
\usepackage{natbib}
\usepackage[nomarkers,notablist,nofiglist]{endfloat}
\usepackage{longtable}
\usepackage{lscape}
\usepackage{multirow}
\usepackage{graphicx}
\usepackage{color}
\usepackage[figuresright]{rotating}
\usepackage{changepage}
\usepackage{booktabs}
\usepackage[citecolor=blue]{hyperref}
\usepackage{gensymb}
\hypersetup {colorlinks=true}
\newcommand{\avg}[1]{\left< #1 \right>} 
 \usepackage{pdf14}

\slugcomment{To appear in Astronomical Journal}

\shorttitle{Light Curves of 51 SN II}
\shortauthors{Galbany et al.}

\begin{document}

\title{$UBVRIz$ Light Curves of 51 Type II Supernovae}

\author{Llu\'is Galbany}
\affil{Millennium Institute of Astrophysics, Universidad de Chile, Chile\\
Departamento de Astronom\'ia, Universidad de Chile, Camino El Observatorio 1515, Las Condes, Santiago, Chile}
\author{Mario Hamuy}
\affil{Departamento de Astronom\'ia, Universidad de Chile, Camino El Observatorio 1515, Las Condes, Santiago, Chile\\
Millennium Institute of Astrophysics, Universidad de Chile, Chile}
\author{Mark M. Phillips}
\affil{Carnegie Observatories, Las Campanas Observatory, Casilla 60, La Serena, Chile.}
\author{Nicholas B. Suntzeff}
\affil{Department of Physics and Astronomy, Texas A\&M University, College Station, TX 77843, USA\\
The George P. and Cynthia Woods Mitchell Institute for Fundamental Physics and Astronomy, College Station, TX 77845}
\author{Jos\'e Maza}
\affil{Departamento de Astronom\'ia, Universidad de Chile, Camino El Observatorio 1515, Las Condes, Santiago, Chile}
\author{Thomas de Jaeger, Tania Moraga, Santiago Gonz\'alez-Gait\'an}
\affil{Millennium Institute of Astrophysics, Universidad de Chile, Chile\\
Departamento de Astronom\'ia, Universidad de Chile, Camino El Observatorio 1515, Las Condes, Santiago, Chile}
\author{Kevin Krisciunas} 
\affil{George P. and Cynthia Woods Mitchell Institute for Fundamental Physics \& Astronomy, Texas A. \& M. University, Department of Physics \& Astronomy, 4242 TAMU, College Station, TX 77843, USA}
\author{Nidia I. Morrell, Joanna Thomas-Osip} 
\affil{Carnegie Observatories, Las Campanas Observatory, Casilla 60, La Serena, Chile.}
\author{Wojtek Krzeminski}
\affil{N. Copernicus Astronomical Center, ul. Bartycka 18, 00-716 Warszawa, Poland}
\author{Luis Gonz\'alez, Roberto Antezana, Marina Wischnjewski$^{\dagger}$}
\affil{Departamento de Astronom\'ia, Universidad de Chile, Camino El Observatorio 1515, Las Condes, Santiago, Chile}
\author{Patrick McCarthy}
\affil{The Observatories of the Carnegie Institution for Science, 813 Santa Barbara Street, Pasadena, CA 91101, USA}
\author{Joseph P. Anderson}
\affil{European Southern Observatory, Alonso de Cordova 3107, Vitacura, Casilla 19001, Santiago, Chile}
\author{Claudia P. Guti\'errez}
\affil{Millennium Institute of Astrophysics, Universidad de Chile, Chile\\
Departamento de Astronom\'ia, Universidad de Chile, Camino El Observatorio 1515, Las Condes, Santiago, Chile\\
European Southern Observatory, Alonso de Cordova 3107, Vitacura, Casilla 19001, Santiago, Chile}
\author{Maximilian Stritzinger} 
\affil{Department of Physics and Astronomy, Aarhus University, Denmark}
\author{Gast\'on Folatelli}
\affil{Instituto de Astrof\'isica de La Plata (IALP, CONICET), Argentina}
\author{Claudio Anguita$^{\dagger}$}
\affil{Departamento de Astronom\'ia, Universidad de Chile, Camino El Observatorio 1515, Las Condes, Santiago, Chile}
\author{Gaspar Galaz}
\affil{Instituto de Astrof\'isica, Pontificia Universidad Cat\'olica de Chile.}
\author{Elisabeth M. Green, Chris Impey}
\affil{Steward Observatory, University of Arizona, 933 N. Cherry Avenue, Tucson, AZ 85721-0065}
\author{Yong-Cheol Kim}
\affil{Astronomy Department, Yonsei University, Seoul, Korea}
\author{Sofia Kirhakos}
\affil{Cerro Tololo Inter-American Observatory, National Optical Astronomy Observatories, Casilla 603, La Serena, Chile\\
Instituto de Pesquisas Espaciais, INPE, Caixa Postal 515, 12201 S\~ao Jos\'e dos Campos, SP, Brazil}
\author{Mathew A. Malkan}
\affil{Astronomy Division, Dept of Physics \& Astronomy, UCLA, LA, CA 90095-1547}
\author{John S. Mulchaey}
\affil{The Observatories of the Carnegie Institution for Science, 813 Santa Barbara Street, Pasadena, CA 91101, USA}
\author{Andrew C. Phillips}
\affil{University of California Observatories}
\author{Alessandro Pizzella}
\affil{Dipartimento di Fisica e Astronomia ``G. Galilei", Universit\`a di Padova, vicolo dell'Osservatorio 3, I-35122 Padova, Italy}
\author{Charles F. Prosser$^{\dagger}$}
\affil{National Optical Astronomy Observatories, 950 North Cherry Avenue, P.O. Box 26732, Tucson, AZ 85726}
\author{Brian P. Schmidt}
\affil{Research School of Astronomy and Astrophysics, The Australian National University, Canberra, ACT 2611, Australia\\
ARC Centre of Excellence for All-sky Astrophysics (CAASTRO)} 
\author{Robert A. Schommer$^{\dagger}$}
\affil{Cerro Tololo Inter-American Observatory, National Optical Astronomy Observatory, La Serena, Chile}
\author{William Sherry}
\affil{Eureka Scientific, Inc., 2452 Delmer Street Suite 100, Oakland, CA 94602-3017}
\author{Louis-Gregory Strolger}
\affil{Space Telescope Science Institute, Science Mission Office, 3700 San Martin Drive, Baltimore MD 21218, USA}
\author{Lisa A. Wells}
\affil{Canada-France-Hawaii Telescope Corp., 64-1238 Mamalahoa Highway, Kamuela HI, 96743}
\and
\author{Gerard M. Williger}
\affil{Dept. of Physics \& Astronomy, Univ. Louisville, Louisville KY 40292 USA}

\email{lgalbany@das.uchile.cl}

\begin{abstract}
We present a compilation of $UBVRIz$ light curves of 51 type II supernovae discovered during the course of four different surveys during 1986 to 2003: the Cerro Tololo Supernova Survey, the Cal\'an/Tololo Supernova Program (C\&T), the Supernova Optical and Infrared Survey (SOIRS), and the Carnegie Type II Supernova Survey (CATS).
The photometry is based on template-subtracted images to eliminate any potential host galaxy light contamination, and calibrated from foreground stars. 
This work presents these photometric data, studies the color evolution using different bands, and explores the relation between the magnitude at maximum brightness and the brightness decline parameter ($s$) from maximum light through the end of the recombination phase.
This parameter is found to be shallower for redder bands and appears to have the best correlation in the $B$ band.
In addition, it also correlates with the plateau duration, being thus shorter (longer) for larger (smaller) $s$ values.
\end{abstract}


\let\thefootnote\relax\footnote{$^{\dagger}$ Deceased.}


\section{Introduction}

It is widely accepted that stars born with masses higher than $\sim$8~$M_\odot$ explode as core-collapse supernovae (CCSN) after some tens of millions of years of evolution. 
At the end of their lives stars born with $\sim$8-9~$M_\odot$ end up with oxygen-neon-magnesium core while higher mass stars end up forming an iron core. 
In both cases the core grows up to reach the Chandrasekhar mass near 1.4 $M_\odot$, and at this point 
the electron degeneracy pressure becomes insufficient to balance gravity and the core is bound to gravitational collapse. 
Collapse is stimulated by partial photo-disintegration of Fe-group nuclei into alpha particles, and by electron capture on protons emitting neutrinos. As a result there is a decrease of the electron density, and hence the pressure at the center of the star is reduced, accelerating the collapse.
This sequence of events is followed by core bounce and subsequently the ejection of the star's envelope, presumably due to energy deposited by neutrinos created in the proto-neutron core (see \citealt{1990SvA....34..163I, 2000Natur.403..727B,2007PhR...442...38J,2012ARNPS..62..407J} for reviews about the explosion mechanisms).

Early-time spectra of CCSN show great diversity \citep{1997ARA&A..35..309F}. 
While the Type~II SN (hereafter SN~II) group consists of spectra dominated by prevalent Balmer spectral features, the Type~I class is characterized by the lack of conspicuous Balmer  features. 
The Type I class is further subdivided according to the presence of He in the spectrum (SN~Ib) or no He lines (SN~Ic). 

The spectral differences among CCSN are thought to be due to the relative ability of SN progenitors to retain their outermost envelopes of unprocessed Hydrogen (H) or Helium (He). 
In this scenario SN~II events, which have the least massive progenitors, are those able to retain a significant fraction of their outer layers prior to explosion.
On the other hand, SN~Ib/c most likely originate from massive stars that lose their H envelope through stellar winds \citep{1993ApJ...411..823W}, mass transfer to a companion star \citep{1995PhR...256..173N}, enhanced mixing \citep{2013ApJ...773L...7F}, or through a combination of these processes.
Supporting evidence for this scenario is available from the detection of SN progenitors in nearby galaxies \citep{2015PASA...32...16S} and statistical analysis of the proximity of CCSN to star forming regions in their host galaxies \citep{2012MNRAS.424.1372A, 2014A&A...572A..38G}.

Historically, SN~II have been sub-classified according to their photometric properties. The majority shows a phase of $\sim$80 days with a ``plateau" of nearly constant luminosity (hence, historically referred as SN~IIP), while a smaller fraction of ``linear'' SN II show a steep initial decline (SN~IIL).
Recent studies have questioned this subdivision and argue that nature provides a continuous 
sequence of objects, ranging from pure `slow decliners' to `fast decliners' \citep{2014ApJ...786...67A, 2015ApJ...799..208S}.
An even smaller fraction of SN~II undergo interaction of their vastly expanding ejecta with  circumstellar material, which can manifest as strong narrow H emission lines in the spectrum, and lead to significant photometric diversity (SN IIn, \citealt{1990MNRAS.244..269S, 2013A&A...555A..10T}). 

SN are not only important in the chemical enrichment of the Universe and the shaping of  galaxies, but also serve as accurate cosmological distance indicators.
Over the past 25 years our group has been systematically studying and collecting photometric and spectroscopic data of all SN  types over the course of the following surveys: 
1) the Cerro Tololo Supernova Survey led by M.M.P. and N.B.S between 1986-1996,
2) the Cal\'an/Tololo Supernova Program (C\&T) led by M.H., J.M., M.M.P, and N.B.S between 1989-1993 \citep{1993AJ....106.2392H}, 
3) the Supernova Optical and Infrared Survey (SOIRS) led by M.H. between 1999-2000 \citep{2001PhDT.......173H}, and 
4) the Carnegie Type II Supernova Survey (CATS) led by M.H., M.M.P, and N.B.S. between 2002-2003.

The purpose of this paper is to report photometric observations of 51 SN~II obtained by these four surveys (excluding SN~1987A's data that were published in great detail by \citealt{1990AJ.....99.1133P} and \citealt{1990AJ.....99.1146H}, and SN 1990E's photometry which was presented in \citealt{1993AJ....105.2236S}), in order to 
 make this dataset available to the community.
This dataset will undoubtedly  contribute to an expanded  understanding of SN~II and improved methods for obtaining precise distances.  
Near-infrared photometry for the current sample will be presented in a separate paper.
These data have been used previously for the study of specific objects \citep{1994AJ....107.1444S, 2009AJ....137...34K, 2009ApJ...703.1612H, 2009ApJ...703.1624M, 2011ApJ...729...61B, 2003MNRAS.338..711Z, 2015MNRAS.450.3137T}.
The subsample of SN~IIP has been used for the determination of distances using the ``Expanding Photosphere Method'' \citep{1994ApJ...432...42S, 2001ApJ...558..615H, 2009ApJ...696.1176J} and the ``Standardized Candle Method'' \citep{2002ApJ...566L..63H, 2004mmu..sympE...2H, 2010ApJ...715..833O, 2014AJ....148..107R}, and for the determination of bolometric corrections \citep{2009ApJ...701..200B}. 
Other studies that have relied on some of the objects in this sample also include: (1) \cite{2003ApJ...582..905H} examined the observed and physical properties of SN II using both photometry and spectroscopy of a selection of 24 SN II; (2)
\cite{ 2014ApJ...786...67A}  performed a characterization of the $V$-band light curves of an expanded sample of SN II; 
and (3) \cite{2014ApJ...786L..15G} have correlated those properties with the H$\alpha$ feature of their spectra.

This paper is organized as follows:  \S~\ref{obs} summarizes our optical observations and describes the data reduction procedures;
 \S~\ref{ana} shows an analysis of the photometric properties of the SN II light-curves, including colors, absolute magnitudes, and the brightness decline parameter $s$.
Finally, a summary and the final conclusions are presented in  \S~\ref{summ}.


\section{Observations}\label{obs}

\begin{figure*}
\centering
\includegraphics*[trim=0.0cm 0.0cm 0.0cm 0.0cm, clip=true,width=\hsize]{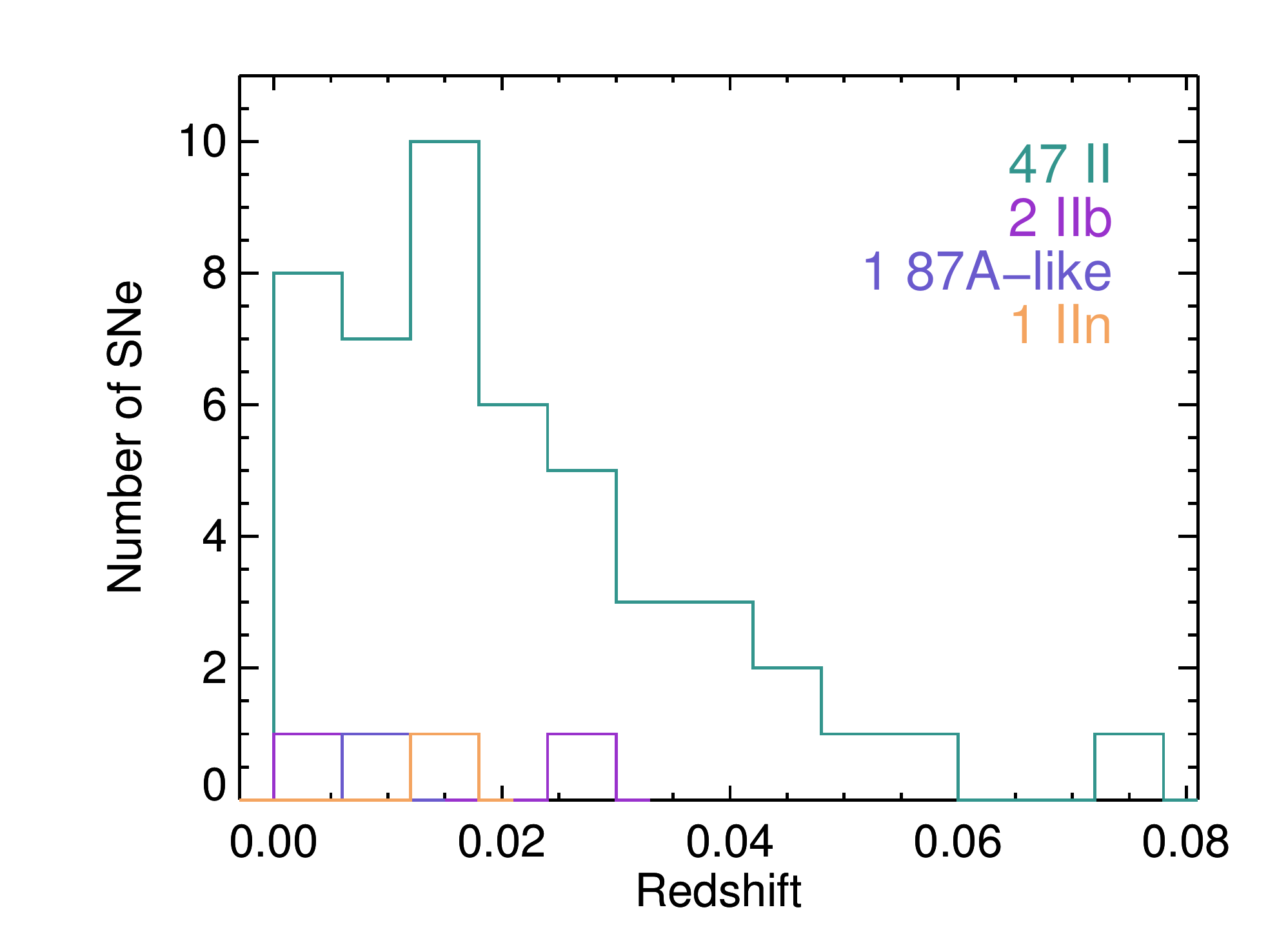}
\caption{Redshift distribution of the 51 SN II presented in this work, separated in normal SN II, IIb, and peculiar events. The median value of the distribution is 0.017, the average value is 0.021, and the standard deviation is 0.016.}
\label{fig.z}
\end{figure*}

A list of the SN~II used in this study is presented in Table \ref{snlist.tab}.
The table includes the following information:
the SN designation and its host-galaxy names; the host-galaxy type;
the SN equatorial coordinates;
the heliocentric redshift of the host galaxy; 
the Galactic extinction, $E(B-V)_{MW}$, from the \cite{2011ApJ...737..103S} dust maps;  
the distance modulus (see  \S~\ref{absmag});
and the survey under which the SN was observed. 
Besides the objects discovered over the course of the C\&T and SOIRS programs using photographic plates (with the Cerro Tololo Curtis-Schmidt Camera and the Cerro El Roble Maksutov Camera of the University of Chile), we also include in the list of follow-up targets SN discovered by others and reported to the IAU Circulars. 
Discovery and classification references for the 51 SN II are listed in Table \ref{refs.tab}.
They are all nearby objects ($z \lesssim$ 0.08, see Figure \ref{fig.z}), selected for our follow-up based on their relatively high apparent brightness and convenient location in the sky (declination $\lesssim$~25$\degree$ North).

As soon as we were notified of a discovery, and whenever we had telescope time allocated to us,  detailed follow-up observations were initiated using various telescopes located at the Cerro Tololo Inter-American Observatory (CTIO), the University of Arizona's Steward Observatory (SO), the Las Campanas Observatory (LCO) of the Carnegie Institution of Science, and the European Southern Observatory (ESO) at La Silla and Paranal.


\subsection{Photometry}

The first object in our list is SN~1986L and it is the only SN  observed with photoelectric techniques (by M.M.P and S.K., using the CTIO 0.9m equipped with a photometer  and  $B$ and $V$ filters).
The remaining SN were observed using a variety of telescopes equipped with CCD detectors and $UBV(RI)_{KC}z$ filters, as indicated in Table \ref{telescopes.tab}.
The observational techniques employed by the C\&T project are presented by \citet{1993AJ....106.2392H}, and the photometric reductions in \cite{1996AJ....112.2408H}. 
The observations and data reduction during the SOIRS project are explained in \cite{2001PhDT.......173H}, whereas the techniques employed during the CATS project can be found in \cite{2009ApJ...703.1612H}. 
We also refer the reader to \citet{2006PASP..118....2H}, which describes the observational procedures of the {\em Carnegie Supernova Project} (CSP), which are nearly identical to those of CATS. 
In fact, CATS was a precursor to the CSP initiated in 2004 with the aim to study SN of all types. 
One main difference between CATS and the original format of the CSP is the latter uses SDSS $ugri$ filters in addition to  Johnson $B$ and $V$ filters for optical imaging \citep{2011AJ....142..156S}.

All photometric reductions were performed with customized IRAF$^1$\let\thefootnote\relax\footnote{$^1$ IRAF is distributed by the National Optical Astronomy Observatories, which are operated by the Association of Universities for Research in Astronomy, Inc., under cooperative agreement with the National Science Foundation.} scripts.
In brief, the photometric reductions begin by subtracting host-galaxy template images from the SN+galaxy images. The templates are high signal-to-noise images (in each filter) of the SN field obtained under good seeing conditions after the SN has faded from detection. 
As a result of this procedure the SN generally end up lying on a smooth background,  allowing us to reliably measure the SN flux with no contamination from the host-galaxy background.
The next step is to compute differential photometry of the SN with respect to a local sequence of stars, calibrated relative to standard star observations obtained over multiple photometric nights.
For this, we measured instrumental magnitudes of the SN and the local sequence stars via PSF fitting when the SN was faint, or simple aperture photometry when the object was bright. 
The transformation of the instrumental magnitudes to the standard $UBV(RI)_{KC}z$ system assume a linear term in magnitude, a color term, and a photometric zeropoint,
\begin{equation}
X = x + CT_X + ZP_X
\end{equation}
where $X$ represents the standard system magnitude, $x$ the instrumental magnitude, the color term $CT_X$ is an average measured over many photometric nights for each telescope/CCD/filter combination, and the zeropoint $ZP_X$ is a fitting parameter determined from all the local standards.
Note that there is no atmospheric extinction term because it is absorbed by the zeropoint when doing differential photometry.
 
The photometric sequences for the 51 SN~II are identified in the finding charts in Figure \ref{charts.fig} and their magnitudes are listed in Table \ref{sequences.tab} along with the standard error of the mean (in units of mmag) and the number of nights on which each star was observed.
In every case, these sequences were derived from observations of Landolt standards (see Appendix D in \citealt{2001ApJ...558..615H} for the definition of the $z$ band and \citealt{2002AJ....124.2100S} for the description of the $z$-band standards).
Table \ref{photometry.tab} lists the resulting $UBVRIz$ magnitudes for the 51 SN. 
The uncertainties are shown in parentheses and the telescope is indicated for each observation. The uncertainties correspond to the photon Poisson statistics, adopting a minimum error of 0.015 mag, which is typical for a single observation of the Landolt standards with CCD detectors.
In total, we provide a dataset with 2,516 photometric points.


\subsection{Spectroscopy}

In addition to broad-band  photometry, several epochs of visual-wavelength spectra were  obtained for this SN set, which are used here  to aid in the determination of the explosion epoch (See \S~\ref{sec:ind}).
The number of epochs per object ranges between 1 and 27, with $\sim$7 spectra being obtained per object on average.
Spectral epochs are shown in Figure \ref{lc.fig} with short vertical brown solid lines. 
All optical spectral sequences, with their reduction and analysis, will be presented in a future publication (Guti\'errez et al. in prep.).


\section{Analysis} \label{ana}

Based on their spectral features and/or light curve morphologies, a handful of SN in our sample have been classified as SN IIb or SN~II peculiar. This includes:
SN~2000cb which shows similar photometric behavior to SN~1987A;
SN~2003bg has been classified as SN IIb and studied by \cite{2009ApJ...703.1612H}; 
SN 2003cv is similar to 2003bg and has also been classified as SN IIb;
SN 2003bj showed signs of SN+CSM interaction in their spectra and has been classified as SN IIn.
They are all plotted using different symbols and sometimes excluded in the following analysis.


\subsection{Individual multi-band light curves} \label{sec:ind}

Multi-band light curves showing their cadence and quality are presented  in Figure \ref{lc.fig}, all referenced to their explosion epoch, which has been determined in a similar fashion as in \cite{ 2014ApJ...786...67A}.
When non-detections are available the intermediate epoch between the last non-detection and the first detection is taken as an approximation of the explosion day, and its error is assumed to be half of this duration. 
In cases with no non-detections available or when the last non-detection is older than 20 days, the explosion epoch has been determined by matching spectral templates to our optical spectra using SNID \citep{2007ApJ...666.1024B} and averaging the epoch of the best fits.
For this, the spectra of SN with well-constrained explosion epoch from non-detections have been incorporated as new templates to SNID (exact details will be given in Guti\'errez et al. in prep.). 
 
All photometric measurements have been corrected for Galactic extinction using dust maps from \cite{2011ApJ...737..103S} assuming an $R_V$=3.1 and a \cite{1989ApJ...345..245C} law.
No correction for SN host galaxy extinction is applied to the data, and neither S- nor K-corrections have been considered due to the similar bands used in the observations and the low redshift range of our data 
(\citealt{2015ApJ...799..208S} and de Jaeger et al. in prep. showed that the K terms are lower than 0.2 mag at redshifts lower than 0.1). In addition, the temporal scale of all light-curves has been corrected for time dilation.
All further analysis presented in the following sections starts with these corrections applied.

In Figure \ref{cov.fig} we show the temporal coverage of our objects sorted by the first (left panel) and last photometric epoch (right panel), all with respect to the estimated explosion date. For the vast majority of objects, the first observation was performed within 20 days from explosion ($\avg{t_{\rm first}} = 14.7 \pm 11.1$ days) and on average the light curves extend through 158.2 $\pm$ 98.7 days, covering the whole recombination phase.
 

\subsection{Color curves}

Our multi-band light curves allow for the study of the color characteristics and its temporal evolution. 
In Figure \ref{col.fig}  the behavior of the $(U-B), (B-V), (B-R), (B-I), (V-R), (V-I)$ and $(R-I)$ colors is presented.
In the top panel,
average values of the color curves binned in 30 days intervals are overplotted, where the horizontal error represents the width of the bin and the vertical error the standard deviation for the objects in each bin.
The bottom panels show each color separately. 

All colors increase steadily at early times during the first few weeks due to the drop in temperature, which shifts the peak of the spectral energy distribution to redder wavelengths. 
This initial slope is more pronounced in colors containing bluer bands or with greater wavelength baselines, because bluer bands are more sensitive to the temperature decline and the increasingly strong line blocking affecting this initial phase \citep{2013MNRAS.433.1745D}.
In the subsequent weeks the increase is less pronounced because the temperature conditions at the photosphere remain similar due to the recombination of H happening during this phase \citep{2003MNRAS.345..111C}.
During the radioactive phase ($\gtrsim$ 100-150 days after the explosion) the color curves become flatter, in part because in this phase the SN II photometric evolution, which depends on the $^{56}$Co decay, is approximately the same in all bands.
At later epochs when approaching the nebular phase ($\lesssim$200 days) all curves start to decrease, the spectrum shows weak continuum and the emission lines start to dominate. 

The range spanned by a given color index decreases in the following order: $(B-I), (B-R), (B-V), (V-I), (V-R)$, and $(R-I)$. 
The same sequence is also seen in the scatter for a color index at a given epoch. 
It has already been shown that scatter in SN II intrinsic color evolution exists, rendering it difficult to determine dust absorption for an individual SN \citep{2013MNRAS.433.1745D, 2015ApJ...806..225P}. 
Keeping this caveat in mind, and assuming that those objects with the bluest colors suffer little to no reddening, one can interpret the color excess as an indication of the amount of extinction \citep{1992ApJ...395..366S}.
This assumption agrees with the fact that the color excesses decrease in the redder bands, implying lower extinction with increasing wavelength.
Further analysis on host galaxy extinction is beyond the scope of this paper, and will be presented in de Jaeger et al. (in prep.).

For the few objects with $U$ and $B$ photometry available we also show in Figure \ref{col.fig} the $(U-B)$ color with blue open squares. 
It is clearly seen how at early epochs it takes negative values showing that right after explosion, SN II emit more intensely in shorter wavelengths than in the optical. 
In the days following the explosion SN II spectra are characterized by a blue and featureless continuum.
On the following weeks, in addition to the temperature decrease, line blanketing affects the UV part of the spectrum making the UV brightness decline steeper than the blue, and producing a rapid $(U-B)$ increase.

In general all objects show similar behavior in all colors, with some special cases. 
We find a few objects further than 2 sigma from the average color curve. 
The three objects with the reddest colors are SN~2003fb, 2003hg and 2003ho. SN~2003hl is intermediate between these three and the rest of the sample. These objects probably suffer important levels of extinction.
On the other hand, SN~1999em (together with SN~1991al, 1993K, 2002ew, 2003B, 2003bn, and 2003ib) is located at the bottom of the curves, which can be interpreted as having little  host-galaxy reddening.
Also shown in solid lines is the color evolution of the four objects that were classified as 1987A-like, SN~IIb, or SN~IIn.
SN~2003cv is also located at the redder end in the color curve evolution at early epochs, while SN~2003bg (also a SN~IIb) follows the average evolution, but is bluer than all other SN II 50 days after explosion.

Although we did not differentiate our sample into SN~IIP and SN~IIL subtypes, we do not see two distinct color behaviors in any of the color curves, therefore confirming that SNII form a continuous class.


\subsection{Absolute magnitudes} \label{absmag}

We calculated absolute magnitudes of all our photometric measurements to study the global behavior of SN II light curves in different bands. 
For that, a measurement of the distance to the SN is needed.
Distances to SN host galaxies with CMB-corrected recession velocities lower than 3000 km s$^{-1}$ are collected from NED and averaged, using only distances based on the Tully-Fisher and Cepheid methods.
For SN with host galaxy velocities higher than 3000 km s$^{-1}$ the distance is measured using the luminosity-distance expression, assuming a Hubble constant, $H_0$, of  68 km s$^{-1}$ Mpc$^{-1}$ and cosmological density parameters $\Omega_M$ = 0.30 and $\Omega_\Lambda$ = 0.70 \citep{2015arXiv150201589P}.
Distance errors are added to the photometric errors in the absolute magnitude error budget.
Figure \ref{allabs.fig} shows all the available MW extinction-corrected absolute magnitude light curves for each of the six bands presented here in separated panels, smoothed using a 3rd order spline polynomial.
All panels show wide ranges in both absolute magnitudes and light-curve morphologies of the SN presented here.

Absolute magnitudes at maximum brightness in all light curves were measured by fitting a second-order polynomial to the early epochs when: (i) at least four points were available, and (ii) the second measurement was brighter than the first (indicating that the first was still part of the rise).
Otherwise  the first photometric point was considered the maximum brightness of the SN.

The average MW extinction-corrected peak absolute magnitudes of all SN in our sample (excluding 87A-like, IIb, and IIn) are the following: 
$\avg{U_{max}}$=--16.06 mag ($\sigma$=1.74, 6 SN); 
$\avg{B_{max}}$=--16.43 mag ($\sigma$=1.19, 47 SN); 
$\avg{V_{max}}$=--16.89 mag ($\sigma$=0.98, 45 SN); 
$\avg{R_{max}}$=--16.96 mag ($\sigma$=1.03, 24 SN); 
$\avg{I_{max}}$=--17.38 mag ($\sigma$=0.95, 46 SN)
($\avg{z_{max}}$ is not reported since there are not enough data to do statistics).
The average absolute magnitude increases by 1.3 mag from $U$ to $I$.
We added a horizontal line and a strip in Figure \ref{allabs.fig} representing the average absolute peak magnitude and its 1$\sigma$ scatter, respectively.
Light curves in our sample fill the strip at peak epochs showing a continuum of peak absolute magnitudes.
Similarly to \cite{2014MNRAS.442..844F, 2014MNRAS.445..554F}, our average magnitudes are slightly lower ($\sim$0.7 mag) than those previously published, but still consistent within the uncertainties.
While only a few objects peak at  magnitudes brighter than the range covered by the 1$\sigma$ strip (including SN~2003bg), a higher number with low-luminosity appear below the strip. 
The range of magnitudes at peak span $\sim$4.5 mag in all bands, from SN~1999br being the faintest and 
SN~2003eg the brightest.
While the broad-line  Type~IIb SN~2003bg is one of the brightest objects in all bands with  data near maximum ($B, V, I$), the other SN~IIn and 1987A-like objects have brightness around the average.

In Figure \ref{allabsn.fig} we show the $BVRI$-band light curves referenced to the epoch of maximum brightness and normalized to peak magnitude. 
Once the differences in maximum brightness among SN II are removed, these panels show how other light-curve parameters compare, such as the duration of the plateau, the post-maximum brightness decline, or the slope of the radioactive tail. 
As pointed out by \cite{2014ApJ...786...67A}, with the $V$-band light curves of the CSP sample, and the sample recently published by  \cite{2015ApJ...799..208S}, we do not distinguish two separate groups among the diversity of light-curves. 
On the other hand, our findings rather show a continuum in all the parameters listed above, and in all the optical bands presented here.


\subsection{Brightness decline} \label{sec:s}

We measure the brightness decline parameter, $s$, defined as the decline rate in magnitudes per 100 days of the post-maximum light-curve until the end of the plateau. 
This is similar to the $s1$ and $s2$ parameters defined by \cite{2014ApJ...786...67A}.
For the reddest bands, the first phase of brightness decline after maximum right before entering into the recombination phase, $s1$, is not clearly seen. Therefore, $s$ is adopted instead of the $s2$ parameter in order to compare multi-band observations in a systematic way. The exact details of this parameter and its behavior in different bands will be a matter of future work.

To measure $s$, it was first necessary  to define the epoch at which the plateau ends for each band separately. Similarly to \cite{2014ApJ...786...67A}, we set this as the epoch at the latest phases of the plateau at which the brightness deviates more than 0.1 mag from the linear fits. 
We stress that (i) for a single SN, this epoch is statistically the same in all bands, and (ii) their distribution is similar for all bands (although the average increases slightly for redder bands) and peaks at 77.5 ($\pm$26.3) days after the explosion epoch.
Photometric decline rates are all measured by fitting a straight line to the defined phase, taking into account photometric errors.
For those objects for which the end of the plateau in the $U$ band cannot be defined, we take that epoch from other bands.

The average values for the decline rates for all SN~II in our sample where the measurement can be confidently performed are: 
$\avg{s_U}$= 8.06 mag 100d$^{-1}$ ($\sigma$=1.87, 6 SN); 
$\avg{s_B}$= 3.17 mag 100d$^{-1}$($\sigma$=1.29, 45 SN); 
$\avg{s_V}$= 1.53 mag 100d$^{-1}$ ($\sigma$=0.91, 45 SN); 
$\avg{s_R}$= 0.92 mag 100d$^{-1}$ ($\sigma$=0.76, 22 SN); 
$\avg{s_I}$= 0.65 mag 100d$^{-1}$ ($\sigma$=1.01, 43 SN)
($\avg{s_z}$ is not reported since there are not enough data to do statistics).
Similarly to the absolute magnitude distributions discussed above, SN decline steeper in the blue bands and the decline gets shallower in the redder bands.

Peculiar SN, IIn and IIb SN have not been considered in the measurement of the average values since their light-curves present different morphologies that are not best described by this characterization.


\subsection{Peak magnitude and brightness decline relation}

In this section we study the relation between the two parameters measured in the previous sections: the peak absolute magnitude and the brightness decline rate. 
This correlation holds promise for the standardization of the SN II absolute peak magnitudes and may enable their use as distance indicators for cosmology \citep{2014ApJ...786...67A}, in a similar way to the luminosity decline-rate relation used for type Ia SN \citep{1993ApJ...413L.105P}.

Here we exclude from our sample of 51 SN I the four IIb, IIn and SN 1987A-like events.
To improve the statistics and the significance of the results presented below we expand our sample of 47 SN II to 114 by including 67 SN II for which photometry is available in the literature, as listed in Table \ref{snlistfig8.tab}.
 
Figure \ref{sdist.fig} presents histograms of the distributions of decline rates, $s$, in each band together with their median values for the complete sample. 
The same trends described in \S\ref{sec:s} are recovered: bluer bands show higher $s$ values, and the median values (represented in dashed vertical lines in the Figure) are lower for redder bands.

In the left panel of Figure \ref{mmaxs.fig}, $M_{max}$ is plotted vs. $s$. 
Filled circles are SN presented in this paper and empty circles are objects from the literature.
These parameters show a trend in the sense that lower luminosity SN decline more slowly, while more luminous events decline more rapidly.
This behavior was previously reported in the $V$ band \citep{2014ApJ...786...67A}, and it is presented here for the first time for the $UBVRI$-band light curves.
We performed linear fits to the data in each band and we found that the one showing the best correlation is the $B$ band (r=--0.59, N=97). This is contradictory to the insignificant correlation found by \cite{2015ApJ...799..215P}, but their result can be due to the small size of their sample.

For those objects for which the measurement of the plateau duration is possible, we show in the right panel of Figure~\ref{mmaxs.fig} its relation with the $s$ parameter. 
Plateau durations cover a range from $\sim$20 to $\sim$80 days in all bands, with some SN having shorter plateaus in the bluer bands, and some with higher values for redder bands.
 SN with deeper declines (higher $s$) are found also to have shorter plateaus, and SN which decline in  brightness  more slowly (lower $s$) have longer plateaus. 
Linear fits to the $BVRI$ data separately give similar slopes of $-$0.03 (mag 100 d$^{-1}$ per day), with increasing correlation factors from $-$0.6 to $-$0.8  for bluer to redder bands.

Both the luminosity and the plateau duration are related to $s$, in a way that SN declining faster have shorter plateaus and brighter magnitudes (see Figure \ref{mmaxs.fig}).
According to \cite{2009ApJ...703.2205K} models, these two parameters basically depend on the kinetic energy of the explosion and the mass of the ejecta (see also \citealt{1993ApJ...414..712P})$^3$\let\thefootnote\relax\footnote{$^3$ Note that the mass of radioactive $^{56}$Ni synthesized in the explosion extends the plateau in time by a few percent \citep{2009ApJ...703.2205K}. It also powers the luminosity after the recombinations phase, and its total mass has been shown to correlate well also with the plateau luminosity \citep{2003ApJ...582..905H, 2015ApJ...799..215P}}.
For larger and/or denser H layers, a higher fraction of energy is lost in the diffusion of the radiation through the envelope, the radiation is trapped for a longer time (thus longer duration  plateau phases), and less energy/radiation escapes and contributes to the luminosity \citep{1993A&A...273..106B}.
The observed relations between luminosity, plateau duration and decline rate (s) indicate that SNe exploding with higher kinetic energies are those resulting from progenitors with smaller and/or less dense H envelope masses at the explosion.
This is in agreement with the current view of massive star evolution, where the progenitors of core collapse SN with reduced H envelopes are stars with larger zero-age main sequence (ZAMS) masses that have experienced a higher degree of mass-loss prior to explosion \citep{2003ApJ...591..288H}.

Finally, the wide range of plateau durations, decline rates, and peak luminosities can also be interpreted as a clear indication of a continuity in the SN II class. 


\section{Summary and Conclusions}\label{summ}

This paper presents a sample of  multi-band, visual-wavelength light curves of 51 SN~II observed from 1986 to 2003 in the course of four different surveys: the Cerro Tololo Supernova Survey, the Cal\'an/Tololo Supernova Program (C\&T), the Supernova Optical and Infrared Survey (SOIRS), and the Carnegie Type II Supernova Survey (CATS).
Near-infrared photometry and optical spectroscopy of this set of SN II will be published in two companion papers.

After determining their explosion dates
and correcting all photometry for Galactic extinction and time dilation,
we investigated their color behavior in different bands, measured their peak absolute magnitudes, and the brightness decline in the recombination phase in all bands.
No evidence of two separate families (SN~IIP and SN~IIL) can be seen in our results, confirming previous reports that there is a continuity in SN II characteristics.

All color curves grow steadily redder during the first few weeks due to a decrease in surface temperature, and reach a maximum around $\sim$100-150 days followed by a shallow color decrease. 
$(U-B)$ colors are found to begin with negative values around maximum followed by a rapid increase to redder colors owing to the temperature decline (cooling) and the increasingly higher line blanketing toward shorter wavelengths.

For a given color index, the scatter among different SN increases for color indices involving bluer bands, supporting the idea that the color diversity could be caused by host-galaxy dust extinction. Going a step further and assuming that all SN have similar intrinsic color curves, we note that SN with higher excess in one color index also have higher excess in other color indices, whereas the bluest SN appear blue in all color indices, lending support to the idea that the color excess is an indication of host-galaxy dust extinction. However, it is possible that part of the color diversity could be due to intrinsic effects. The low luminosity SN 1999br is a clear example of an intrinsically red SN. In a future paper (de Jaeger et al.) we will address this issue.

With all the available MW extinction-corrected absolute magnitude light-curves we find a wide range of magnitudes and light curve morphologies in all $UBVRIz$ bands.
We measured absolute peak magnitudes finding the following mean values: 
$\avg{U_{max}}$=--16.06 $\pm$ 1.74;
$\avg{B_{max}}$=--16.43 $\pm$ 1.19;
$\avg{V_{max}}$=--16.89 $\pm$ 0.98;
$\avg{R_{max}}$=--16.96 $\pm$ 1.03;
$\avg{I_{max}}$=--17.38 $\pm$ 0.95.
Only a few outliers with peak magnitudes brighter and fainter than 1$\sigma$ of the distribution are found. 

We defined the $s$ parameter, which measures the brightness decline rate from maximum light through the end of the recombination phase, and found that this decline parameter is steeper in the blue bands than redder bands.

We added a set of 67 low-z SN II with publicly available photometry to study the absolute magnitude vs. brightness decline parameter $s$ relation. 
From a total sample of 114 SN II, we found a clear correlation in all bands, with the following characteristics: 
(1) more luminous SN have steeper light curves; 
(2) the slope of the correlation decreases with increasing wavelength;
(3) the correlation is higher in the $B$ band.
Finally, we also found a correlation between the $s$ parameter and the plateau duration, being the latter shorter (longer) for larger (smaller) $s$ values.

The complete set of photometry is available electronically$^4$\let\thefootnote\relax\footnote{$^4$ \href{https://github.com/lgalbany/51_SNII_LC}{https://github.com/lgalbany/51\_SNII\_LC}} or can be requested from the authors. Each SN folder in the tarball includes an info file containing its name, subtype, redshift, coordinates, host galaxy name, morphology and MW extinction from \cite{2011ApJ...737..103S}. 

\acknowledgments

We acknowledge the contribution to the observations of 
Elisa Abedrapo, Maria Teresa Acevedo, Sandra dos Anjos, Roberto Avil\'es A., L. Felipe Barrientos, Timothy Ellsworth Bowers, Stephane Brillant, Pablo Candia, Sergio Castell\'on, Carlos Contreras, Arjun Dey, Vannessa Doublier, Jo Ann Eder, Jonathan Elias, Erica Ellingson, Wendy L. Freedman, Catharine Garmany, Ximena G\'omez, Paul J. Green, Olivier R. Hainaut, Leonor Huerta, Daniel Kelson, Rebecca A. Koopmann, Arlo U. Landolt, Andrew Layden, Paul Martini, Philip Massey, Mario Mateo, Mauricio Navarrete, Edward W. Olszewski, Fernando Peralta, Joaqu\'in Perez, Eric Persson, Tim Pickering, Miguel Roth, Eric P. Rubenstein, Maria Teresa Ruiz, Paul C. Schmidtke, Juan C. Seguel, Patrick Seitzer, Robert C. Smith, Ronaldo E. de Souza, Joao E. Steiner, Neil de Grasse Tyson, Stephanie Wachter, Ken-ichi Wakamatsu, Alistair Walker, Doug Welch and Howard K.C. Yee.
Support for LG, MH, TM, SGG, and CPG is provided by the Ministry of Economy, Development, and Tourism's Millennium Science Initiative through grant IC120009, awarded to The Millennium Institute of Astrophysics, MAS. 
LG and SGG acknowledge support by CONICYT through FONDECYT grants 3140566 and 3130680, respectively.
MH acknowledges support provided by 
Fondecyt grants 1920312 and 1060808, 
the Millennium Center for Supernova Science through grant P06-045-F funded by ``Programa Bicentenario de Ciencia y Tecnolog\'ia de CONICYT" and ``Programa Iniciativa Cient\'ifica Milenio del Ministerio de Econom\'ia",
the Carnegie Postdoctoral Fellowship,
and NASA through Hubble Fellowship grant HST-HF-01139.01-A awarded by the Space Telescope Science Institute, which is operated by the Association of Universities for Research in Astronomy, Inc., for NASA, under contract NAS 5-26555.
NBS thanks support from the George P. and Cynthia Woods Mitchell Institute for Fundamental Physics and Astronomy.
MS acknowledges the generous support provided by the Danish Agency for Science and Technology and Innovation realized through a Sapere Aude Level 2 grant.
AL was funded by Grant No. CW-0004-85 from the Space Telescope Science Institute (STScI) at the time of observations.
Based in part on observations at Cerro Tololo Inter-American Observatory, National Optical Astronomy Observatory, which is operated by the Association of Universities for Research in Astronomy (AURA) under a cooperative agreement with the National Science Foundation. 
This paper includes data gathered with the 6.5 meter Magellan Telescopes located at Las Campanas Observatory, Chile. 
Based in part on observations made with ESO Telescopes at the La Silla and Paranal Observatories under programmes 163.H-0285 and 164.H-0376.
We also acknowledge time allocations at the Steward Observatory of the University of Arizona.
This research has made use of the NASA/IPAC Extragalactic Database (NED) which is operated by the Jet Propulsion Laboratory, California Institute of Technology, under contract with the National Aeronautics and Space Administration. 

\bibliographystyle{aa}
\bibliography{cats}

\begin{thebibliography}{192}
\expandafter\ifx\csname natexlab\endcsname\relax\def\natexlab#1{#1}\fi

\bibitem[{{Anderson} {et~al.}(2014){Anderson}, {Gonz{\'a}lez-Gait{\'a}n},
  {Hamuy}, {Guti{\'e}rrez}, {Stritzinger}, {Olivares E.}, {Phillips},
  {Schulze}, {Antezana}, {Bolt}, {Campillay}, {Castell{\'o}n}, {Contreras}, {de
  Jaeger}, {Folatelli}, {F{\"o}rster}, {Freedman}, {Gonz{\'a}lez}, {Hsiao},
  {Krzemi{\'n}ski}, {Krisciunas}, {Maza}, {McCarthy}, {Morrell}, {Persson},
  {Roth}, {Salgado}, {Suntzeff}, \& {Thomas-Osip}}]{2014ApJ...786...67A}
{Anderson}, J.~P., {Gonz{\'a}lez-Gait{\'a}n}, S., {Hamuy}, M., {et~al.} 2014,
  \apj, 786, 67

\bibitem[{{Anderson} {et~al.}(2012){Anderson}, {Habergham}, {James}, \&
  {Hamuy}}]{2012MNRAS.424.1372A}
{Anderson}, J.~P., {Habergham}, S.~M., {James}, P.~A., \& {Hamuy}, M. 2012,
  \mnras, 424, 1372

\bibitem[{{Barbarino} {et~al.}(2015){Barbarino}, {Dall'Ora}, {Botticella},
  {Valle}, {Zampieri}, {Maund}, {Pumo}, {Jerkstrand}, {Benetti}, {Elias-Rosa},
  {Fraser}, {Gal-Yam}, {Hamuy}, {Inserra}, {Knapic}, {LaCluyze}, {Molinaro},
  {Ochner}, {Pastorello}, {Pignata}, {Reichart}, {Ries}, {Riffeser}, {Schmidt},
  {Schmidt}, {Smareglia}, {Smartt}, {Smith}, {Sollerman}, {Sullivan},
  {Tomasella}, {Turatto}, {Valenti}, {Yaron}, \& {Young}}]{2015MNRAS.448.2312B}
{Barbarino}, C., {Dall'Ora}, M., {Botticella}, M.~T., {et~al.} 2015, \mnras,
  448, 2312

\bibitem[{{Barbon} {et~al.}(1973){Barbon}, {Ciatti}, \&
  {Rosino}}]{1973AA....29...57B}
{Barbon}, R., {Ciatti}, F., \& {Rosino}, L. 1973, \aap, 29, 57

\bibitem[{{Benetti} {et~al.}(1991){Benetti}, {Cappellaro}, \&
  {Turatto}}]{1991AA...247..410B}
{Benetti}, S., {Cappellaro}, E., \& {Turatto}, M. 1991, \aap, 247, 410

\bibitem[{{Benetti} {et~al.}(1994){Benetti}, {Cappellaro}, {Turatto}, {della
  Valle}, {Mazzali}, \& {Gouiffes}}]{1994AA...285..147B}
{Benetti}, S., {Cappellaro}, E., {Turatto}, M., {et~al.} 1994, \aap, 285, 147

\bibitem[{{Bersten} {et~al.}(2011){Bersten}, {Benvenuto}, \&
  {Hamuy}}]{2011ApJ...729...61B}
{Bersten}, M.~C., {Benvenuto}, O., \& {Hamuy}, M. 2011, \apj, 729, 61

\bibitem[{{Bersten} \& {Hamuy}(2009)}]{2009ApJ...701..200B}
{Bersten}, M.~C. \& {Hamuy}, M. 2009, \apj, 701, 200

\bibitem[{{Blanton} {et~al.}(1995){Blanton}, {Schmidt}, {Kirshner}, {Ford},
  {Chromey}, \& {Herbst}}]{1995AJ....110.2868B}
{Blanton}, E.~L., {Schmidt}, B.~P., {Kirshner}, R.~P., {et~al.} 1995, \aj, 110,
  2868

\bibitem[{{Blinnikov} \& {Bartunov}(1993)}]{1993A&A...273..106B}
{Blinnikov}, S.~I. \& {Bartunov}, O.~S. 1993, \aap, 273, 106

\bibitem[{{Blondin} \& {Tonry}(2007)}]{2007ApJ...666.1024B}
{Blondin}, S. \& {Tonry}, J.~L. 2007, \apj, 666, 1024

\bibitem[{{Boles} \& {Li}(2003)}]{2003CBET...41....1B}
{Boles}, T. \& {Li}, W. 2003, Central Bureau Electronic Telegrams, 41, 1

\bibitem[{{Bose} {et~al.}(2013){Bose}, {Kumar}, {Sutaria}, {Kumar}, {Roy},
  {Bhatt}, {Pandey}, {Chandola}, {Sagar}, {Misra}, \&
  {Chakraborti}}]{2013MNRAS.433.1871B}
{Bose}, S., {Kumar}, B., {Sutaria}, F., {et~al.} 2013, \mnras, 433, 1871

\bibitem[{{Bouchet} {et~al.}(1991){Bouchet}, {della Valle}, \&
  {Melnick}}]{1991IAUC.5312....2B}
{Bouchet}, P., {della Valle}, M., \& {Melnick}, J. 1991, \iaucirc, 5312, 2

\bibitem[{{Burrows}(2000)}]{2000Natur.403..727B}
{Burrows}, A. 2000, \nat, 403, 727

\bibitem[{{Cappellaro} {et~al.}(1995){Cappellaro}, {Danziger}, {della Valle},
  {Gouiffes}, \& {Turatto}}]{1995AA...293..723C}
{Cappellaro}, E., {Danziger}, I.~J., {della Valle}, M., {Gouiffes}, C., \&
  {Turatto}, M. 1995, \aap, 293, 723

\bibitem[{{Cardelli} {et~al.}(1989){Cardelli}, {Clayton}, \&
  {Mathis}}]{1989ApJ...345..245C}
{Cardelli}, J.~A., {Clayton}, G.~C., \& {Mathis}, J.~S. 1989, \apj, 345, 245

\bibitem[{{Chassagne}(2003)}]{2003IAUC.8085....1C}
{Chassagne}, R. 2003, \iaucirc, 8085, 1

\bibitem[{{Chieffi} {et~al.}(2003){Chieffi}, {Dom{\'{\i}}nguez}, {H{\"o}flich},
  {Limongi}, \& {Straniero}}]{2003MNRAS.345..111C}
{Chieffi}, A., {Dom{\'{\i}}nguez}, I., {H{\"o}flich}, P., {Limongi}, M., \&
  {Straniero}, O. 2003, \mnras, 345, 111

\bibitem[{{Chornock} {et~al.}(2002){Chornock}, {Jha}, {Filippenko}, \&
  {Barris}}]{2002IAUC.8008....2C}
{Chornock}, R., {Jha}, S., {Filippenko}, A.~V., \& {Barris}, B. 2002, \iaucirc,
  8008, 2

\bibitem[{{Chugai} {et~al.}(2005){Chugai}, {Fabrika}, {Sholukhova},
  {Goranskij}, {Abolmasov}, \& {Vlasyuk}}]{2005AstL...31..792C}
{Chugai}, N.~N., {Fabrika}, S.~N., {Sholukhova}, O.~N., {et~al.} 2005,
  Astronomy Letters, 31, 792

\bibitem[{{Ciatti} \& {Rosino}(1977)}]{1977AA....56...59C}
{Ciatti}, F. \& {Rosino}, L. 1977, \aap, 56, 59

\bibitem[{{Clocchiatti} {et~al.}(1996){Clocchiatti}, {Benetti}, {Wheeler},
  {Wren}, {Boisseau}, {Cappellaro}, {Turatto}, {Patat}, {Swartz}, {Harkness},
  {Brotherton}, {Wills}, {Hemenway}, {Cornell}, {Frueh}, \&
  {Kaiser}}]{1996AJ....111.1286C}
{Clocchiatti}, A., {Benetti}, S., {Wheeler}, J.~C., {et~al.} 1996, \aj, 111,
  1286

\bibitem[{{Dall'Ora} {et~al.}(2014){Dall'Ora}, {Botticella}, {Pumo},
  {Zampieri}, {Tomasella}, {Pignata}, {Bayless}, {Pritchard}, {Taubenberger},
  {Kotak}, {Inserra}, {Della Valle}, {Cappellaro}, {Benetti}, {Benitez},
  {Bufano}, {Elias-Rosa}, {Fraser}, {Haislip}, {Harutyunyan}, {Howell},
  {Hsiao}, {Iijima}, {Kankare}, {Kuin}, {Maund}, {Morales-Garoffolo},
  {Morrell}, {Munari}, {Ochner}, {Pastorello}, {Patat}, {Phillips}, {Reichart},
  {Roming}, {Siviero}, {Smartt}, {Sollerman}, {Taddia}, {Valenti}, \&
  {Wright}}]{2014ApJ...787..139D}
{Dall'Ora}, M., {Botticella}, M.~T., {Pumo}, M.~L., {et~al.} 2014, \apj, 787,
  139

\bibitem[{{della Valle} \& {Bianchini}(1992)}]{1992IAUC.5558....3D}
{della Valle}, M. \& {Bianchini}, A. 1992, \iaucirc, 5558, 3

\bibitem[{{Dessart} {et~al.}(2008){Dessart}, {Blondin}, {Brown}, {Hicken},
  {Hillier}, {Holland}, {Immler}, {Kirshner}, {Milne}, {Modjaz}, \&
  {Roming}}]{2008ApJ...675..644D}
{Dessart}, L., {Blondin}, S., {Brown}, P.~J., {et~al.} 2008, \apj, 675, 644

\bibitem[{{Dessart} {et~al.}(2013){Dessart}, {Hillier}, {Waldman}, \&
  {Livne}}]{2013MNRAS.433.1745D}
{Dessart}, L., {Hillier}, D.~J., {Waldman}, R., \& {Livne}, E. 2013, \mnras,
  433, 1745

\bibitem[{{Elias-Rosa} {et~al.}(2003){Elias-Rosa}, {Benetti}, {Marmo},
  {Pastorello}, {Altavilla}, {Navasardyan}, {Riello}, {Turatto}, {Zampieri}, \&
  {Cappellaro}}]{2003IAUC.8187....2E}
{Elias-Rosa}, N., {Benetti}, S., {Marmo}, C., {et~al.} 2003, \iaucirc, 8187, 2

\bibitem[{{Evans} {et~al.}(2003){Evans}, {Bock}, {Krisciunas}, \&
  {Espinoza}}]{2003IAUC.8186....1E}
{Evans}, R., {Bock}, G., {Krisciunas}, K., \& {Espinoza}, J. 2003, \iaucirc,
  8186, 1

\bibitem[{{Evans} {et~al.}(1986){Evans}, {McNaught}, {Cragg}, \&
  {Thompson}}]{1986IAUC.4260....1E}
{Evans}, R., {McNaught}, R., {Cragg}, T., \& {Thompson}, G. 1986, \iaucirc,
  4260, 1

\bibitem[{{Evans} \& {McNaught}(2003)}]{2003IAUC.8150....2E}
{Evans}, R. \& {McNaught}, R.~H. 2003, \iaucirc, 8150, 2

\bibitem[{{Evans} \& {Phillips}(1992)}]{1992IAUC.5625....2E}
{Evans}, R. \& {Phillips}, M.~M. 1992, \iaucirc, 5625, 2

\bibitem[{{Evans} \& {Quirk}(2003)}]{2003IAUC.8042....1E}
{Evans}, R. \& {Quirk}, S. 2003, \iaucirc, 8042, 1

\bibitem[{{Faran} {et~al.}(2014{\natexlab{a}}){Faran}, {Poznanski},
  {Filippenko}, {Chornock}, {Foley}, {Ganeshalingam}, {Leonard}, {Li},
  {Modjaz}, {Nakar}, {Serduke}, \& {Silverman}}]{2014MNRAS.442..844F}
{Faran}, T., {Poznanski}, D., {Filippenko}, A.~V., {et~al.} 2014{\natexlab{a}},
  \mnras, 442, 844

\bibitem[{{Faran} {et~al.}(2014{\natexlab{b}}){Faran}, {Poznanski},
  {Filippenko}, {Chornock}, {Foley}, {Ganeshalingam}, {Leonard}, {Li},
  {Modjaz}, {Serduke}, \& {Silverman}}]{2014MNRAS.445..554F}
{Faran}, T., {Poznanski}, D., {Filippenko}, A.~V., {et~al.} 2014{\natexlab{b}},
  \mnras, 445, 554

\bibitem[{{Filippenko}(1997)}]{1997ARA&A..35..309F}
{Filippenko}, A.~V. 1997, \araa, 35, 309

\bibitem[{{Filippenko} \& {Chornock}(2002)}]{2002IAUC.7988....3F}
{Filippenko}, A.~V. \& {Chornock}, R. 2002, \iaucirc, 7988, 3

\bibitem[{{Filippenko} \& {Foley}(2003)}]{2003IAUC.8214....2F}
{Filippenko}, A.~V. \& {Foley}, R.~J. 2003, \iaucirc, 8214, 2

\bibitem[{{Filippenko} {et~al.}(2003){Filippenko}, {Foley}, \&
  {Serduke}}]{2003IAUC.8189....2F}
{Filippenko}, A.~V., {Foley}, R.~J., \& {Serduke}, F.~J.~D. 2003, \iaucirc,
  8189, 2

\bibitem[{{Foley} {et~al.}(2003){Foley}, {Graham}, {Ganeshalingam}, \&
  {Filippenko}}]{2003IAUC.8060....3F}
{Foley}, R.~J., {Graham}, J., {Ganeshalingam}, M., \& {Filippenko}, A.~V. 2003,
  \iaucirc, 8060, 3

\bibitem[{{Fraser} {et~al.}(2011){Fraser}, {Ergon}, {Eldridge}, {Valenti},
  {Pastorello}, {Sollerman}, {Smartt}, {Agnoletto}, {Arcavi}, {Benetti},
  {Botticella}, {Bufano}, {Campillay}, {Crockett}, {Gal-Yam}, {Kankare},
  {Leloudas}, {Maguire}, {Mattila}, {Maund}, {Salgado}, {Stephens},
  {Taubenberger}, \& {Turatto}}]{2011MNRAS.417.1417F}
{Fraser}, M., {Ergon}, M., {Eldridge}, J.~J., {et~al.} 2011, \mnras, 417, 1417

\bibitem[{{Frey} {et~al.}(2013){Frey}, {Fryer}, \&
  {Young}}]{2013ApJ...773L...7F}
{Frey}, L.~H., {Fryer}, C.~L., \& {Young}, P.~A. 2013, \apjl, 773, L7

\bibitem[{{Gal-Yam} {et~al.}(2011){Gal-Yam}, {Kasliwal}, {Arcavi}, {Green},
  {Yaron}, {Ben-Ami}, {Xu}, {Sternberg}, {Quimby}, {Kulkarni}, {Ofek},
  {Walters}, {Nugent}, {Poznanski}, {Bloom}, {Cenko}, {Filippenko}, {Li},
  {Silverman}, {Walker}, {Sullivan}, {Maguire}, {Howell}, {Mazzali}, {Frail},
  {Bersier}, {James}, {Akerlof}, {Yuan}, {Law}, {Fox}, \&
  {Gehrels}}]{2011ApJ...736..159G}
{Gal-Yam}, A., {Kasliwal}, M.~M., {Arcavi}, I., {et~al.} 2011, \apj, 736, 159

\bibitem[{{Galbany} {et~al.}(2014){Galbany}, {Stanishev}, {Mour{\~a}o},
  {Rodrigues}, {Flores}, {Garc{\'{\i}}a-Benito}, {Mast}, {Mendoza},
  {S{\'a}nchez}, {Badenes}, {Barrera-Ballesteros}, {Bland-Hawthorn},
  {Falc{\'o}n-Barroso}, {Garc{\'{\i}}a-Lorenzo}, {Gomes}, {Gonz{\'a}lez
  Delgado}, {Kehrig}, {Lyubenova}, {L{\'o}pez-S{\'a}nchez}, {de
  Lorenzo-C{\'a}ceres}, {Marino}, {Meidt}, {Moll{\'a}}, {Papaderos},
  {P{\'e}rez-Torres}, {Rosales-Ortega}, \& {van de Ven}}]{2014A&A...572A..38G}
{Galbany}, L., {Stanishev}, V., {Mour{\~a}o}, A.~M., {et~al.} 2014, \aap, 572,
  A38

\bibitem[{{Gall} {et~al.}(2015){Gall}, {Polshaw}, {Kotak}, {Jerkstrand},
  {Leibundgut}, {Rabinowitz}, {Sollerman}, {Sullivan}, {Smartt}, {Anderson},
  {Benetti}, {Baltay}, {Feindt}, {Fraser}, {Gonz{\'a}lez-Gait{\'a}n},
  {Inserra}, {Maguire}, {McKinnon}, {Valenti}, \& {Young}}]{2015AA...582A...3G}
{Gall}, E.~E.~E., {Polshaw}, J., {Kotak}, R., {et~al.} 2015, \aap, 582, A3

\bibitem[{{Gandhi} {et~al.}(2013){Gandhi}, {Yamanaka}, {Tanaka}, {Nozawa},
  {Kawabata}, {Saviane}, {Maeda}, {Moriya}, {Hattori}, {Sasada}, \&
  {Itoh}}]{2013ApJ...767..166G}
{Gandhi}, P., {Yamanaka}, M., {Tanaka}, M., {et~al.} 2013, \apj, 767, 166

\bibitem[{{Ganeshalingam} {et~al.}(2003){Ganeshalingam}, {Graham}, {Pugh}, \&
  {Li}}]{2003IAUC.8134....1G}
{Ganeshalingam}, M., {Graham}, J., {Pugh}, H., \& {Li}, W. 2003, \iaucirc,
  8134, 1

\bibitem[{{Ganeshalingam} \& {Li}(2003{\natexlab{a}})}]{2003CBET...15....1G}
{Ganeshalingam}, M. \& {Li}, W. 2003{\natexlab{a}}, Central Bureau Electronic
  Telegrams, 15, 1

\bibitem[{{Ganeshalingam} \& {Li}(2003{\natexlab{b}})}]{2003IAUC.8179....2G}
{Ganeshalingam}, M. \& {Li}, W. 2003{\natexlab{b}}, \iaucirc, 8179, 2

\bibitem[{{Garnavich} \& {Bass}(2003)}]{2003IAUC.8150....3G}
{Garnavich}, P. \& {Bass}, E. 2003, \iaucirc, 8150, 3

\bibitem[{{Garnavich} {et~al.}(1999){Garnavich}, {Jha}, {Challis}, {Kirshner},
  {Calkins}, {Filippenko}, {Stern}, {Reuland}, \& {Li}}]{1999IAUC.7143....1G}
{Garnavich}, P., {Jha}, S., {Challis}, P., {et~al.} 1999, \iaucirc, 7143, 1

\bibitem[{{Graham} \& {Li}(2002)}]{2002IAUC.8015....1G}
{Graham}, J. \& {Li}, W. 2002, \iaucirc, 8015, 1

\bibitem[{{Guti{\'e}rrez} {et~al.}(2014){Guti{\'e}rrez}, {Anderson}, {Hamuy},
  {Gonz{\'a}lez-Gait{\'a}n}, {Folatelli}, {Morrell}, {Stritzinger}, {Phillips},
  {McCarthy}, {Suntzeff}, \& {Thomas-Osip}}]{2014ApJ...786L..15G}
{Guti{\'e}rrez}, C.~P., {Anderson}, J.~P., {Hamuy}, M., {et~al.} 2014, \apjl,
  786, L15

\bibitem[{{Hamuy}(1993{\natexlab{a}})}]{1993IAUC.5771....1H}
{Hamuy}, M. 1993{\natexlab{a}}, \iaucirc, 5771, 1

\bibitem[{{Hamuy}(1993{\natexlab{b}})}]{1993IAUC.5823....1H}
{Hamuy}, M. 1993{\natexlab{b}}, \iaucirc, 5823, 1

\bibitem[{{Hamuy}(2002{\natexlab{a}})}]{2002IAUC.7987....2H}
{Hamuy}, M. 2002{\natexlab{a}}, \iaucirc, 7987, 2

\bibitem[{{Hamuy}(2002{\natexlab{b}})}]{2002IAUC.7968....1H}
{Hamuy}, M. 2002{\natexlab{b}}, \iaucirc, 7968, 1

\bibitem[{{Hamuy}(2003{\natexlab{a}})}]{2003ApJ...582..905H}
{Hamuy}, M. 2003{\natexlab{a}}, \apj, 582, 905

\bibitem[{{Hamuy}(2003{\natexlab{b}})}]{2003IAUC.8102....4H}
{Hamuy}, M. 2003{\natexlab{b}}, \iaucirc, 8102, 4

\bibitem[{{Hamuy}(2003{\natexlab{c}})}]{2003IAUC.8117....2H}
{Hamuy}, M. 2003{\natexlab{c}}, \iaucirc, 8117, 2

\bibitem[{{Hamuy}(2003{\natexlab{d}})}]{2003IAUC.8045....3H}
{Hamuy}, M. 2003{\natexlab{d}}, \iaucirc, 8045, 3

\bibitem[{{Hamuy}(2004)}]{2004mmu..sympE...2H}
{Hamuy}, M. 2004, Measuring and Modeling the Universe, 2

\bibitem[{{Hamuy} {et~al.}(2009){Hamuy}, {Deng}, {Mazzali}, {Morrell},
  {Phillips}, {Roth}, {Gonzalez}, {Thomas-Osip}, {Krzeminski}, {Contreras},
  {Maza}, {Gonz{\'a}lez}, {Huerta}, {Folatelli}, {Chornock}, {Filippenko},
  {Persson}, {Freedman}, {Koviak}, {Suntzeff}, \&
  {Krisciunas}}]{2009ApJ...703.1612H}
{Hamuy}, M., {Deng}, J., {Mazzali}, P.~A., {et~al.} 2009, \apj, 703, 1612

\bibitem[{{Hamuy} {et~al.}(2006){Hamuy}, {Folatelli}, {Morrell}, {Phillips},
  {Suntzeff}, {Persson}, {Roth}, {Gonzalez}, {Krzeminski}, {Contreras},
  {Freedman}, {Murphy}, {Madore}, {Wyatt}, {Maza}, {Filippenko}, {Li}, \&
  {Pinto}}]{2006PASP..118....2H}
{Hamuy}, M., {Folatelli}, G., {Morrell}, N.~I., {et~al.} 2006, \pasp, 118, 2

\bibitem[{{Hamuy} {et~al.}(2003{\natexlab{a}}){Hamuy}, {Maza}, \&
  {Huerta}}]{2003IAUC.8109....1H}
{Hamuy}, M., {Maza}, J., \& {Huerta}, L. 2003{\natexlab{a}}, \iaucirc, 8109, 1

\bibitem[{{Hamuy} {et~al.}(1993){Hamuy}, {Maza}, {Phillips}, {Suntzeff},
  {Wischnjewsky}, {Smith}, {Antezana}, {Wells}, {Gonzalez}, {Gigoux},
  {Navarrete}, {Barrientos}, {Lamontagne}, {della Valle}, {Elias}, {Phillips},
  {Odewahn}, {Baldwin}, {Walker}, {Williams}, {Sturch}, {Baganoff}, {Chaboyer},
  {Schommer}, {Tirado}, {Hernandez}, {Ugarte}, {Guhathakurta}, {Howell},
  {Szkody}, {Schmidtke}, \& {Roth}}]{1993AJ....106.2392H}
{Hamuy}, M., {Maza}, J., {Phillips}, M.~M., {et~al.} 1993, \aj, 106, 2392

\bibitem[{{Hamuy} {et~al.}(2003{\natexlab{b}}){Hamuy}, {Morrell}, \&
  {Thomas-Osip}}]{2003IAUC.8183....2H}
{Hamuy}, M., {Morrell}, N., \& {Thomas-Osip}, J. 2003{\natexlab{b}}, \iaucirc,
  8183, 2

\bibitem[{{Hamuy} {et~al.}(2003{\natexlab{c}}){Hamuy}, {Phillips}, \&
  {Thomas-Osip}}]{2003IAUC.8088....3H}
{Hamuy}, M., {Phillips}, M., \& {Thomas-Osip}, J. 2003{\natexlab{c}}, \iaucirc,
  8088, 3

\bibitem[{{Hamuy} {et~al.}(1996){Hamuy}, {Phillips}, {Suntzeff}, {Schommer},
  {Maza}, {Antezan}, {Wischnjewsky}, {Valladares}, {Muena}, {Gonzales},
  {Aviles}, {Wells}, {Smith}, {Navarrete}, {Covarrubias}, {Williger}, {Walker},
  {Layden}, {Elias}, {Baldwin}, {Hernandez}, {Tirado}, {Ugarte}, {Elston},
  {Saavedra}, {Barrientos}, {Costa}, {Lira}, {Ruiz}, {Anguita}, {Gomez},
  {Ortiz}, {della Valle}, {Danziger}, {Storm}, {Kim}, {Bailyn}, {Rubenstein},
  {Tucker}, {Cersosimo}, {Mendez}, {Siciliano}, {Sherry}, {Chaboyer},
  {Koopmann}, {Geisler}, {Sarajedini}, {Dey}, {Tyson}, {Rich}, {Gal},
  {Lamontagne}, {Caldwell}, {Guhathakurta}, {Phillips}, {Szkody}, {Prosser},
  {Ho}, {McMahan}, {Baggley}, {Cheng}, {Havlen}, {Wakamatsu}, {Janes},
  {Malkan}, {Baganoff}, {Seitzer}, {Shara}, {Sturch}, {Hesser}, {Hartig},
  {Hughes}, {Welch}, {Williams}, {Ferguson}, {Francis}, {French}, {Bolte},
  {Roth}, {Odewahn}, {Howell}, \& {Krzeminski}}]{1996AJ....112.2408H}
{Hamuy}, M., {Phillips}, M.~M., {Suntzeff}, N.~B., {et~al.} 1996, \aj, 112,
  2408

\bibitem[{{Hamuy} \& {Pinto}(2002)}]{2002ApJ...566L..63H}
{Hamuy}, M. \& {Pinto}, P.~A. 2002, \apjl, 566, L63

\bibitem[{{Hamuy} {et~al.}(2001){Hamuy}, {Pinto}, {Maza}, {Suntzeff},
  {Phillips}, {Eastman}, {Smith}, {Corbally}, {Burstein}, {Li}, {Ivanov},
  {Moro-Martin}, {Strolger}, {de Souza}, {dos Anjos}, {Green}, {Pickering},
  {Gonz{\'a}lez}, {Antezana}, {Wischnjewsky}, {Galaz}, {Roth}, {Persson}, \&
  {Schommer}}]{2001ApJ...558..615H}
{Hamuy}, M., {Pinto}, P.~A., {Maza}, J., {et~al.} 2001, \apj, 558, 615

\bibitem[{{Hamuy} \& {Roth}(2003)}]{2003IAUC.8198....3H}
{Hamuy}, M. \& {Roth}, M. 2003, \iaucirc, 8198, 3

\bibitem[{{Hamuy} {et~al.}(2002){Hamuy}, {Shectman}, \&
  {Thompson}}]{2002IAUC.8001....2H}
{Hamuy}, M., {Shectman}, S., \& {Thompson}, I. 2002, \iaucirc, 8001, 2

\bibitem[{{Hamuy} \& {Suntzeff}(1990)}]{1990AJ.....99.1146H}
{Hamuy}, M. \& {Suntzeff}, N.~B. 1990, \aj, 99, 1146

\bibitem[{{Hamuy}(2001)}]{2001PhDT.......173H}
{Hamuy}, M.~A. 2001, PhD thesis, The University of Arizona

\bibitem[{{Heger} {et~al.}(2003){Heger}, {Fryer}, {Woosley}, {Langer}, \&
  {Hartmann}}]{2003ApJ...591..288H}
{Heger}, A., {Fryer}, C.~L., {Woosley}, S.~E., {Langer}, N., \& {Hartmann},
  D.~H. 2003, \apj, 591, 288

\bibitem[{{Hendry} {et~al.}(2006){Hendry}, {Smartt}, {Crockett}, {Maund},
  {Gal-Yam}, {Moon}, {Cenko}, {Fox}, {Kudritzki}, {Benn}, \&
  {{\O}stensen}}]{2006MNRAS.369.1303H}
{Hendry}, M.~A., {Smartt}, S.~J., {Crockett}, R.~M., {et~al.} 2006, \mnras,
  369, 1303

\bibitem[{{Holoien} {et~al.}(2014){Holoien}, {Prieto}, {Pejcha}, {Stanek},
  {Kochanek}, {Shappee}, {Grupe}, {Morrell}, {Thorstensen}, {Basu}, {Beacom},
  {Bersier}, {Brimacombe}, {Davis}, {Pojmanski}, \&
  {Szczygiel}}]{2014arXiv1411.3322H}
{Holoien}, T.~W.-S., {Prieto}, J.~L., {Pejcha}, O., {et~al.} 2014, ArXiv
  e-prints

\bibitem[{{Hurst} {et~al.}(1999){Hurst}, {Armstrong}, {James}, \&
  {Foulkes}}]{1999IAUC.7275....3H}
{Hurst}, G.~M., {Armstrong}, M., {James}, N., \& {Foulkes}, S. 1999, \iaucirc,
  7275, 3

\bibitem[{{Hutchings} {et~al.}(2002){Hutchings}, {Li}, \&
  {Wood-Vasey}}]{2002IAUC.7964....1H}
{Hutchings}, D., {Li}, W.~D., \& {Wood-Vasey}, W.~M. 2002, \iaucirc, 7964, 1

\bibitem[{{Inserra} {et~al.}(2013){Inserra}, {Pastorello}, {Turatto}, {Pumo},
  {Benetti}, {Cappellaro}, {Botticella}, {Bufano}, {Elias-Rosa}, {Harutyunyan},
  {Taubenberger}, {Valenti}, \& {Zampieri}}]{2013AA...555A.142I}
{Inserra}, C., {Pastorello}, A., {Turatto}, M., {et~al.} 2013, \aap, 555, A142

\bibitem[{{Inserra} {et~al.}(2012){Inserra}, {Turatto}, {Pastorello}, {Pumo},
  {Baron}, {Benetti}, {Cappellaro}, {Taubenberger}, {Bufano}, {Elias-Rosa},
  {Zampieri}, {Harutyunyan}, {Moskvitin}, {Nissinen}, {Stanishev}, {Tsvetkov},
  {Hentunen}, {Komarova}, {Pavlyuk}, {Sokolov}, \&
  {Sokolova}}]{2012MNRAS.422.1122I}
{Inserra}, C., {Turatto}, M., {Pastorello}, A., {et~al.} 2012, \mnras, 422,
  1122

\bibitem[{{Ivanov} \& {Shulman}(1990)}]{1990SvA....34..163I}
{Ivanov}, M.~A. \& {Shulman}, G.~A. 1990, \sovast, 34, 163

\bibitem[{{Janka}(2012)}]{2012ARNPS..62..407J}
{Janka}, H.-T. 2012, Annual Review of Nuclear and Particle Science, 62, 407

\bibitem[{{Janka} {et~al.}(2007){Janka}, {Langanke}, {Marek},
  {Mart{\'{\i}}nez-Pinedo}, \& {M{\"u}ller}}]{2007PhR...442...38J}
{Janka}, H.-T., {Langanke}, K., {Marek}, A., {Mart{\'{\i}}nez-Pinedo}, G., \&
  {M{\"u}ller}, B. 2007, \physrep, 442, 38

\bibitem[{{Jerkstrand} {et~al.}(2015){Jerkstrand}, {Smartt}, {Sollerman},
  {Inserra}, {Fraser}, {Spyromilio}, {Fransson}, {Chen}, {Barbarino},
  {Dall'Ora}, {Botticella}, {Della Valle}, {Gal-Yam}, {Valenti}, {Maguire},
  {Mazzali}, \& {Tomasella}}]{2015MNRAS.448.2482J}
{Jerkstrand}, A., {Smartt}, S.~J., {Sollerman}, J., {et~al.} 2015, \mnras, 448,
  2482

\bibitem[{{Jha} {et~al.}(1999{\natexlab{a}}){Jha}, {Challis}, {Garnavich},
  {Kirshner}, {Calkins}, \& {Stanek}}]{1999IAUC.7296....2J}
{Jha}, S., {Challis}, P., {Garnavich}, P., {et~al.} 1999{\natexlab{a}},
  \iaucirc, 7296, 2

\bibitem[{{Jha} {et~al.}(1999{\natexlab{b}}){Jha}, {Garnavich}, {Challis},
  {Kirshner}, \& {Berlind}}]{1999IAUC.7280....2J}
{Jha}, S., {Garnavich}, P., {Challis}, P., {Kirshner}, R., \& {Berlind}, P.
  1999{\natexlab{b}}, \iaucirc, 7280, 2

\bibitem[{{Jones} {et~al.}(2009){Jones}, {Hamuy}, {Lira}, {Maza},
  {Clocchiatti}, {Phillips}, {Morrell}, {Roth}, {Suntzeff}, {Matheson},
  {Filippenko}, {Foley}, \& {Leonard}}]{2009ApJ...696.1176J}
{Jones}, M.~I., {Hamuy}, M., {Lira}, P., {et~al.} 2009, \apj, 696, 1176

\bibitem[{{Kasen} \& {Woosley}(2009)}]{2009ApJ...703.2205K}
{Kasen}, D. \& {Woosley}, S.~E. 2009, \apj, 703, 2205

\bibitem[{{King}(1999)}]{1999IAUC.7141....1K}
{King}, J.~Y. 1999, \iaucirc, 7141, 1

\bibitem[{{Kirshner} \& {Silverman}(2003)}]{2003IAUC.8042....2K}
{Kirshner}, R. \& {Silverman}, J. 2003, \iaucirc, 8042, 2

\bibitem[{{Klotz} {et~al.}(2002){Klotz}, {Puckett}, {Langoussis}, {Wood-Vasey},
  {Aldering}, {Nugent}, \& {Stephens}}]{2002IAUC.7986....1K}
{Klotz}, A., {Puckett}, T., {Langoussis}, A., {et~al.} 2002, \iaucirc, 7986, 1

\bibitem[{{Krisciunas} {et~al.}(2009){Krisciunas}, {Hamuy}, {Suntzeff},
  {Espinoza}, {Gonzalez}, {Gonzalez}, {Gonzalez}, {Koviak}, {Krzeminski},
  {Morrell}, {Phillips}, {Roth}, \& {Thomas-Osip}}]{2009AJ....137...34K}
{Krisciunas}, K., {Hamuy}, M., {Suntzeff}, N.~B., {et~al.} 2009, \aj, 137, 34

\bibitem[{{Leonard} {et~al.}(2002){Leonard}, {Filippenko}, {Li}, {Matheson},
  {Kirshner}, {Chornock}, {Van Dyk}, {Berlind}, {Calkins}, {Challis},
  {Garnavich}, {Jha}, \& {Mahdavi}}]{2002AJ....124.2490L}
{Leonard}, D.~C., {Filippenko}, A.~V., {Li}, W., {et~al.} 2002, \aj, 124, 2490

\bibitem[{{Li} {et~al.}(2003){Li}, {Puckett}, {Kerns}, \&
  {Marcus}}]{2003IAUC.8214....1L}
{Li}, W., {Puckett}, T., {Kerns}, B., \& {Marcus}, M. 2003, \iaucirc, 8214, 1

\bibitem[{{Li}(1999)}]{1999IAUC.7294....1L}
{Li}, W.~D. 1999, \iaucirc, 7294, 1

\bibitem[{{Liu} {et~al.}(2000){Liu}, {Hu}, {Hang}, {Qiu}, {Zhu}, \&
  {Qiao}}]{2000AAS..144..219L}
{Liu}, Q.-Z., {Hu}, J.-Y., {Hang}, H.-R., {et~al.} 2000, \aaps, 144, 219

\bibitem[{{Llapasset} {et~al.}(2003){Llapasset}, {Yamaoka}, \&
  {Ayani}}]{2003CBET...48....1L}
{Llapasset}, J.-M., {Yamaoka}, H., \& {Ayani}, K. 2003, Central Bureau
  Electronic Telegrams, 48, 1

\bibitem[{{Lloyd Evans} {et~al.}(1986){Lloyd Evans}, {Evans}, \&
  {McNaught}}]{1986IAUC.4262....2L}
{Lloyd Evans}, T., {Evans}, R., \& {McNaught}, R.~H. 1986, \iaucirc, 4262, 2

\bibitem[{{Matheson} {et~al.}(2003{\natexlab{a}}){Matheson}, {Challis},
  {Kirshner}, \& {Berlind}}]{2003IAUC.8225....2M}
{Matheson}, T., {Challis}, P., {Kirshner}, R., \& {Berlind}, P.
  2003{\natexlab{a}}, \iaucirc, 8225, 2

\bibitem[{{Matheson} {et~al.}(2002){Matheson}, {Challis}, {Kirshner}, \&
  {Calkins}}]{2002IAUC.8016....3M}
{Matheson}, T., {Challis}, P., {Kirshner}, R., \& {Calkins}, M. 2002, \iaucirc,
  8016, 3

\bibitem[{{Matheson} {et~al.}(2003{\natexlab{b}}){Matheson}, {Challis},
  {Kirshner}, \& {Calkins}}]{2003IAUC.8134....2M}
{Matheson}, T., {Challis}, P., {Kirshner}, R., \& {Calkins}, M.
  2003{\natexlab{b}}, \iaucirc, 8134, 2

\bibitem[{{Matheson} {et~al.}(2003{\natexlab{c}}){Matheson}, {Challis},
  {Kirshner}, {Calkins}, \& {Berlind}}]{2003IAUC.8136....2M}
{Matheson}, T., {Challis}, P., {Kirshner}, R., {Calkins}, M., \& {Berlind}, P.
  2003{\natexlab{c}}, \iaucirc, 8136, 2

\bibitem[{{Maza} {et~al.}(1999){Maza}, {Hamuy}, {Antezana}, {Gonzalez},
  {Smith}, {Ruiz}, \& {Phillips}}]{1999IAUC.7210....1M}
{Maza}, J., {Hamuy}, M., {Antezana}, R., {et~al.} 1999, \iaucirc, 7210, 1

\bibitem[{{Maza} {et~al.}(1993{\natexlab{a}}){Maza}, {Hamuy}, {Antezana},
  {Valladares}, \& {Aviles}}]{1993IAUC.5812....2M}
{Maza}, J., {Hamuy}, M., {Antezana}, R., {Valladares}, G., \& {Aviles}, R.
  1993{\natexlab{a}}, \iaucirc, 5812, 2

\bibitem[{{Maza} {et~al.}(1992{\natexlab{a}}){Maza}, {Hamuy}, {Antezana},
  {Wells}, \& {Kim}}]{1992IAUC.5499....1M}
{Maza}, J., {Hamuy}, M., {Antezana}, R., {Wells}, L., \& {Kim}, Y.-C.
  1992{\natexlab{a}}, \iaucirc, 5499, 1

\bibitem[{{Maza} {et~al.}(1993{\natexlab{b}}){Maza}, {Hamuy}, {Valladares},
  {Wischnjewsky}, {Anguita}, \& {Antezana}}]{1993IAUC.5693....1M}
{Maza}, J., {Hamuy}, M., {Valladares}, G., {et~al.} 1993{\natexlab{b}},
  \iaucirc, 5693, 1

\bibitem[{{Maza} {et~al.}(1992{\natexlab{b}}){Maza}, {Hamuy}, {Wischnjewsky},
  {Antezana}, {Wells}, \& {Kim}}]{1992IAUC.5496....1M}
{Maza}, J., {Hamuy}, M., {Wischnjewsky}, M., {et~al.} 1992{\natexlab{b}},
  \iaucirc, 5496, 1

\bibitem[{{Mazzali} {et~al.}(2009){Mazzali}, {Deng}, {Hamuy}, \&
  {Nomoto}}]{2009ApJ...703.1624M}
{Mazzali}, P.~A., {Deng}, J., {Hamuy}, M., \& {Nomoto}, K. 2009, \apj, 703,
  1624

\bibitem[{{McNaught} {et~al.}(1992){McNaught}, {Evans}, {Spyromilio}, {Taylor},
  {Hawkins}, \& {Vernon}}]{1992IAUC.5552....1M}
{McNaught}, R.~H., {Evans}, R., {Spyromilio}, J., {et~al.} 1992, \iaucirc,
  5552, 1

\bibitem[{{Miknaitis} {et~al.}(2002){Miknaitis}, {Miceli}, {Garg}, {Chornock},
  {Jha}, {Filippenko}, {Barris}, \& {Hamuy}}]{2002IAUC.8020....1M}
{Miknaitis}, G., {Miceli}, A., {Garg}, A., {et~al.} 2002, \iaucirc, 8020, 1

\bibitem[{{Misra} {et~al.}(2007){Misra}, {Pooley}, {Chandra}, {Bhattacharya},
  {Ray}, {Sagar}, \& {Lewin}}]{2007MNRAS.381..280M}
{Misra}, K., {Pooley}, D., {Chandra}, P., {et~al.} 2007, \mnras, 381, 280

\bibitem[{{Monard}(2002)}]{2002IAUC.7995....2M}
{Monard}, L.~A.~G. 2002, \iaucirc, 7995, 2

\bibitem[{{Monard}(2003)}]{2003IAUC.8186....2M}
{Monard}, L.~A.~G. 2003, \iaucirc, 8186, 2

\bibitem[{{Moore} \& {Li}(2003)}]{2003CBET...40....1M}
{Moore}, M. \& {Li}, W. 2003, Central Bureau Electronic Telegrams, 40, 1

\bibitem[{{Moore} {et~al.}(2003){Moore}, {Li}, \&
  {Boles}}]{2003IAUC.8184....2M}
{Moore}, M., {Li}, W., \& {Boles}, T. 2003, \iaucirc, 8184, 2

\bibitem[{{Morrell} \& {Hamuy}(2003)}]{2003IAUC.8203....2M}
{Morrell}, N. \& {Hamuy}, M. 2003, \iaucirc, 8203, 2

\bibitem[{{Nomoto} {et~al.}(1995){Nomoto}, {Iwamoto}, \&
  {Suzuki}}]{1995PhR...256..173N}
{Nomoto}, K.~I., {Iwamoto}, K., \& {Suzuki}, T. 1995, \physrep, 256, 173

\bibitem[{{Olivares E.} {et~al.}(2010){Olivares E.}, {Hamuy}, {Pignata},
  {Maza}, {Bersten}, {Phillips}, {Suntzeff}, {Filippenko}, {Morrel},
  {Kirshner}, \& {Matheson}}]{2010ApJ...715..833O}
{Olivares E.}, F., {Hamuy}, M., {Pignata}, G., {et~al.} 2010, \apj, 715, 833

\bibitem[{{Papenkova} \& {Li}(2003)}]{2003IAUC.8044....1P}
{Papenkova}, M. \& {Li}, W. 2003, \iaucirc, 8044, 1

\bibitem[{{Papenkova} {et~al.}(2003){Papenkova}, {Li}, {Lotoss/Kait},
  {Schmidt}, {Salvo}, \& {Ford}}]{2003IAUC.8143....2P}
{Papenkova}, M., {Li}, W., {Lotoss/Kait}, {et~al.} 2003, \iaucirc, 8143, 2

\bibitem[{{Papenkova} \& {Li}(2000)}]{2000IAUC.7410....1P}
{Papenkova}, M. \& {Li}, W.~D. 2000, \iaucirc, 7410, 1

\bibitem[{{Pastorello} {et~al.}(2009){Pastorello}, {Valenti}, {Zampieri},
  {Navasardyan}, {Taubenberger}, {Smartt}, {Arkharov}, {B{\"a}rnbantner},
  {Barwig}, {Benetti}, {Birtwhistle}, {Botticella}, {Cappellaro}, {Del
  Principe}, {di Mille}, {di Rico}, {Dolci}, {Elias-Rosa}, {Efimova},
  {Fiedler}, {Harutyunyan}, {H{\"o}flich}, {Kloehr}, {Larionov}, {Lorenzi},
  {Maund}, {Napoleone}, {Ragni}, {Richmond}, {Ries}, {Spiro}, {Temporin},
  {Turatto}, \& {Wheeler}}]{2009MNRAS.394.2266P}
{Pastorello}, A., {Valenti}, S., {Zampieri}, L., {et~al.} 2009, \mnras, 394,
  2266

\bibitem[{{Pastorello} {et~al.}(2004){Pastorello}, {Zampieri}, {Turatto},
  {Cappellaro}, {Meikle}, {Benetti}, {Branch}, {Baron}, {Patat}, {Armstrong},
  {Altavilla}, {Salvo}, \& {Riello}}]{2004MNRAS.347...74P}
{Pastorello}, A., {Zampieri}, L., {Turatto}, M., {et~al.} 2004, \mnras, 347, 74

\bibitem[{{Patat} {et~al.}(1999){Patat}, {Maza}, {Benetti}, \&
  {Cappellaro}}]{1999IAUC.7160....2P}
{Patat}, F., {Maza}, J., {Benetti}, S., \& {Cappellaro}, E. 1999, \iaucirc,
  7160, 2

\bibitem[{{Pejcha} \& {Prieto}(2015{\natexlab{a}})}]{2015ApJ...799..215P}
{Pejcha}, O. \& {Prieto}, J.~L. 2015{\natexlab{a}}, \apj, 799, 215

\bibitem[{{Pejcha} \& {Prieto}(2015{\natexlab{b}})}]{2015ApJ...806..225P}
{Pejcha}, O. \& {Prieto}, J.~L. 2015{\natexlab{b}}, \apj, 806, 225

\bibitem[{{Pennypacker} {et~al.}(1989){Pennypacker}, {Burns}, {Crawford},
  {Friedman}, {Graham}, {Kare}, {Muller}, {Perlmutter}, {Smith}, {Treffers},
  {Williams}, {Basri}, {Bixler}, {Filippenko}, {Foltz}, {Garnett}, {Harkness},
  {Junkkarinen}, {Kennicutt}, {McCarthy}, {Spinrad}, {Wheeler}, {Willick}, \&
  {Wills}}]{1989AJ.....97..186P}
{Pennypacker}, C.~R., {Burns}, M.~S., {Crawford}, F.~S., {et~al.} 1989, \aj,
  97, 186

\bibitem[{{Phillips} \& {Hamuy}(2003)}]{2003IAUC.8130....4P}
{Phillips}, M. \& {Hamuy}, M. 2003, \iaucirc, 8130, 4

\bibitem[{{Phillips} {et~al.}(2003){Phillips}, {Hamuy}, {Roth}, \&
  {Morrell}}]{2003IAUC.8086....2P}
{Phillips}, M., {Hamuy}, M., {Roth}, M., \& {Morrell}, N. 2003, \iaucirc, 8086,
  2

\bibitem[{{Phillips} {et~al.}(1992){Phillips}, {Maza}, {Antezana}, {Wells},
  {Muena}, {Tyson}, {Gal}, \& {Alonso}}]{1992IAUC.5570....2P}
{Phillips}, M., {Maza}, J., {Antezana}, R., {et~al.} 1992, \iaucirc, 5570, 2

\bibitem[{{Phillips}(1992)}]{1992IAUC.5521....1P}
{Phillips}, M.~M. 1992, \iaucirc, 5521, 1

\bibitem[{{Phillips}(1993{\natexlab{a}})}]{1993IAUC.5699....2P}
{Phillips}, M.~M. 1993{\natexlab{a}}, \iaucirc, 5699, 2

\bibitem[{{Phillips}(1993{\natexlab{b}})}]{1993ApJ...413L.105P}
{Phillips}, M.~M. 1993{\natexlab{b}}, \apjl, 413, L105

\bibitem[{{Phillips} {et~al.}(1990){Phillips}, {Hamuy}, {Heathcote},
  {Suntzeff}, \& {Kirhakos}}]{1990AJ.....99.1133P}
{Phillips}, M.~M., {Hamuy}, M., {Heathcote}, S.~R., {Suntzeff}, N.~B., \&
  {Kirhakos}, S. 1990, \aj, 99, 1133

\bibitem[{{Planck Collaboration} {et~al.}(2015){Planck Collaboration}, {Ade},
  {Aghanim}, {Arnaud}, {Ashdown}, {Aumont}, {Baccigalupi}, {Banday},
  {Barreiro}, {Bartlett}, \& et~al.}]{2015arXiv150201589P}
{Planck Collaboration}, {Ade}, P.~A.~R., {Aghanim}, N., {et~al.} 2015, ArXiv
  e-prints

\bibitem[{{Popov}(1993)}]{1993ApJ...414..712P}
{Popov}, D.~V. 1993, \apj, 414, 712

\bibitem[{{Poznanski} {et~al.}(2003){Poznanski}, {Gal-Yam}, {Sharon}, {Maoz},
  {Jannuzi}, {Papenkova}, {Li}, \& {Schwartz}}]{2003IAUC.8058....1P}
{Poznanski}, D., {Gal-Yam}, A., {Sharon}, K., {et~al.} 2003, \iaucirc, 8058, 1

\bibitem[{{Pozzo} {et~al.}(2006){Pozzo}, {Meikle}, {Rayner}, {Joseph},
  {Filippenko}, {Foley}, {Li}, {Mattila}, \& {Sollerman}}]{2006MNRAS.368.1169P}
{Pozzo}, M., {Meikle}, W.~P.~S., {Rayner}, J.~T., {et~al.} 2006, \mnras, 368,
  1169

\bibitem[{{Prieto} {et~al.}(2012){Prieto}, {Lee}, {Drake}, {McNaught},
  {Garradd}, {Beacom}, {Beshore}, {Catelan}, {Djorgovski}, {Pojmanski},
  {Stanek}, \& {Szczygie{\l}}}]{2012ApJ...745...70P}
{Prieto}, J.~L., {Lee}, J.~C., {Drake}, A.~J., {et~al.} 2012, \apj, 745, 70

\bibitem[{{Pritchard} {et~al.}(2014){Pritchard}, {Roming}, {Brown}, {Bayless},
  \& {Frey}}]{2014ApJ...787..157P}
{Pritchard}, T.~A., {Roming}, P.~W.~A., {Brown}, P.~J., {Bayless}, A.~J., \&
  {Frey}, L.~H. 2014, \apj, 787, 157

\bibitem[{{Puckett} {et~al.}(2003){Puckett}, {Toth}, {Schwartz}, {Holvorcem},
  {Wood-Vasey}, {Aldering}, {Nugent}, \& {Kulkarni}}]{2003IAUC.8117....1P}
{Puckett}, T., {Toth}, D., {Schwartz}, M., {et~al.} 2003, \iaucirc, 8117, 1

\bibitem[{{Revnivtsev} {et~al.}(2003){Revnivtsev}, {Tuerler}, {Del Santo},
  {Westergaard}, {Gehrels}, \& {Winkler}}]{2003IAUC.8097....2R}
{Revnivtsev}, M., {Tuerler}, M., {Del Santo}, M., {et~al.} 2003, \iaucirc,
  8097, 2

\bibitem[{{Richmond}(2014)}]{2014JAVSO..42..333R}
{Richmond}, M.~W. 2014, Journal of the American Association of Variable Star
  Observers (JAAVSO), 42, 333

\bibitem[{{Rodr{\'{\i}}guez} {et~al.}(2014){Rodr{\'{\i}}guez}, {Clocchiatti},
  \& {Hamuy}}]{2014AJ....148..107R}
{Rodr{\'{\i}}guez}, {\'O}., {Clocchiatti}, A., \& {Hamuy}, M. 2014, \aj, 148,
  107

\bibitem[{{Salvo} {et~al.}(2003{\natexlab{a}}){Salvo}, {Bessell}, \&
  {Schmidt}}]{2003IAUC.8187....1S}
{Salvo}, M., {Bessell}, M., \& {Schmidt}, B. 2003{\natexlab{a}}, \iaucirc,
  8187, 1

\bibitem[{{Salvo} {et~al.}(2003{\natexlab{b}}){Salvo}, {Schmidt}, \&
  {Tonry}}]{2003IAUC.8098....2S}
{Salvo}, M., {Schmidt}, B., \& {Tonry}, J. 2003{\natexlab{b}}, \iaucirc, 8098,
  2

\bibitem[{{Sanders} {et~al.}(2015){Sanders}, {Soderberg}, {Gezari},
  {Betancourt}, {Chornock}, {Berger}, {Foley}, {Challis}, {Drout}, {Kirshner},
  {Lunnan}, {Marion}, {Margutti}, {McKinnon}, {Milisavljevic}, {Narayan},
  {Rest}, {Kankare}, {Mattila}, {Smartt}, {Huber}, {Burgett}, {Draper},
  {Hodapp}, {Kaiser}, {Kudritzki}, {Magnier}, {Metcalfe}, {Morgan}, {Price},
  {Tonry}, {Wainscoat}, \& {Waters}}]{2015ApJ...799..208S}
{Sanders}, N.~E., {Soderberg}, A.~M., {Gezari}, S., {et~al.} 2015, \apj, 799,
  208

\bibitem[{{Schlafly} \& {Finkbeiner}(2011)}]{2011ApJ...737..103S}
{Schlafly}, E.~F. \& {Finkbeiner}, D.~P. 2011, \apj, 737, 103

\bibitem[{{Schlegel}(1990)}]{1990MNRAS.244..269S}
{Schlegel}, E.~M. 1990, \mnras, 244, 269

\bibitem[{{Schmidt} {et~al.}(1992){Schmidt}, {Kirshner}, \&
  {Eastman}}]{1992ApJ...395..366S}
{Schmidt}, B.~P., {Kirshner}, R.~P., \& {Eastman}, R.~G. 1992, \apj, 395, 366

\bibitem[{{Schmidt} {et~al.}(1994{\natexlab{a}}){Schmidt}, {Kirshner},
  {Eastman}, {Hamuy}, {Phillips}, {Suntzeff}, {Maza}, {Filippenko}, {Ho},
  {Matheson}, {Grashuis}, {Aviles}, {Kirkpatrick}, {Challis}, {Kuijken},
  {Zucker}, {Bolte}, \& {Tyson}}]{1994AJ....107.1444S}
{Schmidt}, B.~P., {Kirshner}, R.~P., {Eastman}, R.~G., {et~al.}
  1994{\natexlab{a}}, \aj, 107, 1444

\bibitem[{{Schmidt} {et~al.}(1994{\natexlab{b}}){Schmidt}, {Kirshner},
  {Eastman}, {Phillips}, {Suntzeff}, {Hamuy}, {Maza}, \&
  {Aviles}}]{1994ApJ...432...42S}
{Schmidt}, B.~P., {Kirshner}, R.~P., {Eastman}, R.~G., {et~al.}
  1994{\natexlab{b}}, \apj, 432, 42

\bibitem[{{Schmidt} {et~al.}(1993){Schmidt}, {Kirshner}, {Schild},
  {Leibundgut}, {Jeffery}, {Willner}, {Peletier}, {Zabludoff}, {Phillips},
  {Suntzeff}, {Hamuy}, {Wells}, {Smith}, {Baldwin}, {Weller}, {Navarette},
  {Gonzalez}, {Filippenko}, {Shields}, {Steidel}, {Perlmutter}, {Pennypacker},
  {Smith}, {Porter}, {Boroson}, {Stathakis}, {Cannon}, {Peters}, {Horine},
  {Freeman}, {Womble}, {Stone}, {Marschall}, {Phillips}, {Saha}, \&
  {Bond}}]{1993AJ....105.2236S}
{Schmidt}, B.~P., {Kirshner}, R.~P., {Schild}, R., {et~al.} 1993, \aj, 105,
  2236

\bibitem[{{Singer} {et~al.}(2003){Singer}, {Beutler}, {Swift}, {Li}, {Yamaoka},
  \& {Itagaki}}]{2003IAUC.8201....1S}
{Singer}, D., {Beutler}, B., {Swift}, B., {et~al.} 2003, \iaucirc, 8201, 1

\bibitem[{{Smartt}(2015)}]{2015PASA...32...16S}
{Smartt}, S.~J. 2015, \pasa, 32, 16

\bibitem[{{Spiro} {et~al.}(2014){Spiro}, {Pastorello}, {Pumo}, {Zampieri},
  {Turatto}, {Smartt}, {Benetti}, {Cappellaro}, {Valenti}, {Agnoletto},
  {Altavilla}, {Aoki}, {Brocato}, {Corsini}, {Di Cianno}, {Elias-Rosa},
  {Hamuy}, {Enya}, {Fiaschi}, {Folatelli}, {Desidera}, {Harutyunyan}, {Howell},
  {Kawka}, {Kobayashi}, {Leibundgut}, {Minezaki}, {Navasardyan}, {Nomoto},
  {Mattila}, {Pietrinferni}, {Pignata}, {Raimondo}, {Salvo}, {Schmidt},
  {Sollerman}, {Spyromilio}, {Taubenberger}, {Valentini}, {Vennes}, \&
  {Yoshii}}]{2014MNRAS.439.2873S}
{Spiro}, S., {Pastorello}, A., {Pumo}, M.~L., {et~al.} 2014, \mnras, 439, 2873

\bibitem[{{Stritzinger} {et~al.}(2002){Stritzinger}, {Hamuy}, {Suntzeff},
  {Smith}, {Phillips}, {Maza}, {Strolger}, {Antezana}, {Gonz{\'a}lez},
  {Wischnjewsky}, {Candia}, {Espinoza}, {Gonz{\'a}lez}, {Stubbs}, {Becker},
  {Rubenstein}, \& {Galaz}}]{2002AJ....124.2100S}
{Stritzinger}, M., {Hamuy}, M., {Suntzeff}, N.~B., {et~al.} 2002, \aj, 124,
  2100

\bibitem[{{Stritzinger} {et~al.}(2011){Stritzinger}, {Phillips}, {Boldt},
  {Burns}, {Campillay}, {Contreras}, {Gonzalez}, {Folatelli}, {Morrell},
  {Krzeminski}, {Roth}, {Salgado}, {DePoy}, {Hamuy}, {Freedman}, {Madore},
  {Marshall}, {Persson}, {Rheault}, {Suntzeff}, {Villanueva}, {Li}, \&
  {Filippenko}}]{2011AJ....142..156S}
{Stritzinger}, M.~D., {Phillips}, M.~M., {Boldt}, L.~N., {et~al.} 2011, \aj,
  142, 156

\bibitem[{{Swift} {et~al.}(2003){Swift}, {Weisz}, {Li}, \&
  {Boles}}]{2003IAUC.8086....1S}
{Swift}, B., {Weisz}, D., {Li}, W., \& {Boles}, T. 2003, \iaucirc, 8086, 1

\bibitem[{{Taddia} {et~al.}(2013){Taddia}, {Stritzinger}, {Sollerman},
  {Phillips}, {Anderson}, {Boldt}, {Campillay}, {Castell{\'o}n}, {Contreras},
  {Folatelli}, {Hamuy}, {Heinrich-Josties}, {Krzeminski}, {Morrell}, {Burns},
  {Freedman}, {Madore}, {Persson}, \& {Suntzeff}}]{2013A&A...555A..10T}
{Taddia}, F., {Stritzinger}, M.~D., {Sollerman}, J., {et~al.} 2013, \aap, 555,
  A10

\bibitem[{{Tak{\'a}ts} {et~al.}(2015){Tak{\'a}ts}, {Pignata}, {Pumo},
  {Paillas}, {Zampieri}, {Elias-Rosa}, {Benetti}, {Bufano}, {Cappellaro},
  {Ergon}, {Fraser}, {Hamuy}, {Inserra}, {Kankare}, {Smartt}, {Stritzinger},
  {Van Dyk}, {Haislip}, {LaCluyze}, {Moore}, \&
  {Reichart}}]{2015MNRAS.450.3137T}
{Tak{\'a}ts}, K., {Pignata}, G., {Pumo}, M.~L., {et~al.} 2015, \mnras, 450,
  3137

\bibitem[{{Tomasella} {et~al.}(2013){Tomasella}, {Cappellaro}, {Fraser},
  {Pumo}, {Pastorello}, {Pignata}, {Benetti}, {Bufano}, {Dennefeld},
  {Harutyunyan}, {Iijima}, {Jerkstrand}, {Kankare}, {Kotak}, {Magill},
  {Nascimbeni}, {Ochner}, {Siviero}, {Smartt}, {Sollerman}, {Stanishev},
  {Taddia}, {Taubenberger}, {Turatto}, {Valenti}, {Wright}, \&
  {Zampieri}}]{2013MNRAS.434.1636T}
{Tomasella}, L., {Cappellaro}, E., {Fraser}, M., {et~al.} 2013, \mnras, 434,
  1636

\bibitem[{{Tsvetkov}(1994)}]{1994AstL...20..374T}
{Tsvetkov}, D.~Y. 1994, Astronomy Letters, 20, 374

\bibitem[{{Tsvetkov}(2006)}]{2006PZ.....26....3T}
{Tsvetkov}, D.~Y. 2006, Peremennye Zvezdy, 26, 3

\bibitem[{{Tsvetkov}(2008)}]{2008PZ.....28....3T}
{Tsvetkov}, D.~Y. 2008, Peremennye Zvezdy, 28, 3

\bibitem[{{Tsvetkov} {et~al.}(2008){Tsvetkov}, {Goranskij}, \&
  {Pavlyuk}}]{2008PZ.....28....8T}
{Tsvetkov}, D.~Y., {Goranskij}, V., \& {Pavlyuk}, N. 2008, Peremennye Zvezdy,
  28, 8

\bibitem[{{Tsvetkov} {et~al.}(2007){Tsvetkov}, {Muminov}, {Burkhanov}, \&
  {Kahharov}}]{2007PZ.....27....5T}
{Tsvetkov}, D.~Y., {Muminov}, M., {Burkhanov}, O., \& {Kahharov}, B. 2007,
  Peremennye Zvezdy, 27, 5

\bibitem[{{Tsvetkov} {et~al.}(2006){Tsvetkov}, {Volnova}, {Shulga}, {Korotkiy},
  {Elmhamdi}, {Danziger}, \& {Ereshko}}]{2006AA...460..769T}
{Tsvetkov}, D.~Y., {Volnova}, A.~A., {Shulga}, A.~P., {et~al.} 2006, \aap, 460,
  769

\bibitem[{{Turatto} {et~al.}(1993){Turatto}, {Cappellaro}, {Benetti}, \&
  {Danziger}}]{1993MNRAS.265..471T}
{Turatto}, M., {Cappellaro}, E., {Benetti}, S., \& {Danziger}, I.~J. 1993,
  \mnras, 265, 471

\bibitem[{{Valenti} {et~al.}(2014){Valenti}, {Sand}, {Pastorello}, {Graham},
  {Howell}, {Parrent}, {Tomasella}, {Ochner}, {Fraser}, {Benetti}, {Yuan},
  {Smartt}, {Maund}, {Arcavi}, {Gal-Yam}, {Inserra}, \&
  {Young}}]{2014MNRAS.438L.101V}
{Valenti}, S., {Sand}, D., {Pastorello}, A., {et~al.} 2014, \mnras, 438, L101

\bibitem[{{Vink{\'o}} {et~al.}(2006){Vink{\'o}}, {Tak{\'a}ts}, {S{\'a}rneczky},
  {Szab{\'o}}, {M{\'e}sz{\'a}ros}, {Csorv{\'a}si}, {Szalai}, {G{\'a}sp{\'a}r},
  {P{\'a}l}, {Csizmadia}, {K{\'o}sp{\'a}l}, {R{\'a}cz}, {Kun}, {Cs{\'a}k},
  {F{\"u}r{\'e}sz}, {DeBond}, {Grunhut}, {Thomson}, {Mochnacki}, \&
  {Koktay}}]{2006MNRAS.369.1780V}
{Vink{\'o}}, J., {Tak{\'a}ts}, K., {S{\'a}rneczky}, K., {et~al.} 2006, \mnras,
  369, 1780

\bibitem[{{Weisz} \& {Li}(2003)}]{2003IAUC.8131....1W}
{Weisz}, D. \& {Li}, W. 2003, \iaucirc, 8131, 1

\bibitem[{{Wells} {et~al.}(1992){Wells}, {Maza}, {Antezana}, {Wakamatsu},
  {Malkan}, \& {Anguita}}]{1992IAUC.5554....1W}
{Wells}, L., {Maza}, J., {Antezana}, R., {et~al.} 1992, \iaucirc, 5554, 1

\bibitem[{{Wells} {et~al.}(1991){Wells}, {Maza}, {Wischnjewsky}, {Antezana},
  {Yee}, \& {Ellingson}}]{1991IAUC.5310....1W}
{Wells}, L., {Maza}, J., {Wischnjewsky}, M., {et~al.} 1991, \iaucirc, 5310, 1

\bibitem[{{Williams} {et~al.}(1993){Williams}, {Martin}, {Schmidtke},
  {Phillips}, {Maza}, \& {Wischnjewsky}}]{1993IAUC.5733....1W}
{Williams}, A., {Martin}, R., {Schmidtke}, P.~C., {et~al.} 1993, \iaucirc,
  5733, 1

\bibitem[{{Winzer}(1974)}]{1974JRASC..68...36W}
{Winzer}, J.~E. 1974, \jrasc, 68, 36

\bibitem[{{Wood} \& {Andrews}(1974)}]{1974MNRAS.167...13W}
{Wood}, R. \& {Andrews}, P.~J. 1974, \mnras, 167, 13

\bibitem[{{Wood-Vasey} {et~al.}(2004){Wood-Vasey}, {Aldering}, {Lee}, {Loken},
  {Nugent}, {Perlmutter}, {Siegrist}, {Wang}, {Antilogus}, {Astier}, {Hardin},
  {Pain}, {Copin}, {Smadja}, {Gangler}, {Castera}, {Adam}, {Bacon},
  {Lemonnier}, {P{\'e}contal}, {P{\'e}contal}, \&
  {Kessler}}]{2004NewAR..48..637W}
{Wood-Vasey}, W.~M., {Aldering}, G., {Lee}, B.~C., {et~al.} 2004, \nar, 48, 637

\bibitem[{{Wood-Vasey} {et~al.}(2003{\natexlab{a}}){Wood-Vasey}, {Aldering}, \&
  {Nugent}}]{2003IAUC.8105....1W}
{Wood-Vasey}, W.~M., {Aldering}, G., \& {Nugent}, P. 2003{\natexlab{a}},
  \iaucirc, 8105, 1

\bibitem[{{Wood-Vasey} {et~al.}(2003{\natexlab{b}}){Wood-Vasey}, {Aldering}, \&
  {Nugent}}]{2003IAUC.8104....2W}
{Wood-Vasey}, W.~M., {Aldering}, G., \& {Nugent}, P. 2003{\natexlab{b}},
  \iaucirc, 8104, 2

\bibitem[{{Wood-Vasey} {et~al.}(2003{\natexlab{c}}){Wood-Vasey}, {Aldering},
  {Nugent}, \& {Chassagne}}]{2003IAUC.8082....1W}
{Wood-Vasey}, W.~M., {Aldering}, G., {Nugent}, P., \& {Chassagne}, R.
  2003{\natexlab{c}}, \iaucirc, 8082, 1

\bibitem[{{Wood-Vasey} {et~al.}(2002{\natexlab{a}}){Wood-Vasey}, {Aldering},
  {Nugent}, \& {Li}}]{2002IAUC.8006....3W}
{Wood-Vasey}, W.~M., {Aldering}, G., {Nugent}, P., \& {Li}, K.
  2002{\natexlab{a}}, \iaucirc, 8006, 3

\bibitem[{{Wood-Vasey} {et~al.}(2003{\natexlab{d}}){Wood-Vasey}, {Aldering},
  {Nugent}, {Mulchaey}, \& {Phillips}}]{2003IAUC.8088....2W}
{Wood-Vasey}, W.~M., {Aldering}, G., {Nugent}, P., {Mulchaey}, J., \&
  {Phillips}, M. 2003{\natexlab{d}}, \iaucirc, 8088, 2

\bibitem[{{Wood-Vasey} {et~al.}(2003{\natexlab{e}}){Wood-Vasey}, {Aldering},
  {Nugent}, {Papenkova}, \& {Li}}]{2003IAUC.8101....2W}
{Wood-Vasey}, W.~M., {Aldering}, G., {Nugent}, P., {Papenkova}, M., \& {Li}, W.
  2003{\natexlab{e}}, \iaucirc, 8101, 2

\bibitem[{{Wood-Vasey} {et~al.}(2002{\natexlab{b}}){Wood-Vasey}, {Farris},
  {Weisz}, \& {Li}}]{2002IAUC.7967....1W}
{Wood-Vasey}, W.~M., {Farris}, B., {Weisz}, D., \& {Li}, W.~D.
  2002{\natexlab{b}}, \iaucirc, 7967, 1

\bibitem[{{Woodings} {et~al.}(1999){Woodings}, {Martin}, {Williams}, {Verveer},
  \& {Biggs}}]{1999IAUC.7158....1W}
{Woodings}, S., {Martin}, R., {Williams}, A., {Verveer}, A., \& {Biggs}, J.
  1999, \iaucirc, 7158, 1

\bibitem[{{Woosley} {et~al.}(1993){Woosley}, {Langer}, \&
  {Weaver}}]{1993ApJ...411..823W}
{Woosley}, S.~E., {Langer}, N., \& {Weaver}, T.~A. 1993, \apj, 411, 823

\bibitem[{{Zampieri} {et~al.}(2003){Zampieri}, {Pastorello}, {Turatto},
  {Cappellaro}, {Benetti}, {Altavilla}, {Mazzali}, \&
  {Hamuy}}]{2003MNRAS.338..711Z}
{Zampieri}, L., {Pastorello}, A., {Turatto}, M., {et~al.} 2003, \mnras, 338,
  711

\bibitem[{{Zhang} {et~al.}(2014){Zhang}, {Wang}, {Mazzali}, {Bai}, {Zhang},
  {Bersier}, {Huang}, {Fan}, {Mo}, {Wang}, {Yi}, {Wang}, {Xin}, {Liangchang},
  {Zhang}, {Lun}, {Wang}, {He}, \& {Walker}}]{2014ApJ...797....5Z}
{Zhang}, J., {Wang}, X., {Mazzali}, P.~A., {et~al.} 2014, \apj, 797, 5

\bibitem[{{Zhang} {et~al.}(2006){Zhang}, {Wang}, {Li}, {Zhou}, {Ma}, {Jiang},
  \& {Chen}}]{2006AJ....131.2245Z}
{Zhang}, T., {Wang}, X., {Li}, W., {et~al.} 2006, \aj, 131, 2245

\end{thebibliography}

\begin{figure*}
\centering
\includegraphics*[trim=0.0cm 0.0cm 0.0cm 0.0cm, clip=true,width=0.9\hsize]{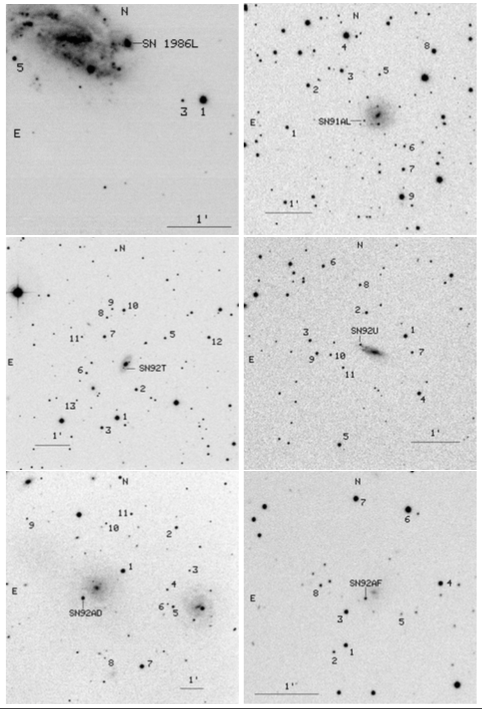}
\caption{$V$-band images of the 51 SNe. North is up and east is to the left. Both the supernova and the comparison stars used to derive differential photometry of the SN are labeled. The scale is shown with an horizontal line near the bottom. Here, supernovae 1986L, 1991al, 1992T, 1992U, 1992ad, and 1992af are shown.
\label{charts.fig}}
\end{figure*}
\begin{figure*}
\renewcommand{\thefigure}{\arabic{figure} (Cont.)}
\addtocounter{figure}{-1}
\centering
\includegraphics*[trim=0.0cm 0.0cm 0.0cm 0.0cm, clip=true,width=0.9\hsize]{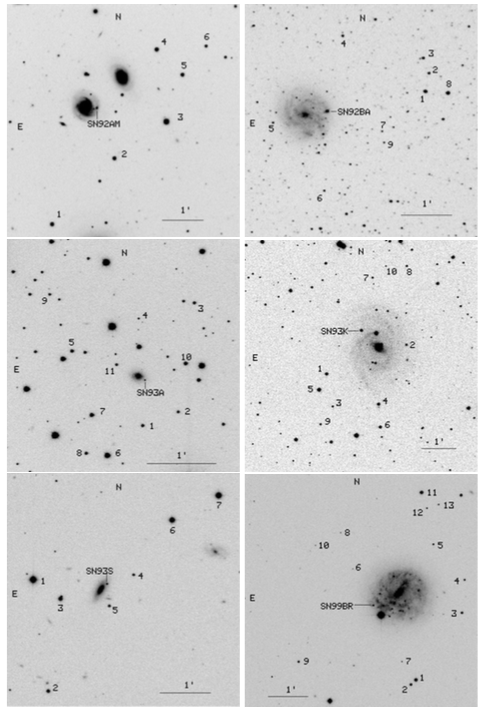}
\caption{Supernovae 1992am, 1992ba, 1993A, 1993K, 1993S, and 1999br.}
\end{figure*}
\begin{figure*}
\renewcommand{\thefigure}{\arabic{figure} (Cont.)}
\addtocounter{figure}{-1}
\centering
\includegraphics*[trim=0.0cm 0.0cm 0.0cm 0.0cm, clip=true,width=0.9\hsize]{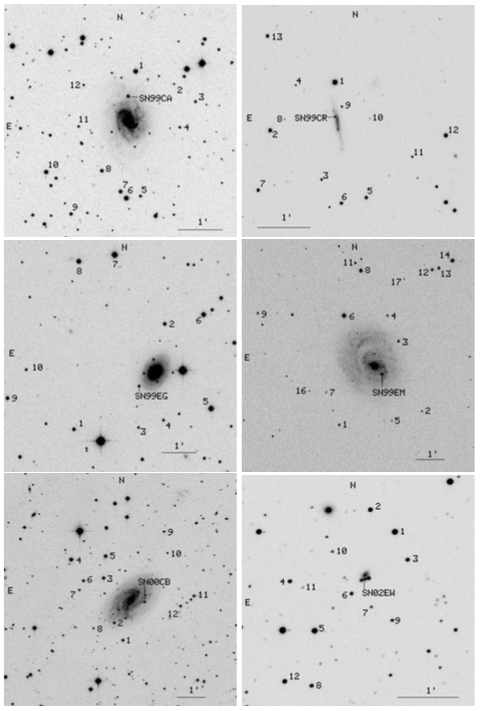}
\caption{Supernovae 1999ca, 1999cr, 1999eg, 1999em, 2000cb, and 2002ew.}
\end{figure*}
\begin{figure*}
\renewcommand{\thefigure}{\arabic{figure} (Cont.)}
\addtocounter{figure}{-1}
\centering
\includegraphics*[trim=0.0cm 0.0cm 0.0cm 0.0cm, clip=true,width=0.9\hsize]{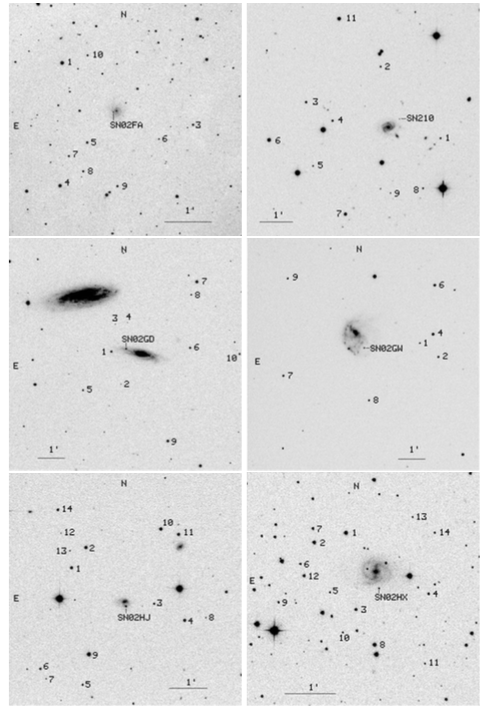}
\caption{Supernovae 2002fa, 0210, 2002gd, 2002gw, 2002hj, and 2002hx.}
\end{figure*}
\begin{figure*}
\renewcommand{\thefigure}{\arabic{figure} (Cont.)}
\addtocounter{figure}{-1}
\centering
\includegraphics*[trim=0.0cm 0.0cm 0.0cm 0.0cm, clip=true,width=0.9\hsize]{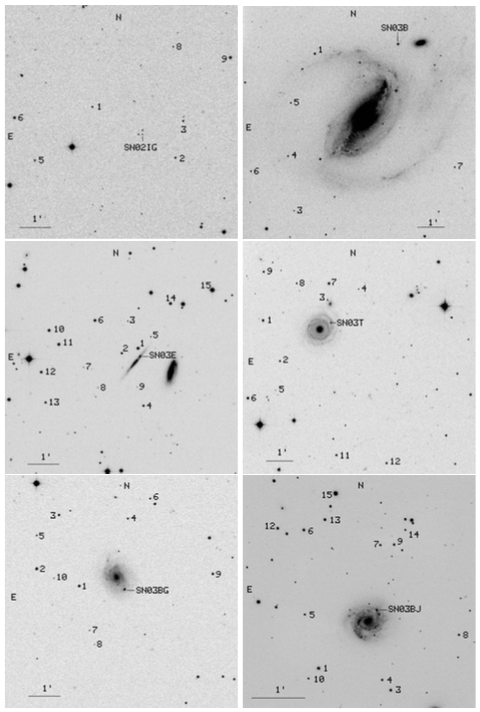}
\caption{Supernovae 2002ig, 2003B, 2003E, 2003T, 2003bg, and 2003bj.}
\end{figure*}
\begin{figure*}
\renewcommand{\thefigure}{\arabic{figure} (Cont.)}
\addtocounter{figure}{-1}
\centering
\includegraphics*[trim=0.0cm 0.0cm 0.0cm 0.0cm, clip=true,width=0.9\hsize]{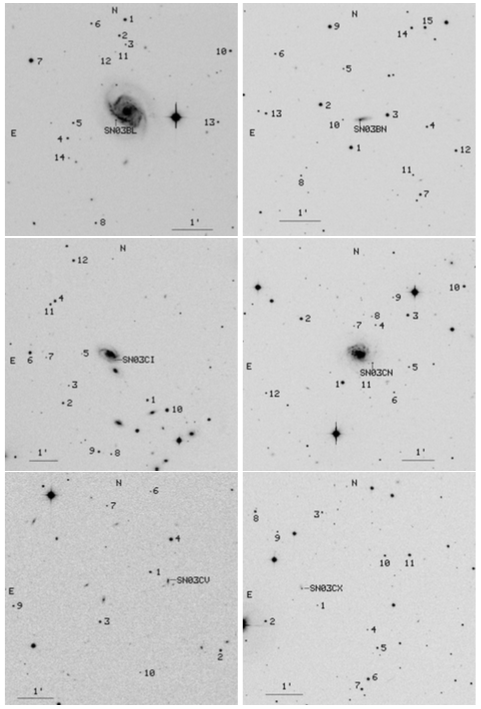}
\caption{Supernovae 2003bl, 2003bn, 2003ci, 2003cn, 2003cv, and 2003cx.}
\end{figure*}
\begin{figure*}
\renewcommand{\thefigure}{\arabic{figure} (Cont.)}
\addtocounter{figure}{-1}
\centering
\includegraphics*[trim=0.0cm 0.0cm 0.0cm 0.0cm, clip=true,width=0.9\hsize]{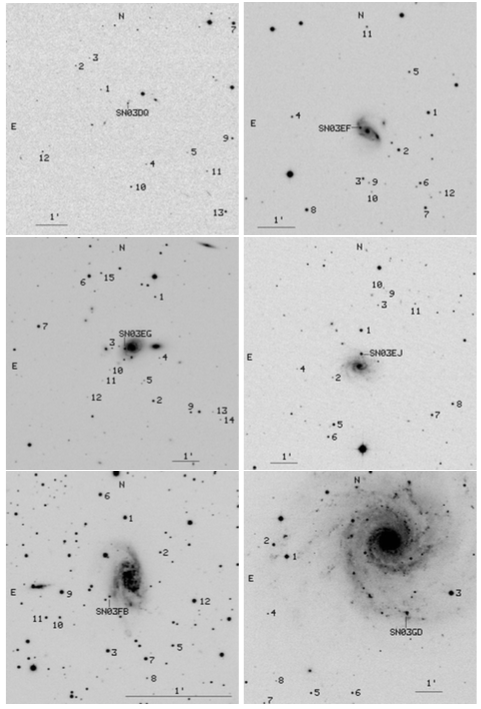}
\caption{Supernovae 2003dq, 2003ef, 2003eg, 2003ej, 2003fb, and 2003gd.}
\end{figure*}
\begin{figure*}
\renewcommand{\thefigure}{\arabic{figure} (Cont.)}
\addtocounter{figure}{-1}
\centering
\includegraphics*[trim=0.0cm 0.0cm 0.0cm 0.0cm, clip=true,width=0.9\hsize]{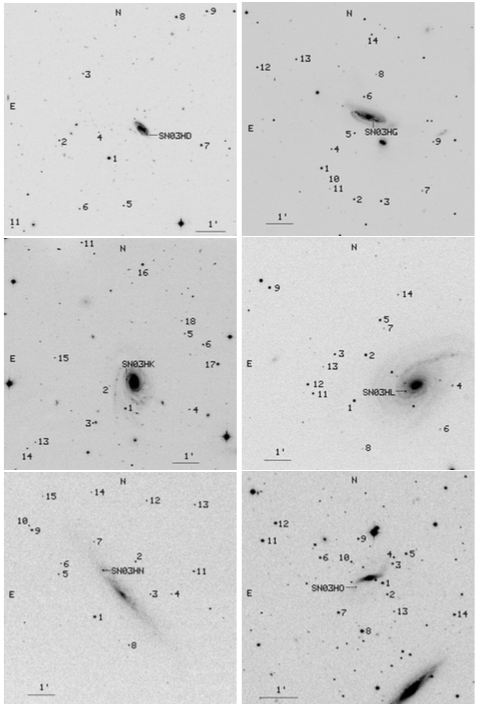}
\caption{Supernovae 2003hd, 2003hg, 2003gk, 2003hl, 2003hn, and 2003ho.}
\end{figure*}
\begin{figure*}
\renewcommand{\thefigure}{\arabic{figure} (Cont.)}
\addtocounter{figure}{-1}
\centering
\includegraphics*[trim=0.0cm 0.0cm 0.0cm 0.0cm, clip=true,width=0.9\hsize]{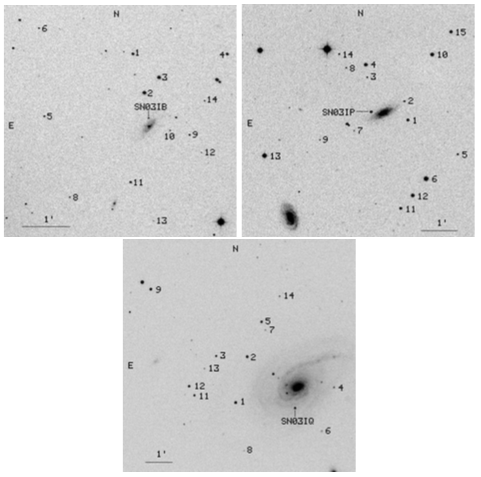}
\caption{Supernovae 2003ib, 2003ip, and 2003iq.}
\end{figure*}

\begin{figure*}
\centering
\includegraphics*[width=0.9\hsize]{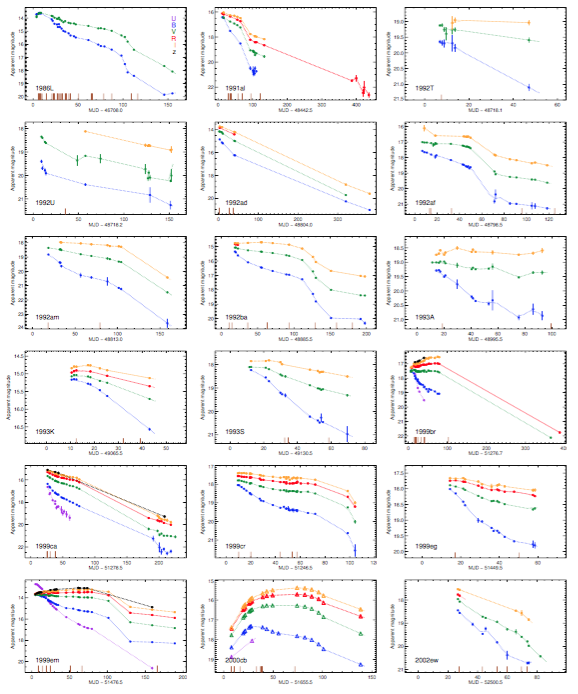}
\caption{MW Extinction-corrected apparent $UBVRIz$ light-curves of the 51 SNe II. The explosion epoch in each panel is that estimated in Guti\'errez et al. (in prep.). Photometric errors are plotted and are usually smaller than the symbol. Type IIb, IIn, and SN 1987A-like SNe II are plotted with unfilled triangles. Lines correspond to spline fits of the data. Vertical brown tick marks represent the epochs of the available spectra.}
\label{lc.fig}
\end{figure*}
\begin{figure*}
\renewcommand{\thefigure}{\arabic{figure} (Cont.)}
\addtocounter{figure}{-1}
\centering
\includegraphics*[width=0.9\hsize]{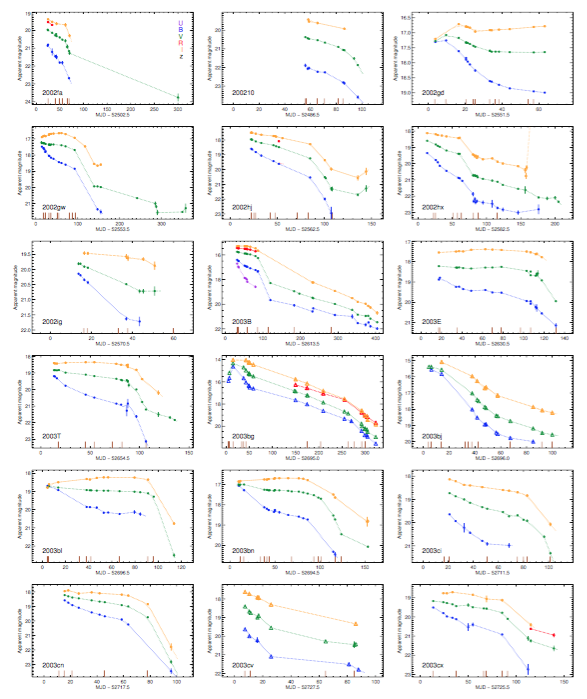}
\caption{}
\end{figure*}
\begin{figure*}
\renewcommand{\thefigure}{\arabic{figure} (Cont.)}
\addtocounter{figure}{-1}
\centering
\includegraphics*[width=0.9\hsize]{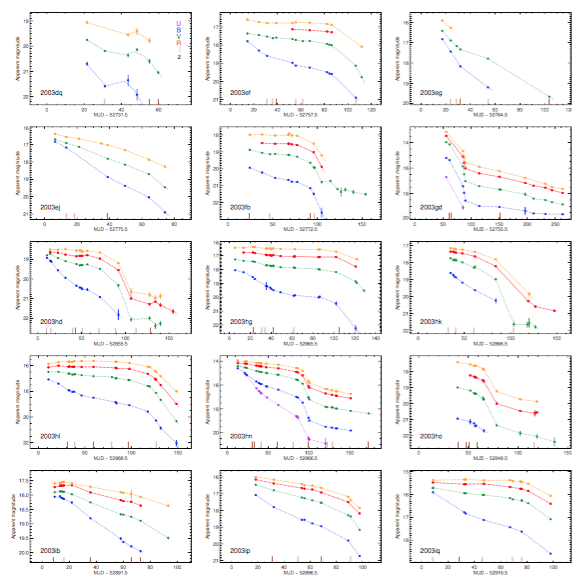}
\caption{}
\end{figure*}

\begin{figure*}
\centering
\includegraphics*[trim=1.5cm 0.0cm 1.0cm 0.0cm, clip=true,width=0.49\hsize]{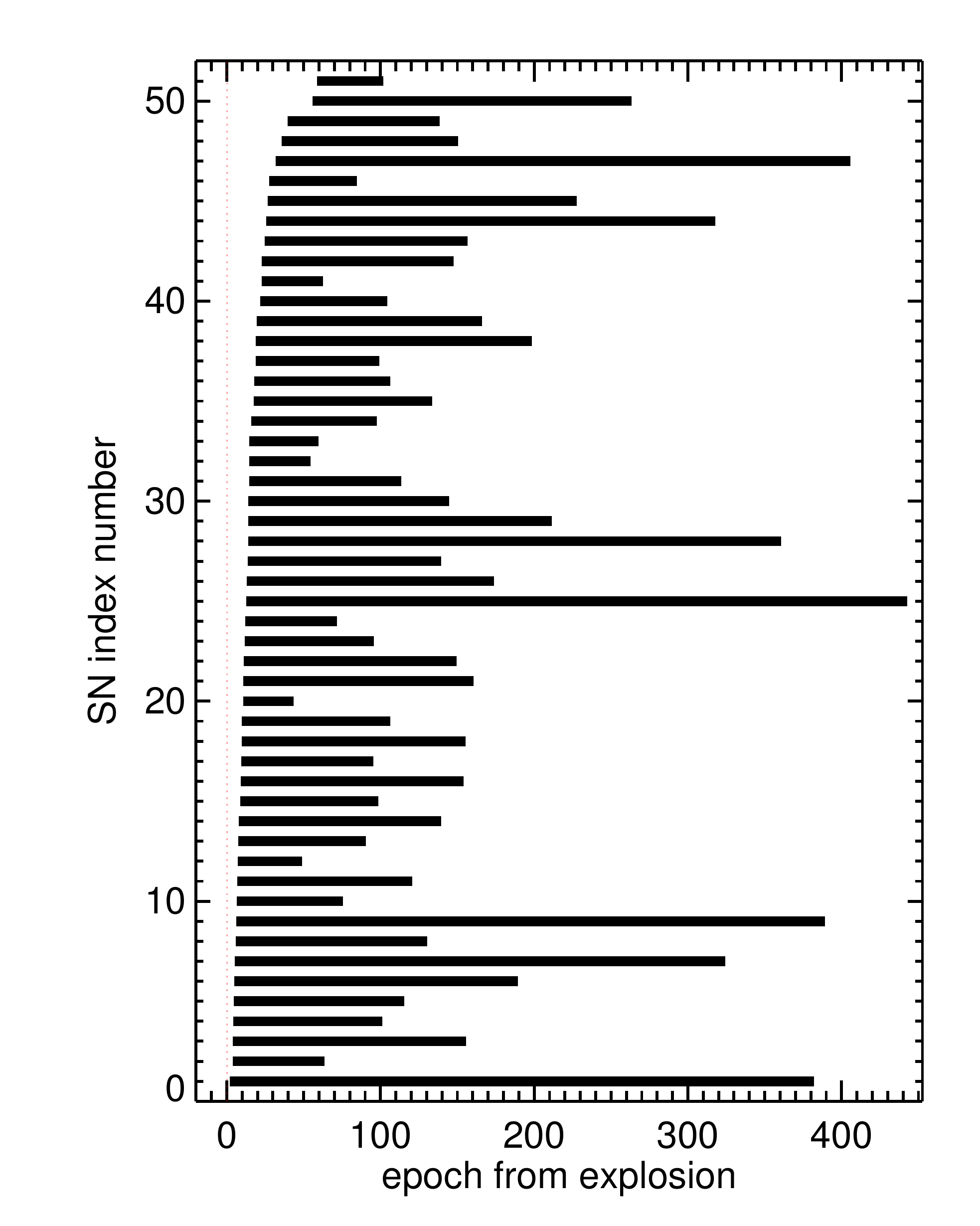}
\includegraphics*[trim=2.5cm 0.0cm 0.0cm 0.0cm, clip=true,width=0.49\hsize]{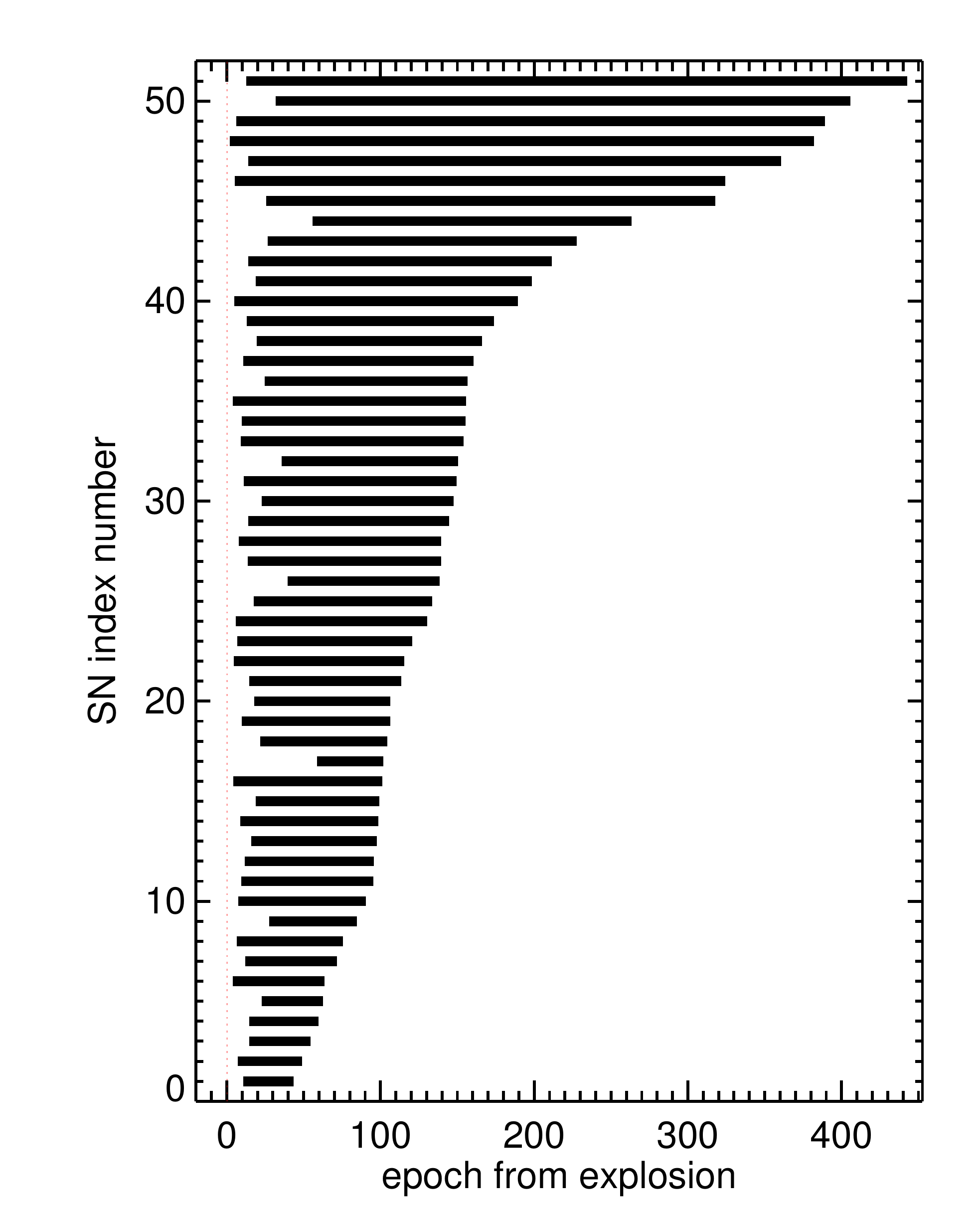}
\caption{Light-curve coverage for the 51 SN II presented in this work, sorted by increasing first photometric observation (left) and by last photometric epoch (right), all measured from explosion. Each horizontal bar represents the coverage of one SN. The vertical red dotted line represents the explosion day. The average epoch of the first observation is 14.7 $\pm$ 11.1 days (median 11.7 days) and the last observation is 158.2 $\pm$ 98.7 days (median 139.5 days).}
\label{cov.fig}
\end{figure*}

\begin{figure*}
\centering
\includegraphics*[trim=1.2cm 0.0cm 0.3cm 0.7cm, clip=true,width=0.93\hsize]{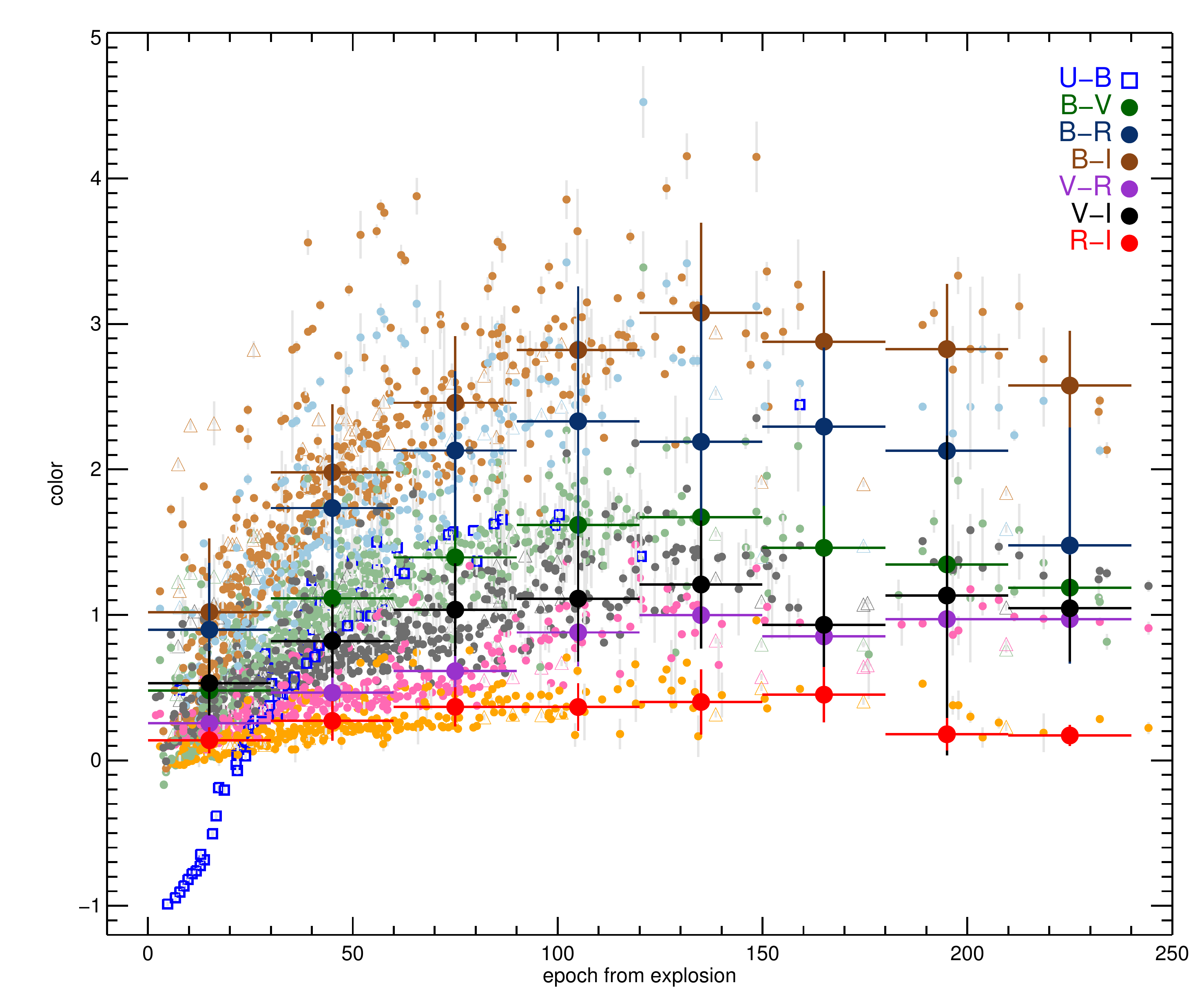}
\includegraphics*[trim=0cm 0cm 0cm 0cm, clip=true,width=0.93\hsize]{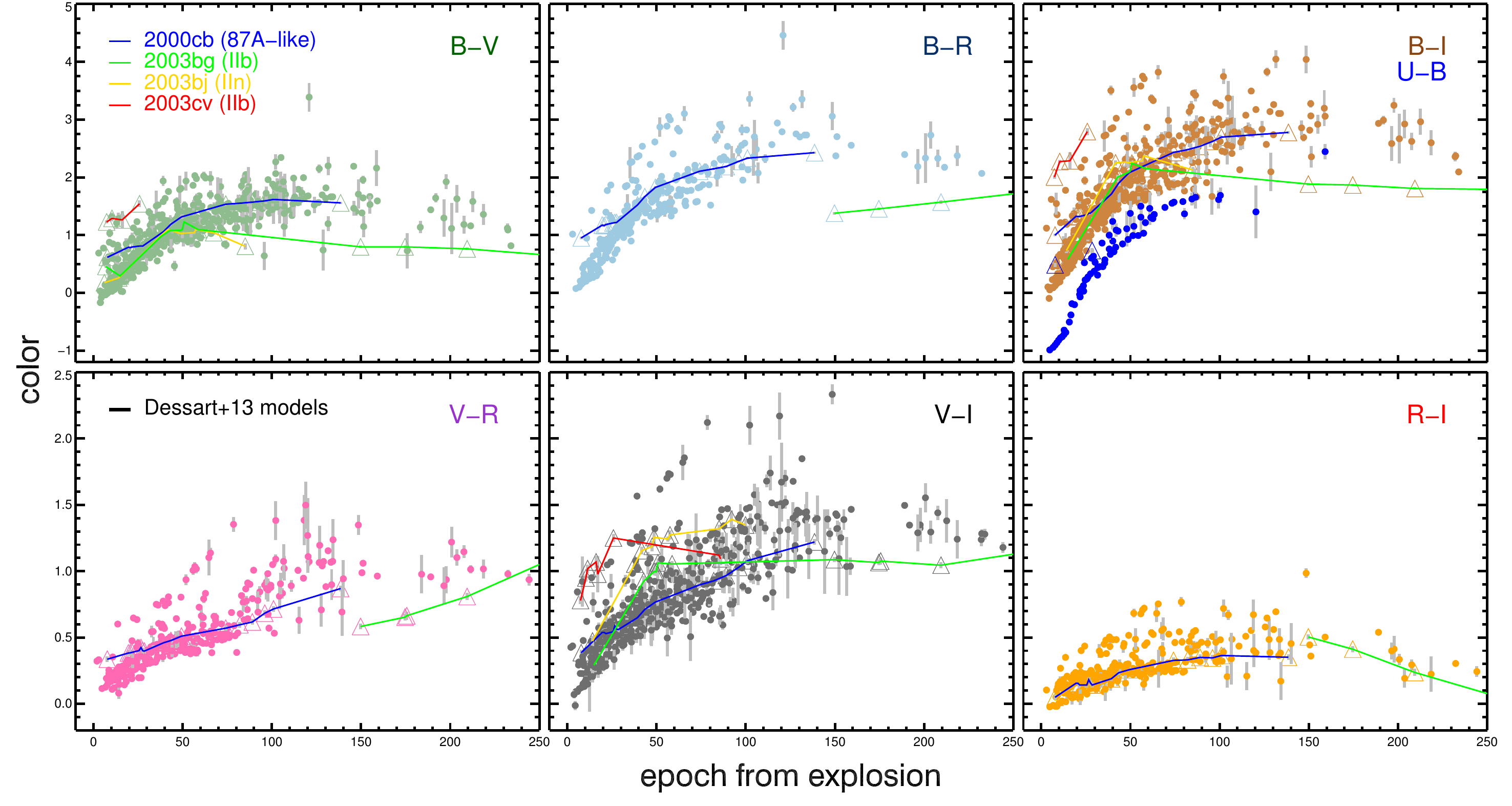}
\caption{Color evolution diagrams. {\it Top:} Small dots represent individual measurements and big dots indicate average colors in bins of 30 days. All of them increase until reaching a maximum around 100-150 days from explosion, and then saturate or start decreasing. Blue open squares are the $(U-B)$ color data. {\it Bottom:} Individual color panels. Color measurements of the SN 1987A-like, IIb, and IIn are plotted with empty triangles, and in $(B-I)$ they are shown in solid lines.}
\label{col.fig}
\end{figure*}

\begin{figure*}
\centering
\includegraphics*[trim=0.0cm 1.0cm 0.3cm 0.5cm, clip=true,width=0.335\hsize]{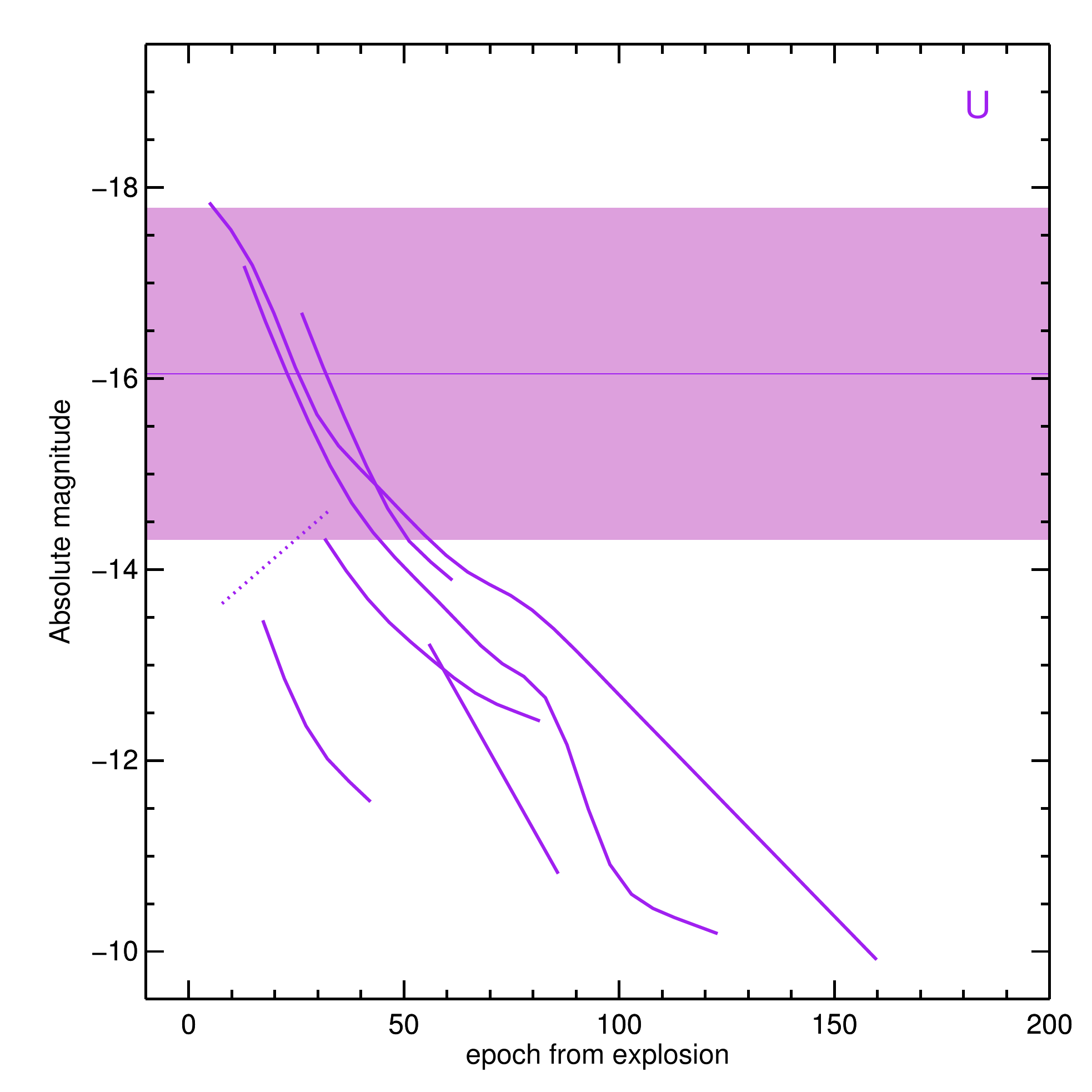}
\includegraphics*[trim=1.5cm 1.0cm 0.3cm 0.5cm, clip=true,width=0.31\hsize]{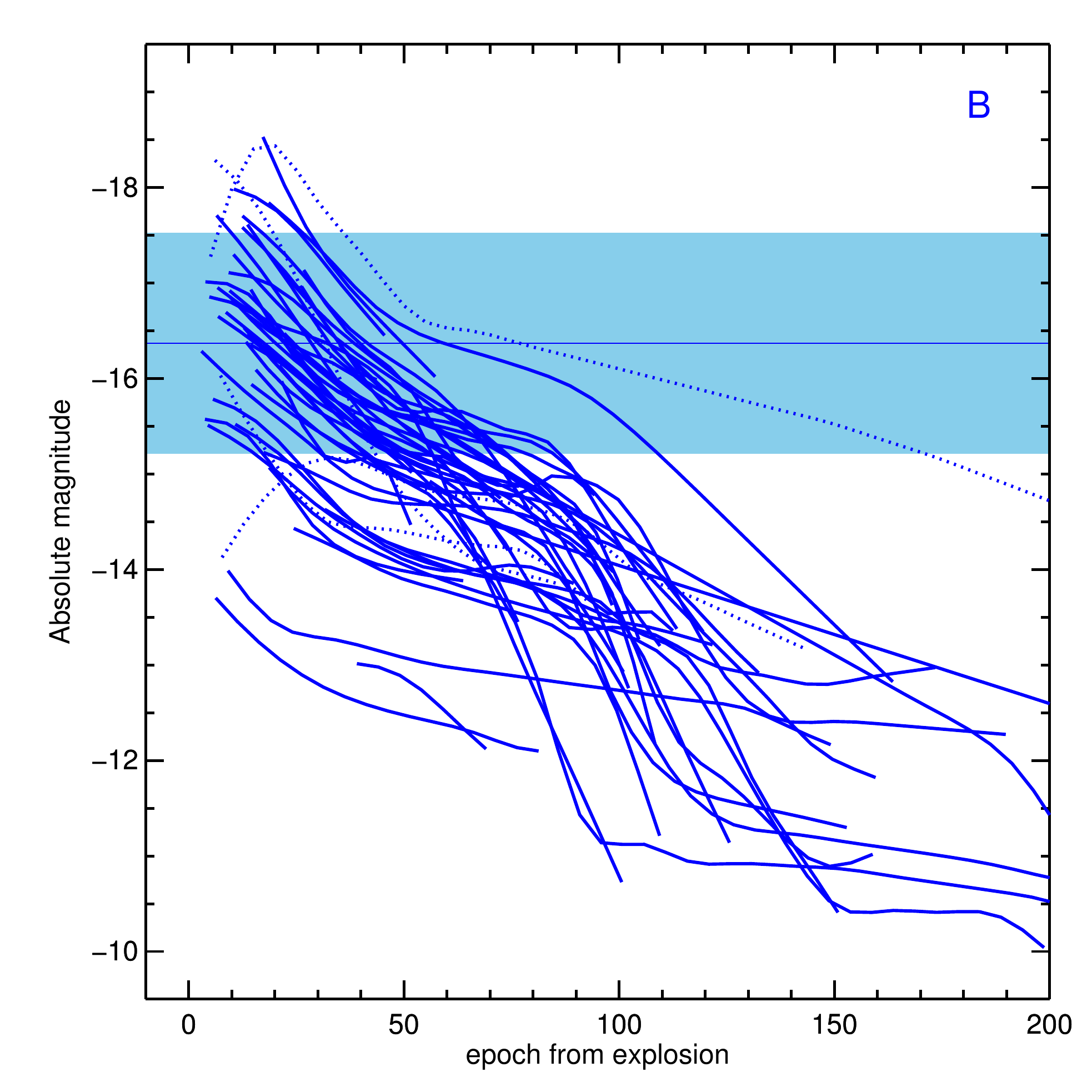}
\includegraphics*[trim=1.5cm 1.0cm 0.3cm 0.5cm, clip=true,width=0.31\hsize]{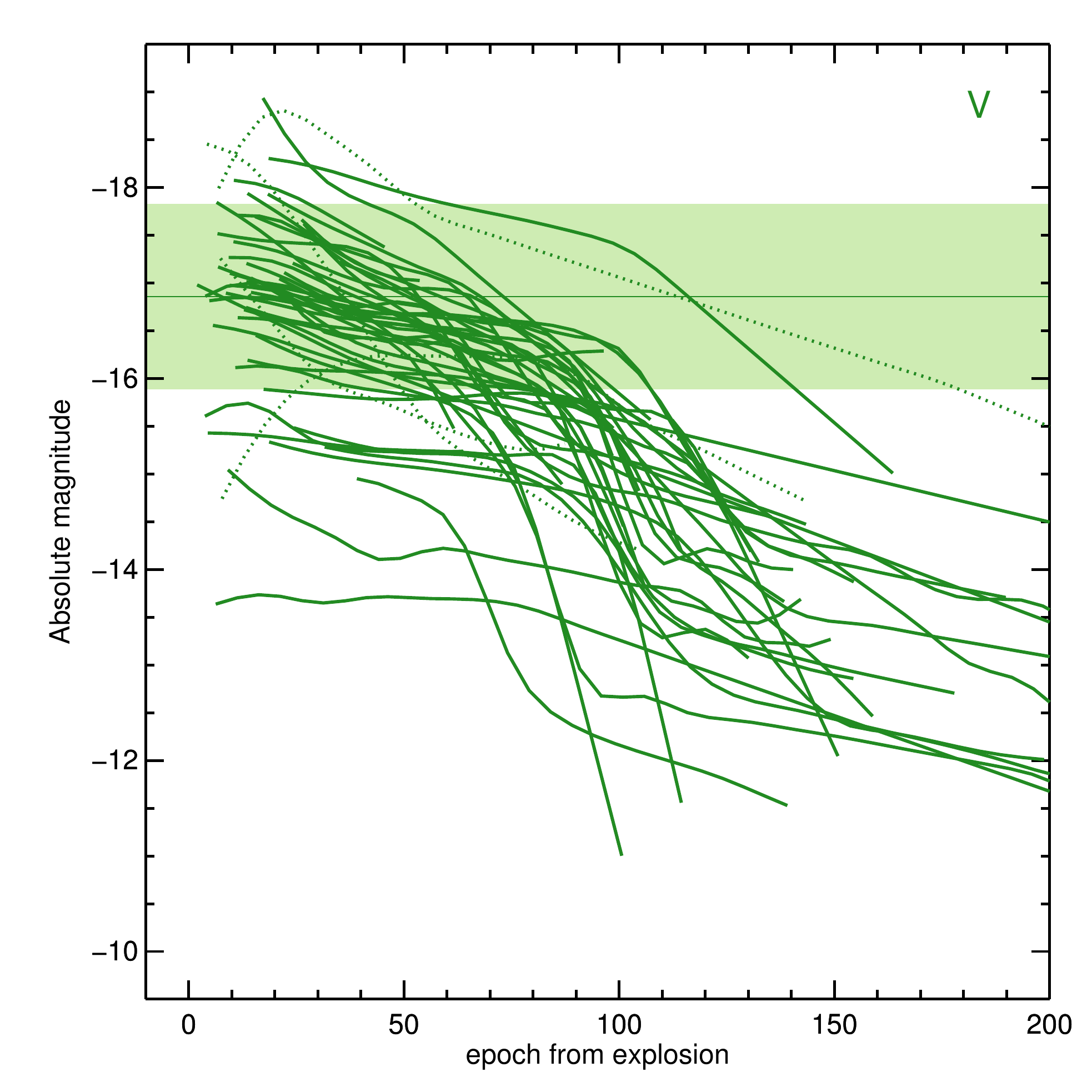}\\
\includegraphics*[trim=0.0cm 0.0cm 0.3cm 0.5cm, clip=true,width=0.335\hsize]{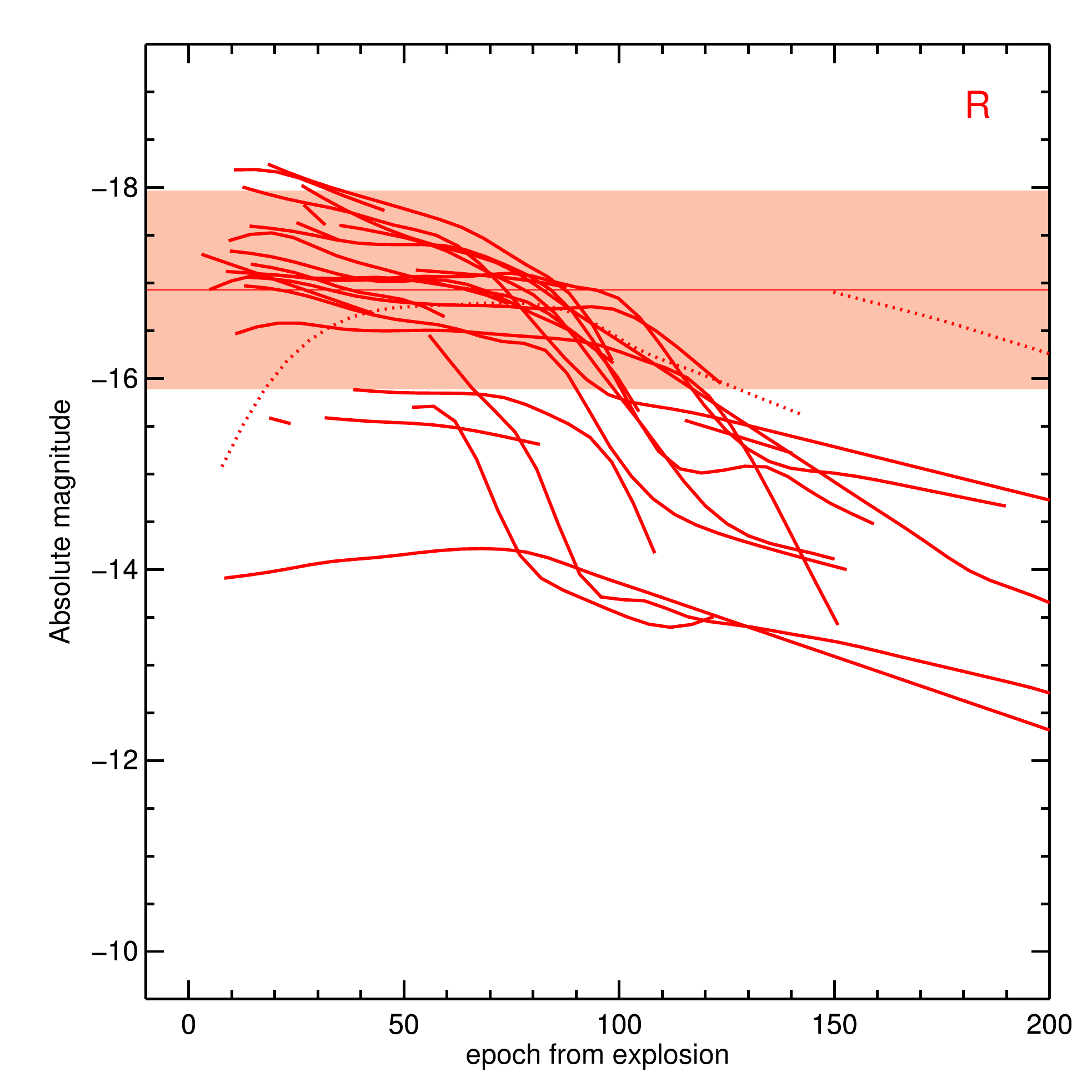}
\includegraphics*[trim=1.5cm 0.0cm 0.3cm 0.5cm, clip=true,width=0.31\hsize]{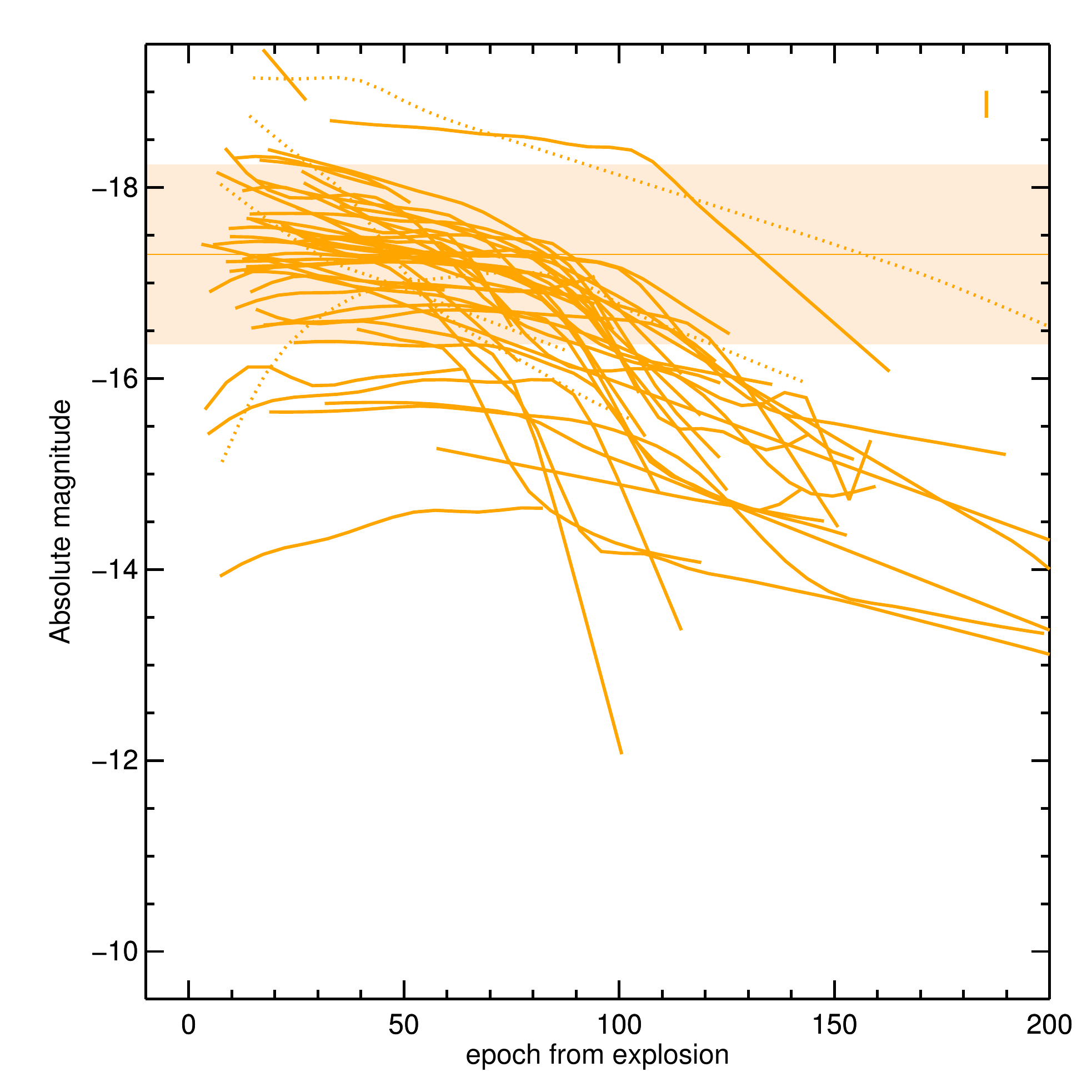}
\includegraphics*[trim=1.5cm 0.0cm 0.3cm 0.5cm, clip=true,width=0.31\hsize]{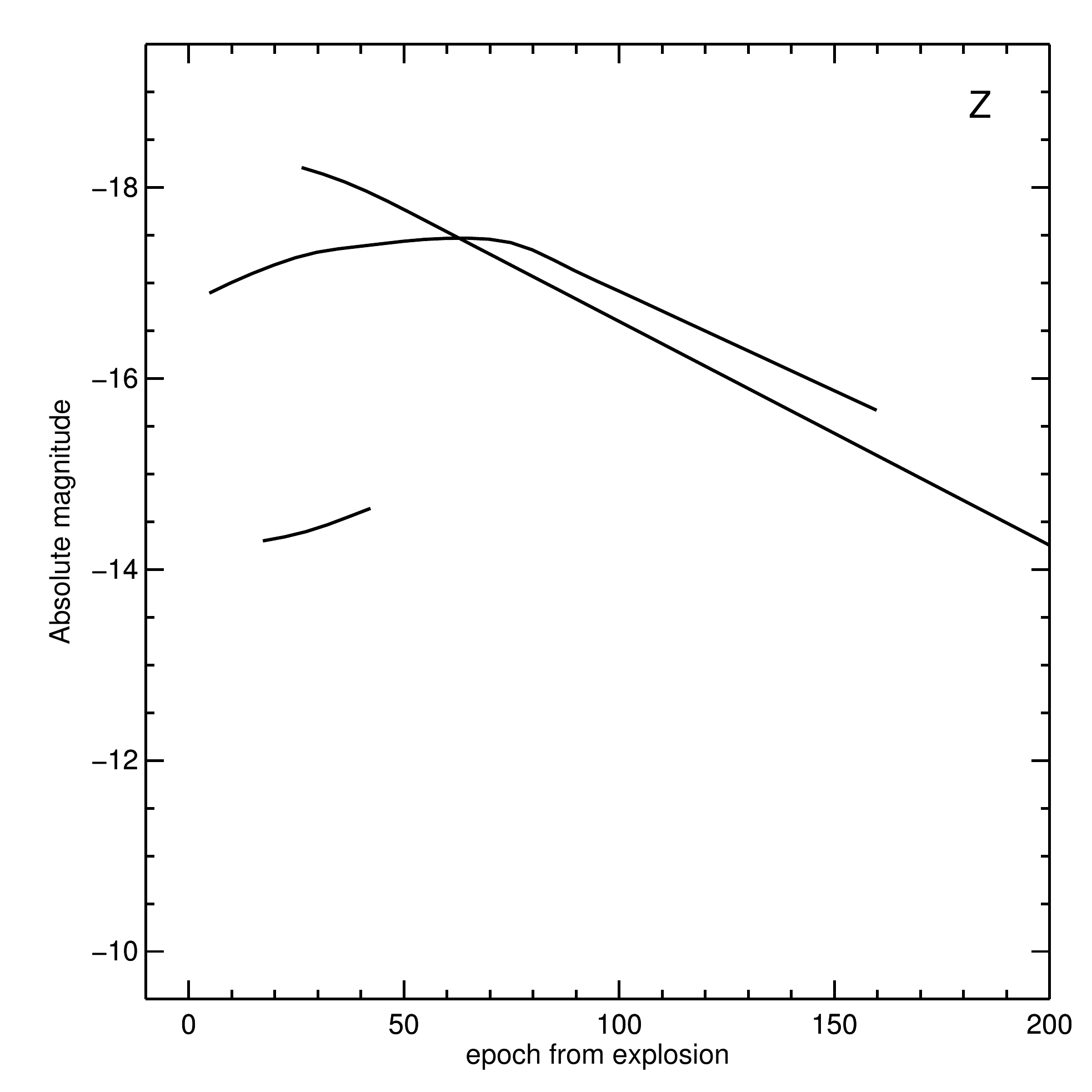}
\caption{Absolute magnitude light-curves for all SN presented here overplotted in separate panels by different bands. Curves have been smoothed using 3rd order spline polynomials. Type IIb, IIn, and peculiar SN II are shown with dotted lines. The horizontal lines and the colored stripes correspond to the peak average absolute magnitudes and their 1$\sigma$ deviations. }
\label{allabs.fig}
\end{figure*}

\begin{figure*}
\centering
\includegraphics*[trim=1.1cm 1.5cm 0.3cm 0.5cm, clip=true,width=0.49\hsize]{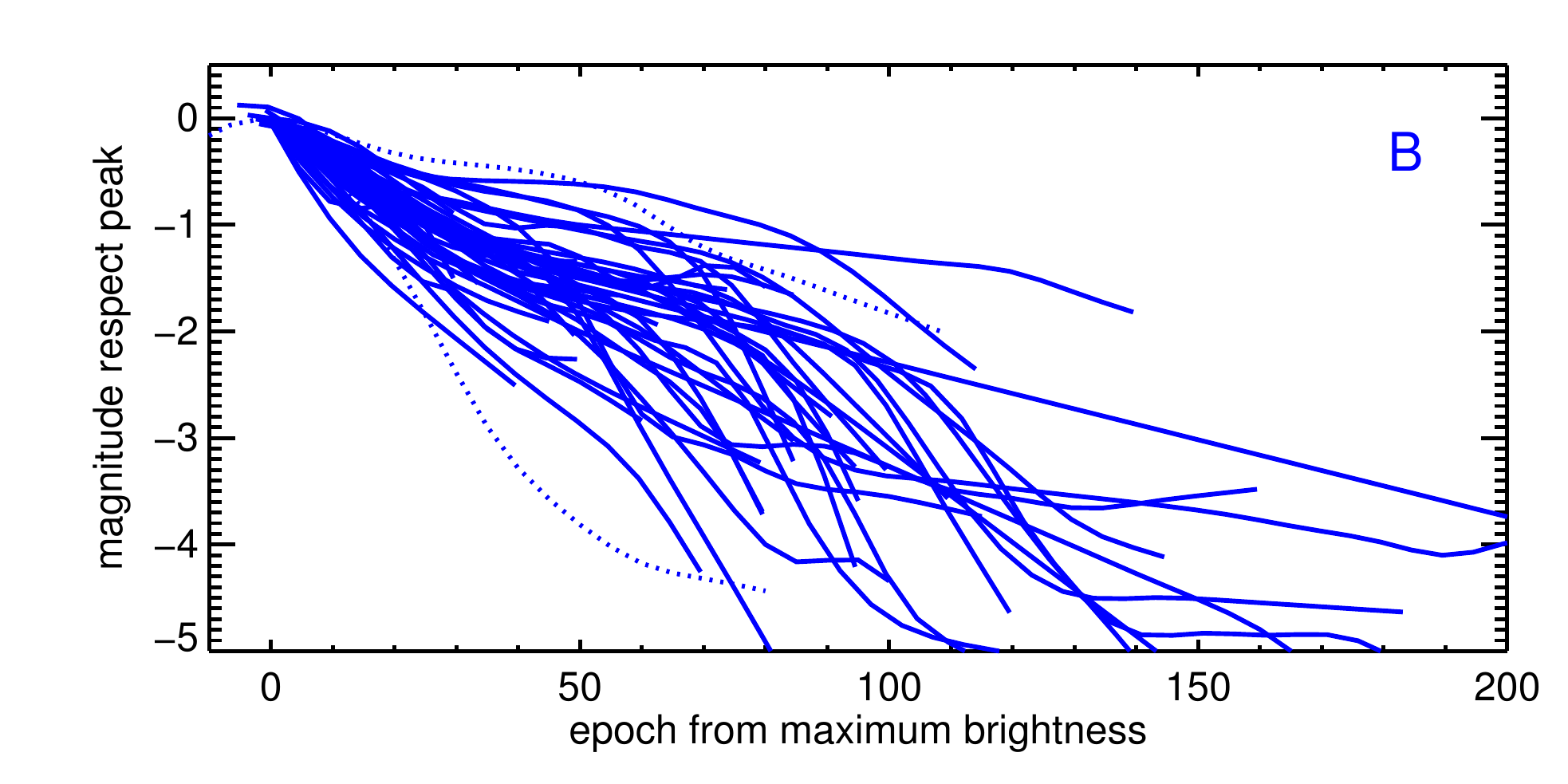}
\includegraphics*[trim=1.8cm 1.5cm -0.4cm 0.5cm, clip=true,width=0.49\hsize]{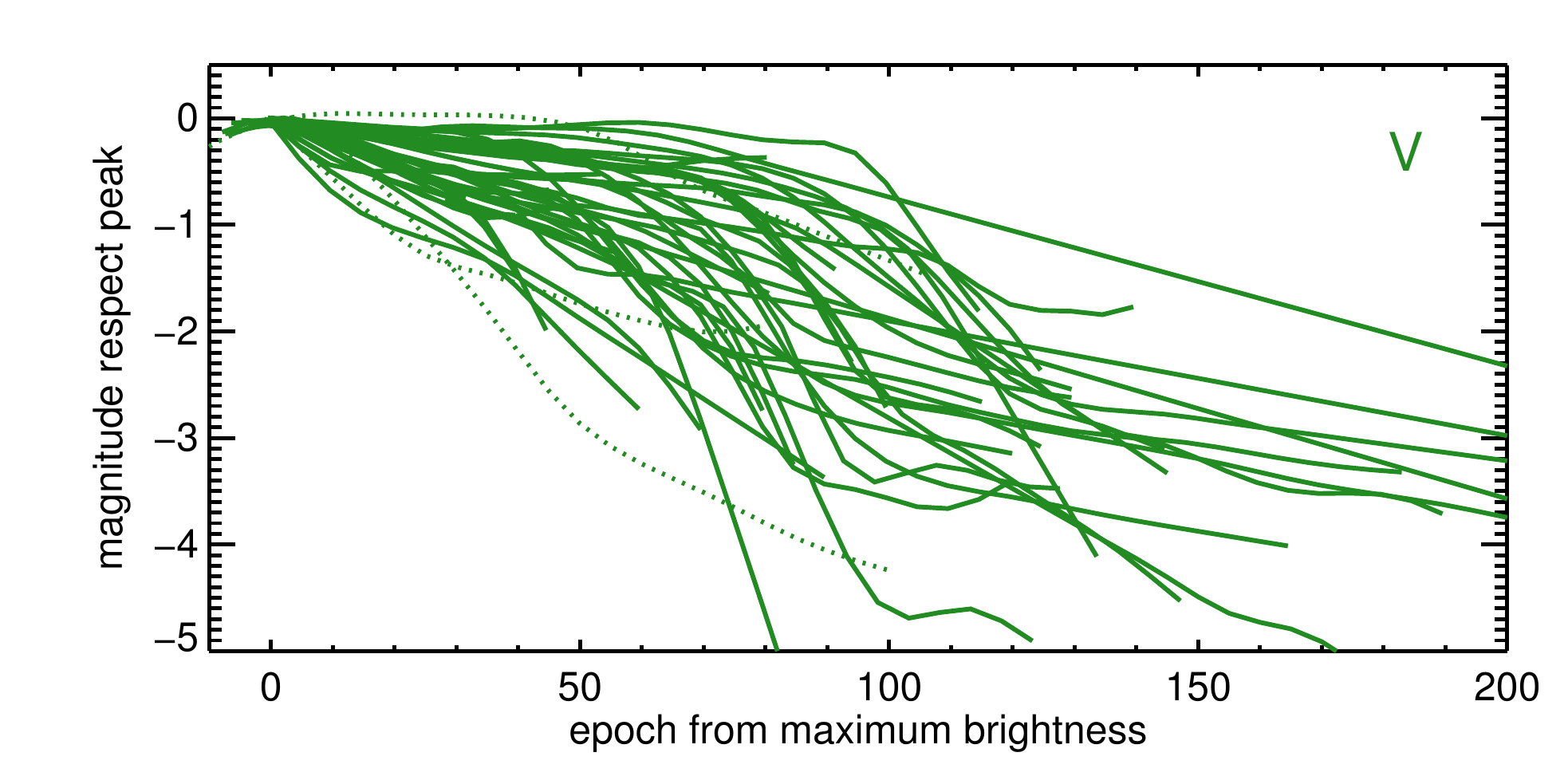}\\
\includegraphics*[trim=1.1cm 0.0cm 0.3cm 0.5cm, clip=true,width=0.49\hsize]{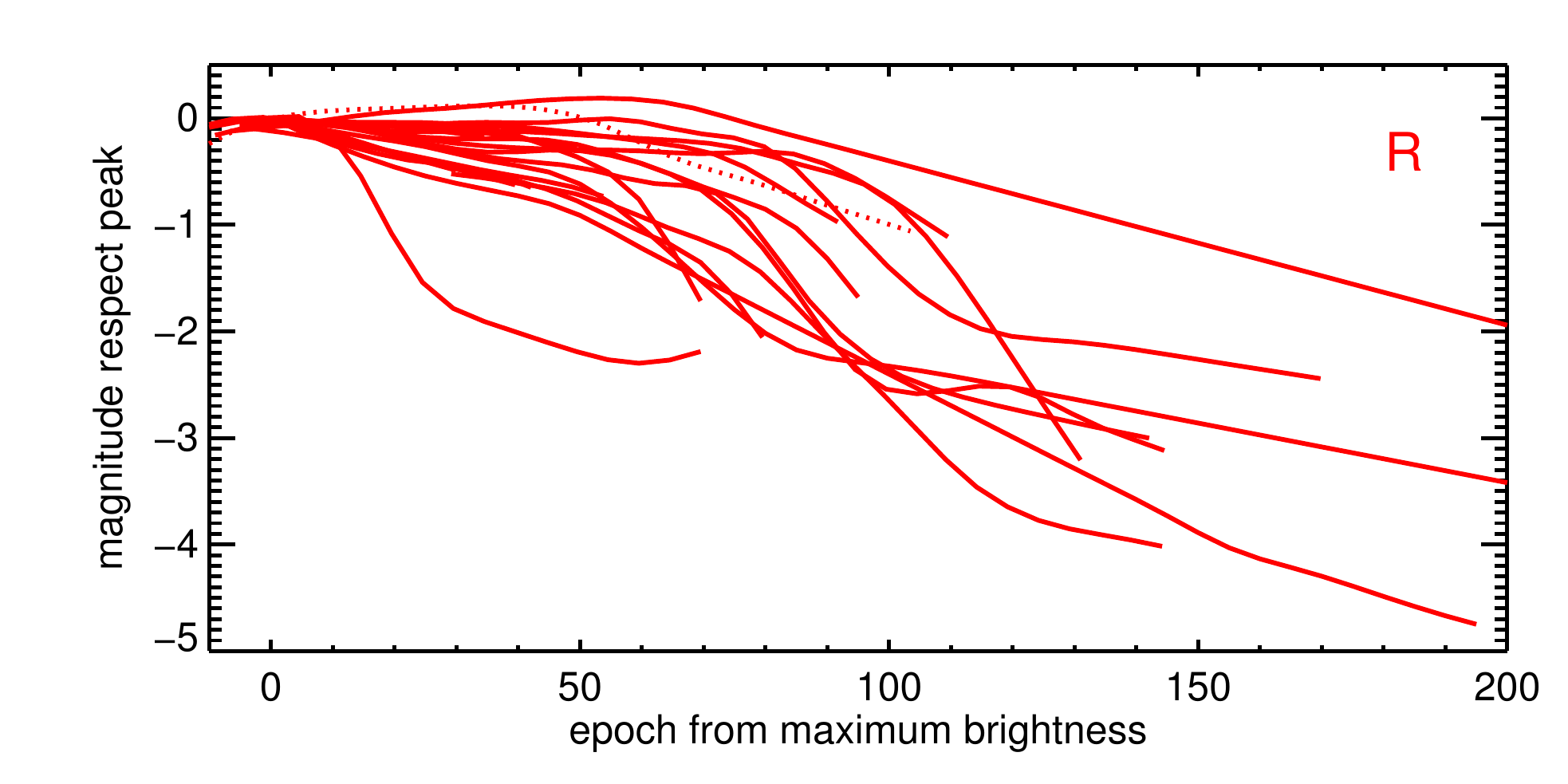}
\includegraphics*[trim=1.8cm 0.0cm -0.4cm 0.5cm, clip=true,width=0.49\hsize]{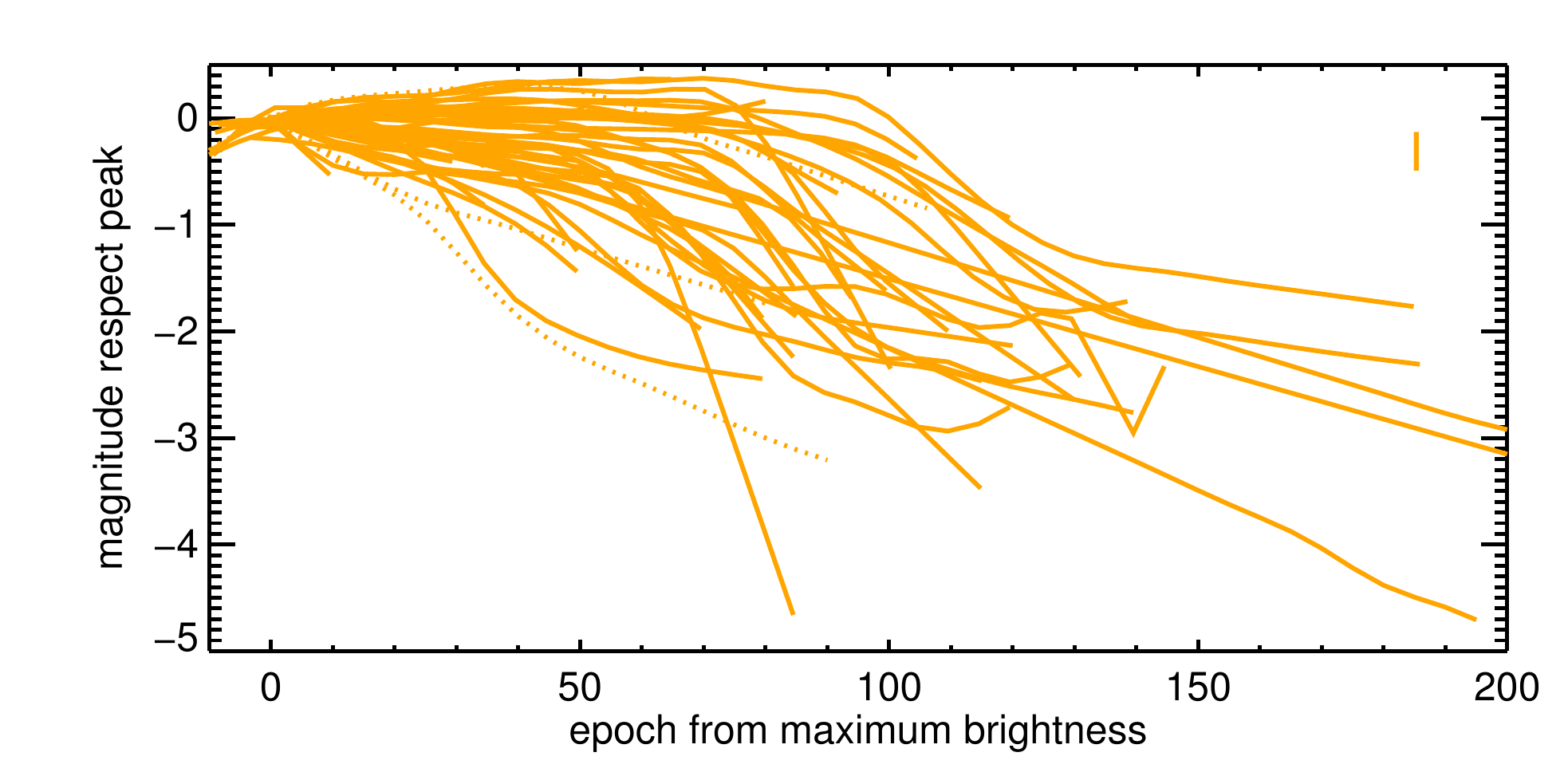}
\caption{Magnitude light-curves referenced to the epoch of the maximum brightness and normalized to the peak. Type IIb, IIn, and peculiar SN II are plotted with dotted lines. All panels show a continuous distribution between fast and slow decliners in all bands.}
\label{allabsn.fig}
\end{figure*}

\begin{figure*}
\centering
\includegraphics*[trim=1.6cm 1.0cm 0.5cm 0.9cm, clip=true,width=0.7\hsize]{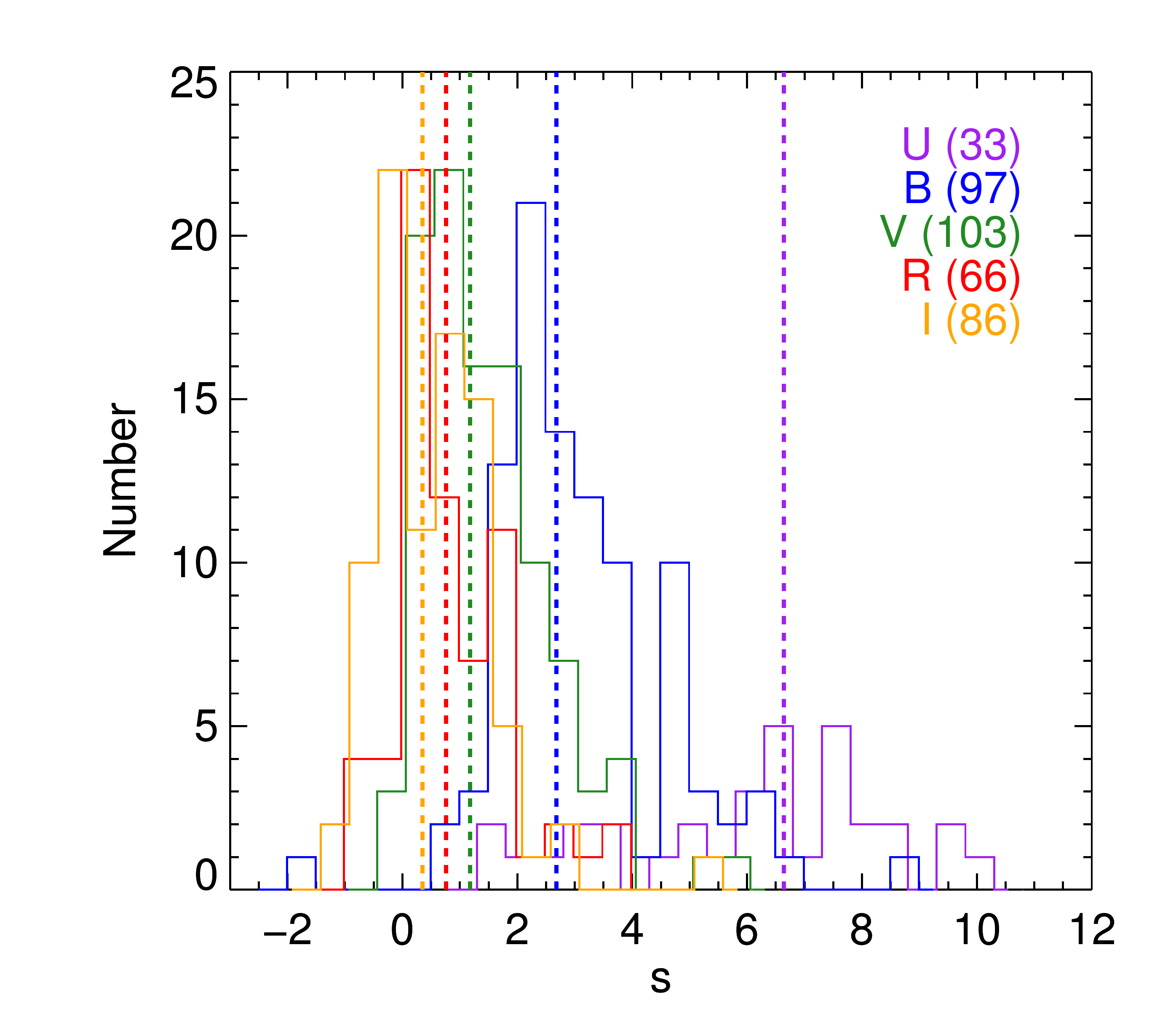}
\caption{Distribution of the slopes of the plateau in each filter for the expanded sample of 114 SN II. 
The median of the distribution decreases with redder filters.}
\label{sdist.fig}
\end{figure*}

\begin{figure*}
\centering
\includegraphics*[trim=1.6cm 1.0cm 0.5cm 0.9cm, clip=true,width=0.49\hsize]{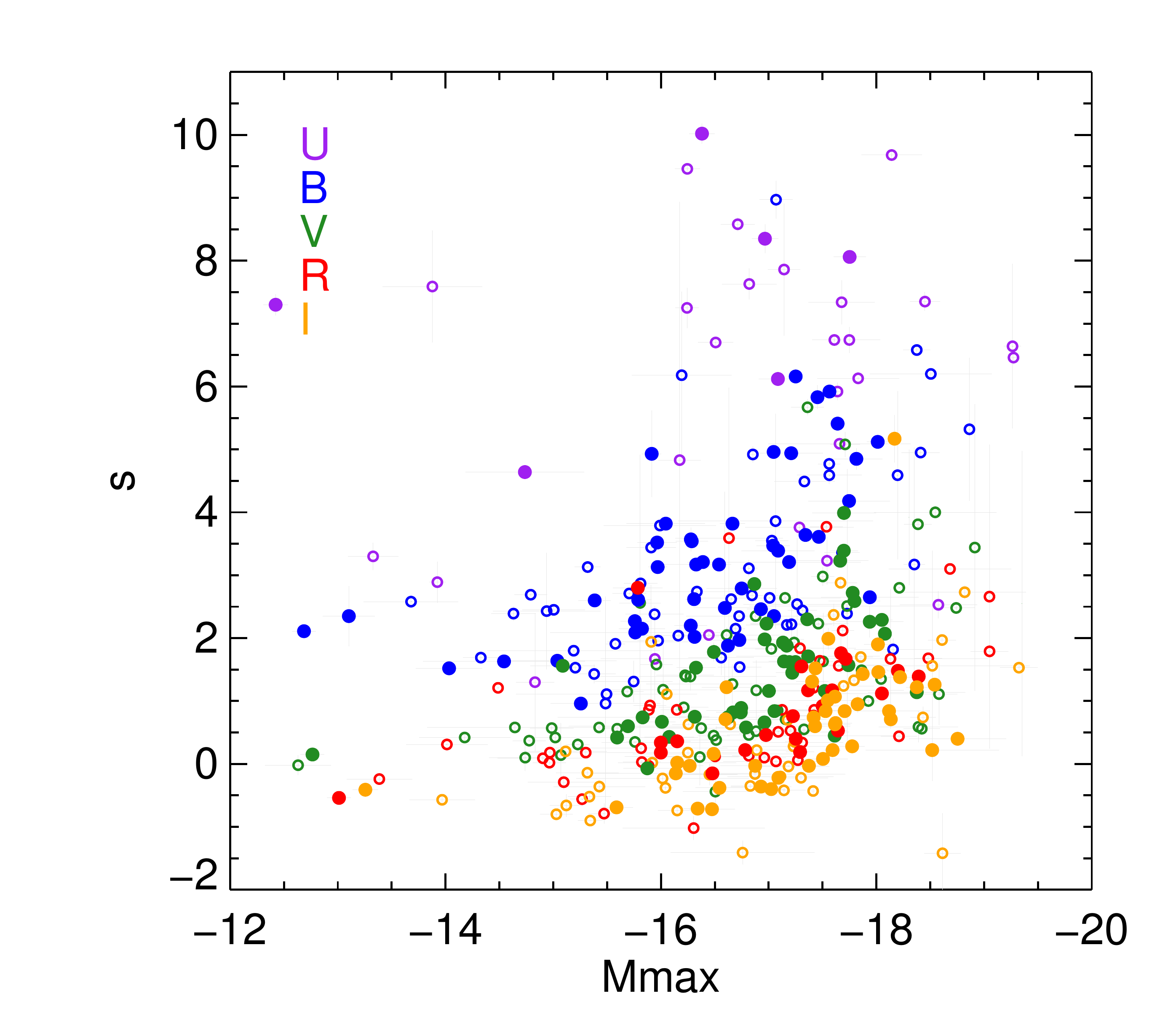}
\includegraphics*[trim=1.6cm 1.0cm 0.5cm 0.9cm, clip=true,width=0.49\hsize]{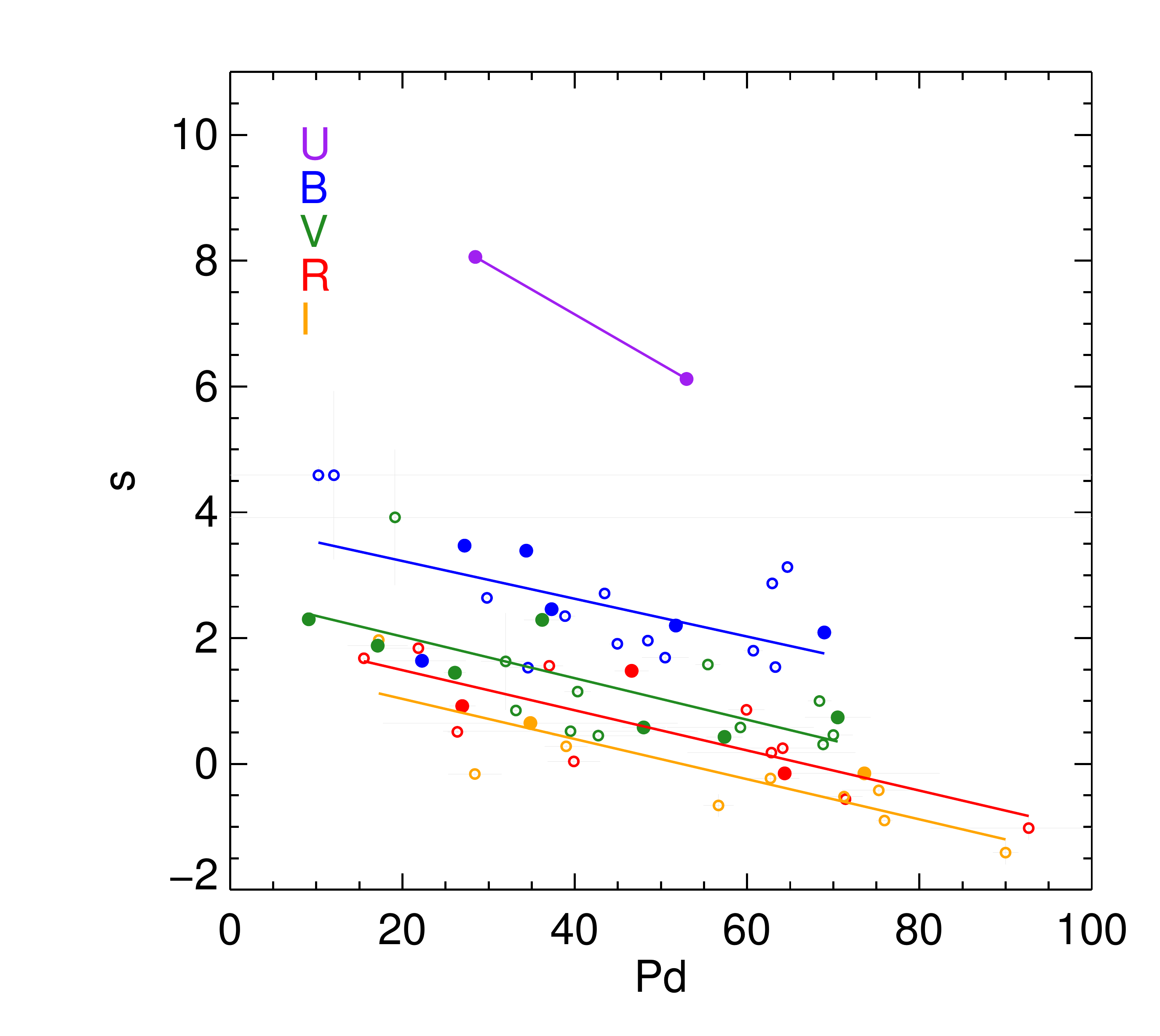}
\caption{Left: Slope of the plateau vs. peak absolute magnitude for all $UBVRI$ bands. Filled circles correspond to SN II presented in this work, while open circles are other objects from the literature. Peculiar objects such as SN IIb, IIn, and 1987A-like are excluded from this analysis. 
Right: Relation between the plateau duration and the post-maximum brightness decline. A trend can be seen in all bands indicating shorter plateaus for faster declining SN. Solid lines indicate linear fits to the points.}
\label{mmaxs.fig}
\end{figure*}

\begin{deluxetable} {lcllrcllc}
\tabletypesize{\scriptsize} 
\tablecolumns{7}
\tablenum{1}
\tablewidth{0pc}
\tablecaption{Type II Supernovae \label{snlist.tab}}
\tablehead{
\colhead{SN name} &
\colhead{Host Galaxy} &
\colhead{Host type\tablenotemark{a}}  &
\colhead{RA(J2000)} &
\colhead{DEC(J2000)} &
\colhead{$z_{helio}$\tablenotemark{b}} &
\colhead{$E(B-V)$} &
\colhead{DM} &
\colhead{Survey\tablenotemark{c}}}
\startdata                                                                                                                          
1986L &NGC 1559                    &SB(s)cd      &04 17 31.2&--62 47 07&0.00435    &0.030 & 30.72(0.34)\tablenotemark{d}&1\\     
1991al&2MASX J19422191--5506275    &S            &19 42 24.0&--55 06 23&0.01525$^*$&0.051 & 34.12(0.14)                 &2\\     
1992T &2MASX J13425875--3153105    &SB           &13 43 01.5&--31 53 36&0.03898$^*$&0.048 & 36.30(0.05)                 &2\\     
1992U &ESO 074-- G 004             &SBc          &20 40 46.5&--70 41 33&0.01086$^*$&0.051 & 33.39(0.06)                 &2\\     
1992ad&NGC 4411B                   &SAB(s)cd     &12 26 49.6& +08 52 39&0.00424    &0.030 & 31.13(0.80)\tablenotemark{d}&2\\     
1992af&ESO 340-- G 038             &Sc           &20 30 40.2&--42 18 35&0.01847    &0.052 & 34.52(0.01)                 &2\\     
1992am&MCG --01--04--039           &S            &01 25 02.7&--04 39 01&0.04799$^*$&0.049 & 36.66(0.05)                 &2\\     
1992ba&NGC 2082                    &SB(r)b       &05 41 47.1&--64 18 01&0.00395    &0.058 & 30.41(0.80)\tablenotemark{d}&2\\
\enddata
\tablenotetext{a}{From NASA/IPAC Extragalactic Database (NED).}
\tablenotetext{b}{From NED except those marked with $^*$ which have been measured here.}
\tablenotetext{c}{(1) Cerro Tololo Supernova Survey; (2) Cal\'an/Tololo Supernova Program (C\&T); 
(3) Supernova Optical and Infrared Survey (SOIRS); (4) Carnegie Type II Supernovae Survey (CATS).} 
\tablenotetext{d}{From NED using only Tully-Fisher measurements. Errors are standard deviation of the mean.}
\tablenotetext{e}{No Tully-Fisher measurements available. DM measured used SN II data.}
\tablenotetext{f}{Although its redshift is lower than 0.01, it has been measured using Tully-Fisher measurements.}
\tablenotetext{g}{Peculiar 1987A-like SN.}
\tablenotetext{h}{Type IIb SN.}
\end{deluxetable}

\begin{deluxetable} {ccc}
\tablecolumns{3}
\tablenum{2}
\tablewidth{0pc}
\tablecaption{Discovery and classification references for the 51 Type II Supernovae \label{refs.tab}}
\tablehead{
\colhead{SN name} &
\colhead{Discovery} &
\colhead{Classification}}
\startdata                                                                                                                          
1986L & \cite{1986IAUC.4260....1E} & \cite{1986IAUC.4262....2L} \\
1991al& \cite{1991IAUC.5310....1W} & \cite{1991IAUC.5312....2B} \\
1992T & \cite{1992IAUC.5496....1M} &       ---\tablenotemark{a}                   \\
1992U & \cite{1992IAUC.5499....1M} & \cite{1992IAUC.5521....1P} \\
1992ad& \cite{1992IAUC.5552....1M} & \cite{1992IAUC.5552....1M} \\
1992af& \cite{1992IAUC.5554....1W} & \cite{1992IAUC.5558....3D} \\
1992am& \cite{1992IAUC.5570....2P} & \cite{1992IAUC.5570....2P} \\
1992ba& \cite{1992IAUC.5625....2E} & \cite{1992IAUC.5625....2E} \\
\enddata
\tablenotetext{a}{Not reported in the literature. Classification provided here for the first time.}
\tablenotetext{b}{The CATS survey performed the follow up of SN 200210, which was discovered by the SN Factory \citep{2004NewAR..48..637W} and was never reported to the International Astronomical Union (IAU) to provide an official designation.}
\end{deluxetable}

\begin{deluxetable} {ccc}
\tablecolumns{3}
\tablenum{3}
\tablewidth{0pc}
\tablecaption{Telescope and Instruments Employed for Photometry \label{telescopes.tab}}
\tablehead{
\colhead{Observatory} &
\colhead{Telescope} &
\colhead{Instrument} }
\startdata

Cerro Tololo   & 0.9m          & Photoelectric Photometer \\
Cerro Tololo   & 0.9m          & CCD Camera     \\
Cerro Tololo   & 1.0m          & CCD Camera     \\
Cerro Tololo   & 1.5m          & CCD Camera     \\
Cerro Tololo   & Blanco 4m     & CCD Camera     \\
Las Campanas   & Swope 1m      & CCD Camera     \\
Las Campanas   & du Pont 2.5m  & CCD Camera     \\
Las Campanas   & Baade 6.5m    & CCD Camera     \\
Las Campanas   & Clay  6.5m    & CCD Camera     \\
La Silla       & 1.54m         & CCD Camera     \\
La Silla       & NTT 3.5m      & CCD Camera     \\
Steward        & 2.3m          & CCD Camera     \\
Steward        & 1.6m          & CCD Camera     \\

\enddata



\end{deluxetable}



\begin{deluxetable} {lccccccc}
\tablecolumns{8}
\tablenum{4}
\tablewidth{0pc}
\tablecaption{$UBVRIZ$ Photometric Sequences \label{sequences.tab}}
\tablehead{
\colhead{Star} &
\colhead{$U$} &
\colhead{$B$} &
\colhead{$V$} &
\colhead{$R$} &
\colhead{$I$} &
\colhead{$z$} &
\colhead{$n$} }
\startdata

\cutinhead{SN 1986L}
1 &    \nodata     &   13.530(020) &   12.970(010) &   \nodata     &   \nodata     &   \nodata     & 12      \\
2 &    \nodata     &   15.710(050) &   14.770(020) &   \nodata     &   \nodata     &   \nodata     & 12      \\
3 &    \nodata     &   17.930(050) &   17.160(020) &   \nodata     &   \nodata     &   \nodata     & 12      \\
4 &    \nodata     &   16.360(050) &   15.500(020) &   \nodata     &   \nodata     &   \nodata     & 12      \\
5 &    \nodata     &   13.970(020) &   13.390(010) &   \nodata     &   \nodata     &   \nodata     & 12      \\
\cutinhead{SN 1991al}
1 &   \nodata     &   17.908(008) &   17.036(012) &   16.560(008) &   16.074(010) &   \nodata     &  5 \\
2 &   \nodata     &   17.633(010) &   16.770(006) &   16.273(010) &   15.808(008) &   \nodata     &  5 \\
3 &   \nodata     &   17.428(008) &   16.619(011) &   16.144(012) &   15.710(009) &   \nodata     &  5 \\
\enddata
\tablecomments{Uncertainties given in parenthesis in thousandths of a magnitude. $n$ is the number of nights on which each star was observed.}
\end{deluxetable}



\begin{deluxetable} {lccccccc}
\tablecolumns{8}
\tablenum{5}
\tablewidth{0pc}
\tablecaption{$UBVRIz$ Photometry for 51 Type II Supernovae \label{photometry.tab}}
\tablehead{
\colhead{JD-2,400,000} &
\colhead{$U$} &
\colhead{$B$} &
\colhead{$V$} &
\colhead{$R$} &
\colhead{$I$} &
\colhead{$z$} &
\colhead{Telescope} }
\startdata

\cutinhead{SN 1986L}
46712.4 &    \nodata   & 13.836(015) & 14.013(017) &   \nodata   &   \nodata   &   \nodata    & CTIO-0.9m  \\
46712.4 &    \nodata   & 13.815(015) & 13.940(017) &   \nodata   &   \nodata   &   \nodata    & CTIO-4m   \\
46714.3 &    \nodata   & 13.771(015) & 13.756(017) &   \nodata   &   \nodata   &   \nodata    & CTIO-0.9m \\
46715.3 &    \nodata   & 13.712(015) & 13.711(017) &   \nodata   &   \nodata   &   \nodata    & CTIO-0.9m \\
46716.2 &    \nodata   & 13.764(015) & 13.683(017) &   \nodata   &   \nodata   &   \nodata    & CTIO-0.9m \\
46717.3 &    \nodata   & 13.755(015) & 13.680(017) &   \nodata   &   \nodata   &   \nodata    & CTIO-0.9m \\
46729.3 &    \nodata   & 14.201(015) & 13.912(017) &   \nodata   &   \nodata   &   \nodata    & CTIO-0.9m \\
46730.4 &    \nodata   & 14.313(015) & 13.940(017) &   \nodata   &   \nodata   &   \nodata    & CTIO-0.9m \\
\enddata
\tablecomments{Uncertainties given in parentheses in thousandths of a magnitude.}
\end{deluxetable}


\begin{deluxetable}{lccclcc}
\tabletypesize{\scriptsize} 
\tablecolumns{3}
\tablenum{6}
\tablewidth{0pc}
\tablecaption{Type II Supernovae used in Fig. 8 \label{snlistfig8.tab}}
\tablehead{
\colhead{SN name} &
\colhead{bands} &
\colhead{Photometry reference}}
\startdata                                                                                                
1968L      &$UBV  $&\cite{1974MNRAS.167...13W}                                        \\ 
1970G      &$UBV  $&\cite{1973AA....29...57B,1974JRASC..68...36W}                     \\
1973R      &$ BV  $&\cite{1977AA....56...59C}                                         \\
1986I      &$  VRI$&\cite{1989AJ.....97..186P}                                        \\
1988A      &$ BVR $&\cite{1991AA...247..410B,1993MNRAS.265..471T}                     \\
1988H      &$ BVR $&\cite{1993MNRAS.265..471T}                                        \\
1989C      &$ BVR $&\cite{1993MNRAS.265..471T}                                        \\
1990E      &$ BVRI$&\cite{1993AJ....105.2236S,1994AA...285..147B}                     \\
\enddata
\end{deluxetable}

\end{document}